# From Marx's fundamental equalities to the solving of the transformation problem

# -

# Coherence of the model

Norbert Ankri & Païkan Marcaggi

UNIS, INSERM, Aix-Marseille Université

France

Correspondence should be addressed to Norbert Ankri: norbert.ankri@univ-amu.fr

## SUMMARY

Recently, V. Laure van Bambeke used an original approach to solve the famous problem of transformation of values into production prices by considering that capital reallocation to each department (branch) was part of the problem and required to comply with both Marx's fundamental equalities: between the total sum of values and the total sum of prices, and between the total surplus value and the total profit (in price). Here, we confirm the validity of this consideration in relation with the satisfaction of demand (social need which is able to pay for the given product). However, V. Laure van Bambeke's method of solving an overdetermined system of equations implies that compliance with Marx's fundamental equalities can only be approached, which suggests that Marx's conception would apply approximatively, rather than as a strict law. Here, on the contrary, we show that the transformation problem is solvable from a determined (two-branch models) or an underdetermined system of equations enabling to obtain exact solutions through an algorithm we provide, with no approximation needed. For systems with three branches or more, the solution of the transformation problem belongs to an infinite ensemble and may be determined by an array of factors (e.g., taste of consumers, advertisement, competition between capitalists, politics), accounting for the observed high competition-driven market fluidity. Furthermore, we show that the transformation problem is solvable in the absence of fixed capital, supporting that dealing with the latter is not essential and cannot be seen as a potential flaw of the approach. In these particular cases, the rate of profit can be determined from the eigenvalue of the sociotechnical coefficient matrix, without any information required about the capital allocation to the various branches. Our algorithm enables simulations illustrating how the transient rise in the rate of profit predicted by the Okishio theorem is consistent with the tendency of the rate of profit to fall (TRPF) subsequent to capital reallocation, and how the TRPF is governed by the increase of organic composition, in value. We establish that the long-standing transformation problem is not such a problem since it is easily solved through our algorithm, whatever the number of branches considered. This emphasizes the high coherence of Marx's conception, and its impressive relevance regarding issues such as the TRPF, which have remained intensely debated.

## AIDE-MÉMOIRE

The capitalist's profit is made from the price value of the commodities produced and is proportional to the capital engaged in price. According to the labour theory of value), the value of commodities is proportional to the amount of labour socially necessary for their production and profit has its origin in labour, which is the only mechanism that creates value (1). On the basis of a model of the market economy in commodity production branches (departments), the organic composition of a branch is defined by the ratio of constant capital (the sum of fixed capital, i.e. including machines, buildings, raw materials, etc.) to variable capital (the cost of labour power). The labour theory of value seems to predict that the lower the organic composition of a branch, the higher its rate of profit, since the proportion of value-creating labour is greater. This intuitive prediction is contradicted by the facts, which tend to show that rates of profit are generally similar in all industries and that capital flows preferentially to the most capitalistic branches (with a high organic composition). To resolve this apparent paradox, Marx assumed that commodities are sold at a "market production price" different from their value, and proposed a theory of value and exploitation that is grounded in two conservation laws that we will refer to as *Marx's fundamental equalities*: the total sum of the prices of all the commodities produced is equal to the total sum of all their values, and the sum of profits expressed in price is equal to the sum of profits expressed in value (for convenience, we refer to profit in price and profit in value as profit and surplus value, respectively).

To date, the mathematical method for converting values into a market production prices has not been clearly established, leading to a long-standing debate famously referred to as the " transformation problem " of value into a market production price. The many methods proposed have all been subjected to more or less restrictive constraints or have had to abandon one or the other of the two fundamental equalities. Surprisingly, mobility of capital across sectors, though implicitly considered by Marx as part of the problem (see below), has most often been ignored. Recently, V. Laure van Bambeke, following a path already glimpsed by others (2, 3), has shown that taking this mobility and the consequent capital reallocation into account is necessary to capture demand (solvent social need) and solve the transformation problem, in accordance with Marx's original idea (4). Here, we follow this path. However, V. Laure van Bambeke' approach is based on solving an overdetermined system of equations by the Moore-Penrose method, which implies not strict compliance with Marx's fundamental equalities in the course of a trial-and-error process. This suggests that Marx's conception might not apply as a law. In contrast, we show a novel approach that enables to solve the transformation problem while complying with Marx's fundamental equalities at all times, by an algorithm which is far more efficient than the tedious trial-and-error process.

Furthermore, our approach is more direct and easily generalizable to any number of branches, with a reduced computation time. Additionally, in contrast to V. Laure van Bambeke's analysis, we show that considering fixed capital is not necessary for the solving. By showing that the transformation problem is equally well solved in the absence of fixed capital, a somewhat artificial case since a capitalist without fixed capital is not the one observed, we perform an ideal thought experiment that allows us to dismiss any objections regarding the various ways of dealing with fixed capital, and we provide an explanation for the fact that in these particular systems, the rate of profit can be determined from the matrix of socio-technical coefficients alone, without taking into account the allocation of capital between the different branches.

Our approach enables to show why complying with both Marx's fundamental equalities is the first and unavoidable condition for a coherent solution that takes into account the solvent social need. Both fundamental equalities are strictly complied with, even in case of unequal rates of profit in the different branches (either because of a monopoly or because the dynamics have brought the system into a state of disequilibrium). Finally, our approach enables to account for a tendency of the average rate of profit to fall (TRPF) as the organic composition expressed in value rises, even in the absence of fixed capital.

## INTRODUCTION

To begin this introduction we will quote the physicist Albert Einstein (5) :

> By using the means of production, the worker produces new goods which become the property of the capitalist. The essential point about this process is the relation between what the worker produces and what he is paid, both measured in terms of real value. Insofar as the labor contract is "free", what the worker receives is determined not by the real value of the goods he produces, but by his minimum needs and by the capitalists' requirements for labor power in relation to the number of workers competing for jobs. It is important to understand that even in theory the payment of the worker is not determined by the value of his product.

If the opinion of a great scientist is not acceptable as an infallible and authoritative argument, even less so in a field that is not his own, we have to consider the personality of Albert Einstein, who is unlikely to express logical thoughts on issues that he did not previously fully understand. Einstein, who always showed an extraordinary intuition in various fields of physical sciences, writes here about real value, a notion which refers to the theory of value in the Marxist sense and which has been judged superfluous by the majority of contemporary economists. Thus, the one who annihilated the useless notions of luminiferous aether and absolute movement and contributed to the emergence of space-time by revealing the relative character of space or time considered separately, endorses the Marxist theory of value which makes surplus value the hidden source of profit (although quite visible when measured in terms of sweat and timed work of employees on assembly lines at Toyota, Ford, Renault, PSA, Ford, Volvo, or in the warehouses of Amazon). Given his generaly accepted clarity of mind, Albert Einstein's statement prompts to take another look into the coherence of Marx's theory of value, which we believe has been wrongly discredited and questioned.

The two tables below are proposed by Marx in Volume III of Capital (1).

| Capital | Rate of surplus value | Surplus value | Rate of profit | Value of ommodities | Cost of production |
|---|---|---|---|---|---|
| **I.** 80 *c* + 20 *v* | 100% | 20 | 20% | 90 | 70 |
| **II.** 70 *c* + 30 *v* | 100% | 30 | 30% | 111 | 81 |
| **III.** 60 *c* + 40 *v* | 100% | 40 | 40% | 131 | 91 |
| **IV.** 85 *c* + 15 *v* | 100% | 15 | 15% | 70 | 55 |
| **V.** 95 *c* + 5 *v* | 100% | 5 | 5% | 20 | 15 |
| 390 *c* + 110 *v* | | 110 | | | Total |
| 78 *c* + 22 *v* | | 22 | 22% | | Mean |

**Table 1A** (Capital, Volume III chapter IX (1))

In this first table five branches are represented, the proportion between constant capital C (machines and raw materials) and variable capital V (wages corresponding to the consumer goods necessary for the reproduction of labour power) is different between them. The rates of exploitation (or rates of surplus value) are identical, at 100%, which is the simplest assumption of similar exploitation of labour power in all branches. The table gives, for each branch, the cost of production, i.e. the portion of total capital is value used over one production cycle. For each branch, the value of commodities is given by the sum of the cost of production and the surplus value. The surplus value, which can be defined as the difference in value between the goods produced and the goods intended for workers' consumption, is only possible because the labour time socially necessary for the production of these goods is greater than that spent on the production of the goods intended for their consumption. The "internal” rate of profit (named rate of profit in Table 1A), not to be confused with the rate of profit in prices, see Table 1B) is the ratio of surplus-value to capital invested. Since the rate of surplus value is identical in each branch, the surplus values realized lead to different internal rates of profit in each branch, the average rate of profit being 22%. In order for this average rate of profit to apply to each branch (capitalists share the total surplus-value in proportion to their investment), Marx introduced the concept that prices of commodities do not equal their values. Compared to its corresponding value, the price of commodity produced by a branch is higher when the branch’s C/V ratio is above average or lower when the branch’s C/V ratio is below average. This is illustrated in the table below. The sum of prices’ deviations from values is zero.

| Capital | Surplus value | Value of commodities | Cost of production | Price of ommodities | Rate of profit | Price to value difference |
|---|---|---|---|---|---|---|
| **I.** 80 *c* + 20 *v* | 20 | 90 | 70 | 92 | 22 % | +2 |
| **II.** 70 *c* + 30 *v* | 30 | 111 | 81 | 103 | 22 % | -8 |
| **III.** 60 *c* + 40 *v* | 40 | 131 | 91 | 113 | 22 % | -18 |
| **IV.** 85 *c* + 15 *v* | 15 | 70 | 55 | 77 | 22 % | +7 |
| **V.** 95 *c* + 5 *v* | 5 | 20 | 15 | 37 | 22 % | +17 |

**Table 1B** (Capital, Volume III chapter IX (1)). This table does not show the column of rates of surplus value, but the latter is unchanged. In Marx’s citation below, cost of production and price of commodities are referred to as “cost-price” and “price of production”, respectively.

The only problem is that capitalists do not buy the various materials from which commodities are composed at their value (as the table suggests) but at their market production price. In other words, this last table, to be accurate, should show costs of production and capital in price, not in value. This difficulty had been fully perceived by Marx, as shown by this passage from Chapter IX of Volume III of Capital (1):

> Since the price of production may differ from the value of a commodity, it follows that the cost-price of a commodity containing

> this price of production of another commodity may also stand above or below that portion of its total value derived from the value of the means of production consumed by it. It is necessary to remember this modified significance of the cost-price, and to bear in mind that there is always the possibility of an error if the cost-price of a commodity in any particular sphere is identified with the value of the means of production consumed by it. Our present analysis does not necessitate a closer examination of this point.

Marx makes it clear that he considers this error as a detail that does not justify the interruption of the thread of his reasoning. As we shall show below, in fact, this error can be solved without undermining the author's conclusions. However, it turns out that previous attempts to correct this "error" (6, 7) have led to the invalidation of two laws considered fundamental by the author of Capital, which we shall call Marx's fundamental equalities: 1) The equality between the sum of the values of the commodities of the whole economy and the sum of their prices. 2) The equality between the sum of the surplus-values (in value) of all the branches and the sum of profits (in price). These equalities are not simple normalizations but the expression of the law of conservation of value, value created by surplus-value alone within the capitalist mode of production and which must satisfy the demand, i.e. the solvent social need (the social need which is able to pay for the given product). [1]

Surprisingly, none of the works mentioned above considers that mobility of capital across sectors is an integral part of the problem of transformation, considering that this phenomenon has already produced its effect through the equalization of the rate of profit. This mobility of capital seems, however, to be an implicit fact for Marx, who explains that it ensures the balance of the rates of profit between branches (excerpt from Volume III, chapter XXII of Capital (1)):

> If prices of commodities in one sphere are below or above the price of production (wherein we deliberately leave aside the fluctuations attendant upon the various phases of the industrial cycle in each and every enterprise) the balance is effected through the expansion or curtailment of production, *i.e.*, the expansion or curtailment of the masses of commodities thrown on the market by industrial capitals – caused by inflow or outflow of capital to and from individual spheres of production.

---

[1] Throughout the text, we name “solvent social need” what has also been named “social need which is able to pay for the given product” by Rubin (8) who defines it as « what Marx called the "quantitatively definite social need" for a given product » (p474 in (9)), « the "amount of social want"» (p136 in (9)), « the "given quantity of social want" » (p138 in (9)). Rubin further adds a note: «By social need, Marx often meant the quantity of products which are sought on the market. But these terminological differences do not concern us here. Our aim is not to define given terms, but to distinguish various concepts, namely: 1) value per unit of commodity; 2) the quantity of units of a commodity which is sought at the market at a given value; 3) the multiplication of the value per unit of commodity times the number of units which are sought on the market at a given value. What is important here is to emphasize that the volume of social need for products of a given kind is not independent of the value per unit of the commodity, and presupposes that value. »

The tables Marx uses to illustrate his theory are to be considered as illustrations of a hypothetical state of equilibrium after the transfer of capital between branches. The fact of having assigned the same capital of 100 for each branch in Tables 1A and 1B is a simplification that does not take into account the solvent social need, the satisfaction of which, as we shall show, requires reallocation of capital conditioned upon the compliance with both fundamental equalities.

By taking into account the mobility of capital across branches as a mechanism leading to a uniform rate of profit, V. Laure van Bambeke has shown that a solution of the transformation problem is *compatible* with the compliance with both fundamental equalities, but the proposed solving method does not involve strict compliance. Here, we describe a more elementary solving method for which both fundamental equalities are strictly complied with, at all times.

In the capitalist system considered, capital is free to flow between branches. To take an electrical analogy, there is a state of high "conductance" between the different branches. In electrostatics, when calculating the electric field at a point in space, one can certainly consider an arbitrary configuration of electric charges, and it is then assumed that these charges are held in their place by some constraints. But as soon as we consider that these charges are free in a conductive medium, as for example on the surface of a Faraday cage, we no longer have the right to place them wherever we want. The distribution of the charges and the calculation of the resulting field (zero inside the cage) become dependent problems. Looking at the problem the other way round, things can sometimes seem surprising, even magical. Thus, in a famous physics book (10), one can read this description of the electric field in a closed metal box (Faraday cage):

> There is a highly nonuniform distribution of charge over the surface of the box. Now the field everywhere in space, *including the interior of the box,* is the sum of the field of this charge distribution and the fields of the external sources. It seems hardly credible that the surface charge has so cleverly arranged itself on the box that its field precisely *cancels* the field of the external sources at every point inside the box. Yet this must indeed be what has happened, in view of the above proof.

Yet, there is no magic in this state of things, which is very well explained by a physical law. It is with this type of "backwards" reasoning that M. Husson (11) concludes that the analysis by V. Laure van Bambeke (12) "makes no economic sense" since it would mean building "in an arbitrary way, a hypothetical economy compatible with the remarkable identities".[2]

In the problem we are dealing with, we show that considering the allocation of capital resulting from its unimpeded mobility across branches is inseparable from solving the transformation problem, the solution of which being the transformation coefficients of values into prices as well as a specific capital allocation for each branch that takes into account the solvent social need. Contrary to a surprisingly widespread view (11), the capital allocation between branches can in no way constitute an exogenous factor, it is primarily determined by the satisfaction of the solvent social need. This does not prevent capital allocation from being secondarily determined by other factors (e.g., taste of consumers, advertisement, competition between capitalists, politics) since the higher the branch number, the more there is possible "play" within the increasingly extending ensemble of solutions of the transformation problem.

[2] Marx's fundamental equalities are called by Husson "remarkable identities". The text is translated from the following original in French: « n'a pas de sens économique »… « de manière arbitraire, une économie hypothétique compatible avec les identités remarquables ».

## THE MODELS IN THIS STUDY

The problem of transformation was introduced by Marx in order to explain how branches of different organic composition tend to exhibit a similar rate of profit, which we will call the average or *uniform rate of profit*. The postulate that the only mechanism for the creation of value is surplus value (made on labour which is paid less than the value it produces) leads to the intuitive expectation that branches whose capital is more distributed towards variable capital (V) and which therefore generate more surplus value should generate a higher rate of profit. But this is not what one observes. In the real world, each branch tends towards the same average (or uniform) rate of profit. To explain this apparent paradox, Marx proposed a mismatch between values and market production prices and gave the keys to mathematically solve the transformation of values into prices. This problem of transformation has faced numerous criticisms, and it was not until the work of V. Laure van Bambeke (4) that it was shown that it was solvable as soon as it was accepted that capital should flow between branches until it satisfies an equilibrium corresponding to the occurring of a uniform rate of profit and the satisfaction of the solvent social need. V. Laure van Bambeke's approach puts back in the saddle the problem of transformation posed by Marx, but it is not entirely satisfying because (i) it does not solve the problem when the fixed capital is zero; (ii) it uses the Moore-Penrose method of approximation which implies that compliance with the fundamental equalities is only approximated, at least until the convergence of rates of profit to a uniform rate of profit is reached, in this sense it does not apply Marx's theory of value as a strict law; (iii) its resolution method, based on successive and parsimonious transfers of capital from one branch to another until an "acceptable" error is found for all the equations, is tedious and hard to apply when the number of branches is larger than two.

Here, considering successively models of economy with two, three, four and five branches, we provide a direct mathematical resolution of the transformation problem, which strictly complies with both fundamental equalities, and which can be generalized to any given number of branches. Our approach, when applied to an ideal case with no fixed capital and no profit, provides a mean to grasp the meaning of Marx's fundamental equalities, which embody the conservation of value and the adequacy between the production and the solvent social need.

The initial figures displayed in tables are arranged by rows (production branches of each type of commodity) corresponding to some capital allocations between branches which do not necessarily comply with the fundamental equalities. By dividing figures of a row, corresponding to a sector of activity, by the total capital committed for this sector, we define the intangible parameters of the model, defined as the *socio-technical coefficients*. For a given commodity produced (output), these coefficients indicate the necessary proportions (ratio of values) of the commodities that constitute it (input). The socio-technical coefficients are a function of the nature of the commodities, the advances in technology and the social parameters that impact on the level of performance of the labour power.

Solving the transformation problem means, for models of economy with two or more branches, determining for each commodity, the *transformation coefficient* of value into price and the *capital allocation* between branches, which enable the realization of an identical rate of profit by each branch (uniform rate) while complying with Marx's fundamental equalities. The algorithm we propose for this solving further allows for the possibility of constraints that would impose a given difference between these rates of profit (the occurrence of monopolies, for example).

We are in line with the pioneering work of V. Laure van Bambeke (4, 12, 13) and we have adopted, for the two- and three-branch models, a fixed capital that is imported, which is bought by the capitalists at its price and progressively transmits its value to the commodities produced in the form of an annual amortization equal to the ratio of the price of the fixed capital to the number of production cycles. For these models, the total value transmitted by this fixed capital is considered equal to its price. We next consider models with a branch producing the machines (fixed capital, the price of which may then differ from its value).

Moreover, we also consider the situation in which fixed capital is zero, and we show that the transformation problem remains solvable.

The two-branch model is unique in the sense that it allows for only one solution (one pair of transformation coefficients and one single capital allocation between branches).

The branches of the various models are defined as follows.

• Two-branch model: C (raw materials), V (labour power).

• Three-branch model: E (energy), C, V.

• Four-branch models:

a) Branches E, C, V, L (luxury): the commodities produced by the luxury branch L are not used by the first three core branches.

b) Branches M (machines), E, C, V: the machines are produced by branch M, within the economic system under consideration. Their amortization is still following the same accounting rules as before, but in this case their price (which does not necessarily equal their value) is taken into account.

• Five-branch model: M, E, C, V, L. We add a luxury branch in which the commodities produced are not consumed by the first four core branches.

For the various models, in addition to theoretical demonstrations of the method to solve the transformation, we provide a detailed algorithm (see Appendix) and the runtime of a program written in Labview (complementary open access document on HAL). This program produces results with a precision of 14 significant digits. There is no theoretical limit to the precision of the results. The tables we provide as examples display results with about ten significant digits to illustrate the precision of the method. In comparison, the work of V. Laure van Bambeke displays results with six significant digits obtained from manipulating numbers with at least eight significant digits, the calculation of which is difficult since, even with only three branches, the number of possible transfers is already very large.

# A- Transformation problem for a two-branch economy model

## 1. Two-branch model with fixed capital

### a) *Definitions*

In this minimal model, production is divided into two branches: C and V. Branch C produces raw materials. Branch V produces commodities that are consumed by the workers in both branches.

In each branch, the capital invested is subdivided into three subtypes of capital.

We note that for branch i, the total invested capital $K_i$ is subdivided into $F_i$, $C_i$ and $V_i$ :

$$\boldsymbol{K_i} = \boldsymbol{F_i} + \boldsymbol{C_i} + \boldsymbol{V_i}$$

$F_i$ : fixed capital, i.e. the capital invested in the purchase of the infrastructure and the machines. This capital is amortized over a number $\boldsymbol{n}$ of cycles. $\boldsymbol{D_i} = {}^{\boldsymbol{F_i}}/_{\boldsymbol{n}}$ defines the quantity of value transmitted by the fixed capital at each cycle.

$C_i$ : circulating constant capital, for example the capital needed at each cycle for the purchase of raw materials. The sum of the circulating constant capital and the fixed capital constitutes the constant capital.

$V_i$ : variable capital, defined as the capital needed at each cycle to reproduce the labour power.

The proportion of non-variable capital in branch i is defined as the **organic composition** of branch i: $CO_i = \frac{F_i+C_i}{V_i}$

The total production of branch i produces surplus value, denoted $\boldsymbol{PL_i}$, dependent on the rate of exploitation of labour, denoted $\boldsymbol{e_i}$.[3] This exploitation rate depends on the struggle between workers and their employers and the resulting balance of power. Our initial assumption is that this balance of power equilibrates across branches and that labour is exploited on average at the same exploitation rate denoted $e$. (Assuming different exploitation rates for different branches does not change the general conclusions of the transformation.)

Thus, ∀i, $\boldsymbol{PL_i} = e\boldsymbol{V_i}$

The total production of branch i at each cycle is noted $W_i$ :

$$\boldsymbol{W_i} = \boldsymbol{D_i} + \boldsymbol{C_i} + \boldsymbol{V_i} + \boldsymbol{PL_i} = \boldsymbol{D_i} + \boldsymbol{C_i} + (\boldsymbol{1} + e)\boldsymbol{V_i}$$

In Capital, Volume III, Marx postulates that commodities are exchanged according to market production prices that differ from values. The transformation of value into price is solved by a transformation coefficient specific to the type of commodity. This coefficient is denoted $x_i$ for branch i. In other words, for branch i, the price of total output is $\boldsymbol{x_i W_i}$.

---

[3] Note that the $PL$ symbol is after the French translation of surplus value, i.e. "plus value".

In the two-branch model, coefficient $x_1$ applies to raw materials and coefficient $x_2$ applies to the goods produced by branch 2. In this essential model, wages enable the purchase of the goods produced by branch 2, so the coefficient $x_2$ also applies to the workers' wages.

Thus, the total capital *in price* invested in the branch i, noted $K_{pi}$, is written :

$$\boldsymbol{K_{pi}} = \boldsymbol{F_i} + x_1 \boldsymbol{C_i} + x_2 \boldsymbol{V_i}$$

The profit in price of branch i is noted $\boldsymbol{S_i}$ and defined by the following equation which gives the total production in price of branch i, at each cycle:

$$\boldsymbol{x_i W_i} = \boldsymbol{D_i} + \boldsymbol{x_1 C_i} + \boldsymbol{x_2 V_i} + \boldsymbol{S_i}$$

### b) *Marx's fundamental equalities*

According to Marx, the transformation coefficients of value into price are constrained by a real equilibrium of the economic system in value. Marx proposed two equalities corresponding to these constraints.

The first equality postulates that the sum of profits (prices) is equal to the sum of capital gains (values).

$\sum_i \boldsymbol{S_i} = \sum_i \boldsymbol{PL_i}$ **fundamental equality I**

The second equality postulates that the sum of capitals committed in price is equal to the sum of capitals committed in value. That is: $\sum \boldsymbol{K_{pi}} = \sum \boldsymbol{K_i}$

This equality can also be written:

$$\sum(\boldsymbol{F_i} + x_1 \boldsymbol{C_i} + x_2 \boldsymbol{V_i}) = \sum(\boldsymbol{F_i} + \boldsymbol{C_i} + \boldsymbol{V_i})$$

$$\sum(x_1 \boldsymbol{C_i} + x_2 \boldsymbol{V_i}) = \sum(\boldsymbol{C_i} + \boldsymbol{V_i})$$

$$\sum(\boldsymbol{D_i} + x_1 \boldsymbol{C_i} + x_2 \boldsymbol{V_i}) = \sum(\boldsymbol{D_i} + \boldsymbol{C_i} + \boldsymbol{V_i})$$

and, taking into account the fundamental equality I:

$$\sum(\boldsymbol{D_i} + x_1 \boldsymbol{C_i} + x_2 \boldsymbol{V_i} + \boldsymbol{S_i}) = \sum(\boldsymbol{D_i} + \boldsymbol{C_i} + \boldsymbol{V_i} + \boldsymbol{PL_i})$$

This amounts to postulating that the sum of production in price is equal to the sum of production in value, and can also be written:

$\sum_i \boldsymbol{x_i W_i} = \sum_i \boldsymbol{W_i}$ **fundamental equality II**

c) *The transformation in equations*

Use of terms normalized relative to capital

In what follows, we solve the transformation problem, i.e. we determine the coefficients $x_i$ and the allocation of capital between branches while complying with the two fundamental equalities. The simplest case to consider is the same rate of profit (uniform rate) for each branch. This rate of profit, denoted $r$, is such that, for each branch $\boldsymbol{S_i = rK_{pi}}$.

The total capital $K_T$ is allocated as $K_1$ and $K_2$ in branches 1 and 2 respectively. This allocation is not fixed, it depends on the transformation. In contrast, the organic composition of the branches and the **socio-technical coefficients** $\boldsymbol{f_i = {}^{F_i}/_{K_i}}$ , $\boldsymbol{c_i = {}^{C_i}/_{K_i}}$ et $\boldsymbol{v_i = {}^{V_i}/_{K_i}}$ are fixed by the means of the technique, the nature of the considered commodities, their composition and the degree of qualification of the workforce.

For each branch, $\quad \boldsymbol{f_i + c_i + v_i} = 1$ (1a)

For the following demonstration, we have normalized, relative to $K_i$, the various terms involved:

$$\boldsymbol{w_i = {}^{W_i}/_{K_i}\,,\ d_i = {}^{D_i}/_{K_i}\,,\ pl_i = {}^{PL_i}/_{K_i}\,,\ s_i = {}^{S_i}/_{K_i}}$$

Therefore, the relations defined in the previous paragraph can be written:

$$pl_i = \boldsymbol{e}v_i \tag{2a}$$

$$d_i = {}^{f_i}/_{\boldsymbol{n}} \tag{3a}$$

$$w_i = d_i + c_i + v_i + pl_i = d_i + c_i + (1+e)v_i \tag{4a}$$

$$x_i w_i = d_i + x_1 c_i + x_2 v_i + s_i \tag{5a}$$

$$s_i = r(nd_i + x_1 c_i + x_2 v_i) \tag{6a}$$

We set $k_1 = {}^{K_1}/_{K_T}$ and $k_2 = {}^{K_2}/_{K_T}$ , with $K_T = K_1 + K_2$ .

Fundamental equality II and determination of $\boldsymbol{k_i}$ as a function of $\boldsymbol{x_i}$

For the two-branch model, the fundamental equality II reads:

$$k_1 w_1 + k_2 w_2 = k_1 x_1 w_1 + k_2 x_2 w_2$$

The following system of equations is used to determine $k_1$ and $k_2$ as a function of $x_1$ and $x_2$.

$$\left.\begin{aligned} &k_1 + k_2 = 1 \\ &k_1 w_1(1 - x_1) + k_2 w_2(1 - x_2) = 0 \end{aligned}\right\}$$

Assuming the denominator of the determinant is non-zero, the solutions are:

$$k_1 = \frac{\begin{vmatrix} 1 & 1 \\ 0 & w_2(1-x_2) \end{vmatrix}}{\begin{vmatrix} 1 & 1 \\ w_1(1-x_1) & w_2(1-x_2) \end{vmatrix}}$$

$$k_2 = \frac{\begin{vmatrix} 1 & 1 \\ w_1(1-x_1) & 0 \end{vmatrix}}{\begin{vmatrix} 1 & 1 \\ w_1(1-x_1) & w_2(1-x_2) \end{vmatrix}}$$

### System of equations of the branches in prices and determination of $x_i$ as a function of $r$

$x_1$ and $x_2$ can be determined as a function of $r$ (or $r_1$ and $r_2$ if the rates of profit are postulated not equal) by using equations (5a) and (6a) which yield the system of equations of the branches in prices:

$$[c_1(1+r) - w_1]x_1 + v_1(1+r)x_2 = -(1+nr)d_1$$

$$c_2(1+r)x_1 + [v_2(1+r) - w_2]x_2 = -(1+nr)d_2$$

Therefore $k_1$ and $k_2$ can be determined as a function of $r$ likewise.

### Fundamental equality I and determination of $r^*$

We define the z-function as the difference between total profit and total surplus value:

$$z(x_1, x_2) = k_1\big(x_1(w_1 - c_1) - x_2 v_1 - (d_1 + pl_1)\big) + k_2\big(x_2(w_2 - v_2) - x_1 c_2 - (d_2 + pl_2)\big) \quad (7a)$$

The z-function can therefore also be expressed as a function of $r$. When it cancels, it enables the determination of $r$ for which fundamental equality I is verified.

The algorithm detailed in the Appendix enables to determine the value of $r$ for which the z-function cancels. We show in Chapter İ ("Variation of the z-function") that, regardless of the number of branches, the solution $r^*$ of the equation $z(r) = 0$ is the first cancellation of the z-function for decreasing z.

### Geometrical interpretation

In the space of $K$, which is the plane $(K_1, K_2)$, this solution corresponds to the intersection of the two lines of equations 8a and 9a below. The line of equation 7a is merged with the one of equation 9a.

8a) $K_1 + K_2 = K_T$

9a) $K_1 w_1(1 - x_1) + K_2 w_2(1 - x_1) = 0$

7a) $K_1[(w_1 - c_1)\, x_1 - v_1\, x_2 - (d_1 + pl_1)] + K_2[(w_2 - v_2)\, x_2 - c_2\, x_1 - (d_2 + pl_2)] = 0$

### d) *Conclusion*

Whatever the socio-technical coefficients of the two branches, a value $r^*$ of the rate of profit is found while complying with the two fundamental equalities. Since, in this model, surplus value is the only profit-creating mechanism, the internal rate of profit (calculated in terms of value) is greater for the less capitalistic branch (low organic composition). But in price, in the real world, a comparable rate of profit is observed in all branches (uniform rate). Marx had described this process by formulating that "capitalists are brothers" (14), i.e. they share the total surplus-value, and this sharing translates into a uniform rate of profit (in price). In other words, in price, a highly capitalistic branch can have the same rate of profit as a branch with a low organic composition, even if in value terms it is the latter branch that generates proportionally more surplus-value. Note also that the solution for the pair $(k_1, k_2)$ is unique. In other words, there is only one possible way to allocate capital between branches. This makes the two-dimensional case a special case, in contrast to cases with higher dimensions for which the number of solutions is infinite (see below).

### e) *Example*

We start from the configuration of the two-branch example used by V. Laure van Bambeke (4) with $K_T$ = 715 and n = 5 cycles :

| **INIT VALUES** | F | C | V | PL | W | K |
|---|---|---|---|---|---|---|
| BRANCH 1 C | 125 | 200 | 90 | 60 | 375 | 415 |
| BRANCH 2 V | 100 | 80 | 120 | 80 | 300 | 300 |
| TOTAL | 225 | 280 | 210 | 140 | 675 | 715 |

**Table 2A**

We calculate with our algorithm the only capital allocation and transformation coefficients which constitute the transformation solution. This solution provides the following tables.

| **VALUES** | F | C | V | PL | W | K |
|---|---|---|---|---|---|---|
| BRANCH 1 C | 129.6884522 | 207.5015236 | 93.37568562 | 62.25045708 | 389.0653567 | 430.565661462 |
| BRANCH 2 V | 94.81144618 | 75.84915694 | 113.7737354 | 75.84915694 | 284.4343385 | 284.434338538 |
| TOTAL | 224.4998984 | 283.3506805 | 207.149421 | 138.099614 | 673.4996953 | 715 |

**Table 2B**

| **PRICES** | F | C | V | S | W | Kp |
|---|---|---|---|---|---|---|
| BRANCH 1 C | 129.6884522 | 222.2667889 | 84.28714183 | 84.25860798 | 416.7502291 | 436.242382948 |
| BRANCH 2 V | 94.81144618 | 81.24638441 | 102.6997865 | 53.84100604 | 256.7494662 | 278.757617052 |
| TOTAL | 224.4998984 | 303.5131733 | 186.9869283 | 138.099614 | 673.4996953 | 715 |

**Table 2C**

The difference between the total sum of profits $S_T$ and the total sum of surplus values $PL_T$ as well as the difference between the total value of production $W_T$ and its price $Wp_T$ are such that:

$$|\boldsymbol{S_T - PL_T}| < \mathbf{10^{-14}}\ ;\ \ |\boldsymbol{W_T - Wp_T}| < \mathbf{10^{-14}}$$

| | $x_i$ | $r_i$ |
|---|---|---|
| BRANCH 1 C | 1.071157382 | 0.193146313 |
| BRANCH 2 V | 0.902666912 | 0.193146313 |

This allocation gives a uniform rate of profit r* of **0.1931463133178**

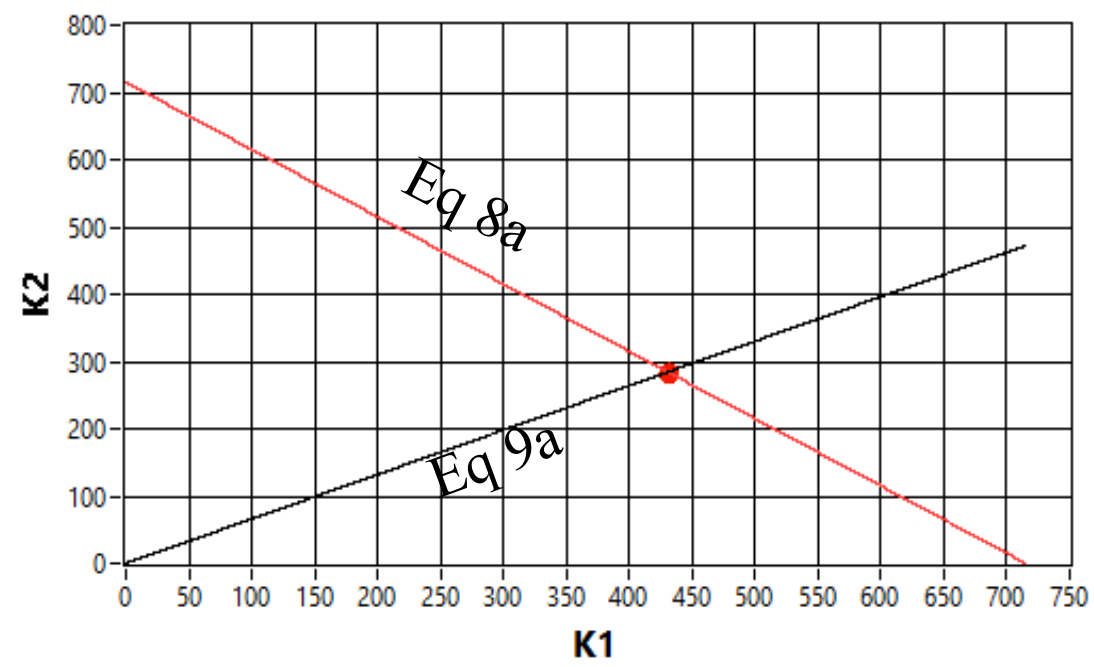

**Figure 1**

## 2. Condition to fulfill the solvent social need (to meet the demand)

The fulfilling of the solvent social need forces the production of commodity i to cover the consumption of that same commodity i in both branches.[4] This notion does not pose any difficulty when all the goods in the system are produced by the same system (closed system). As soon as fixed capital is introduced into the model (4, 12, 13, 15), the system must be considered as open: fixed capital is invested in commodities (e.g. machines) that are not produced in the system but imported. In a closed system, the different labour times that constitute the commodities can add up. In an open system such as the one with fixed capital, even if the imported commodities (defined as D) have a transformation coefficient fixed at 1, their value corresponds to labour time performed in a different economic reference frame. These imported commodities do not correspond to labour time in the system under consideration, they are joint values. In order to study the needs in terms of value in commodities produced within the system, it is necessary, as in the case of a change of reference frame in kinematics, to subtract from the total value W, the value D contributed by the other system. In other words, in order to satisfy the needs while ensuring the equilibrium of their balance of payments, the capitalists of each branch must produce a production surplus from which the invested fixed capital has been deducted. This inequality reads:

---

[4] We can also define the satisfaction of the social need in terms of value, in which case the value produced by a branch must be at least equal to the sum of the values used for this commodity. This is illustrated in chapter G, for a 5 branch model with simple reproduction (see definition in part B-1-c) either in value or in price.

$$k_1(x_1 w_1 - d_1) \geq x_1(k_1 c_1 + k_2 c_2)$$

$$k_2(x_2 w_2 - d_2) \geq x_2(k_1 v_1 + k_2 v_2)$$

## 3. Simplified case (no fixed capital, no surplus value) and the sense of the fundamental equalities

### a) *The transformation in equations*

The simplified case of the two-branch model with no fixed capital and no profit is useful to clearly see the link between the fundamental equalities and the satisfaction of the solvent social need.

The system of equations of the branches in price becomes:

$$[c_1 - w_1]x_1 + v_1 x_2 = 0$$

$$c_2 x_1 + [v_2 - w_2]x_2 = 0$$

We have $w_1 = c_1 + v_1$ and $w_2 = c_2 + v_2$, we deduce $x_1 = x_2$

Fundamental equality II becomes:

$$k_1 w_1 + k_2 w_2 = k_1 x_1 w_1 + k_2 x_2 w_2$$

Since $x_1 = x_2$, we have $x_1 = x_2 = 1$

Fulfilling the solvent social need at each cycle requires:

$$k_1 w_1 \geq k_1 c_1 + k_2 c_2$$

$$k_2 w_2 \geq k_1 v_1 + k_2 v_2$$

Fundamental equality I (7a) becomes:

$$k_1(x_1(w_1 - c_1) - x_2 v_1) + k_2(x_2(w_2 - v_2) - x_1 c_2) = 0$$

which can also read $$[k_1 w_1 - (k_1 c_1 + k_2 c_2)]x_1 + [k_2 w_2 - (k_1 v_1 + k_2 v_2)]x_2 = 0$$

This formulation of fundamental equality I shows that it is satisfied only under the condition that:

$$k_1 w_1 = k_1 c_1 + k_2 c_2$$

$$k_2 w_2 = k_1 v_1 + k_2 v_2$$

In other words, the demand is met exactly at each cycle. In this way, the model without fixed capital and without surplus value reveals the link between fundamental equality I and the adequacy of the production to the demand.

This system of equations may also be written:

$$k_1(w_1 - c_1) - k_2 c_2 = 0$$

$$-k_1 v_1 + k_2(w_2 - v_2) = 0$$

This system allows an infinity of solutions (zero determinant since $w_1 = c_1 + v_1$ et $w_2 = c_2 + v_2$), which means that these two equations are redundant. By keeping only one of them, and taking into account the equality $K_1 + K_2 = K_T$, a unique solution is found for $(k_1, k_2)$ :

$$k_1 = \frac{c_2}{v_1 + c_2} \quad \text{and} \quad k_2 = \frac{v_1}{v_1 + c_2}$$

### b) *Example*

Starting values with capital allocation (50/50):

| **INIT VALUES** | F | C | V | PL | W=K |
|---|---|---|---|---|---|
| BRANCH 1 C | 0 | 131.9788674 | 59.39049034 | 0 | 191.36935774 |
| BRANCH 2 V | 0 | 76.54774311 | 114.8216146 | 0 | 191.36935774 |
| TOTAL | 0 | 208.52661051 | 174.21210494 | 0 | 382.73871545 |

Following the transformation, the displayed values illustrate compliant capital allocation:

| **VALUES or PRICES** | F | C | V | PL | W=K |
|---|---|---|---|---|---|
| BRANCH 1 C | 0 | 148.6363944 | 66.88637748 | 0 | 215.52277188 |
| BRANCH 2 V | 0 | 66.88637748 | 100.3295662 | 0 | 167.21594368 |
| TOTAL | 0 | 215.52277188 | 167.21594368 | 0 | 382.7387154 |

**Table 2D**

$K_1 = 215.523 \quad K_2 = 167.216$

## 4. Note on capital allocation between branches in the absence of surplus value

In the absence of surplus value, $x_1 = x_2 = 1$

The fulfillment of the solvent social need can be written:

$$k_1(w_1 - d_1) \geq k_1 c_1 + k_2 c_2$$

$$k_2(w_2 - d_2) \geq k_1 v_1 + k_2 v_2$$

Fundamental equality I can be written as

$$[k_1(w_1 - d_1) - (k_1 c_1 + k_2 c_2)] + [k_2(w_2 - d_2) - (k_1 v_1 + k_2 v_2)] = 0$$

Therefore, it is complied with provided that

$$k_1(w_1 - d_1) = k_1 c_1 + k_2 c_2$$

$$k_2(w_2 - d_2) = k_1 v_1 + k_2 v_2$$

Posing $w'_i = (w_i - d_i)$, we recover the same system of equations as in the previous case.

Taking into account the equality $k_1 + k_2 = 1$, we find the same unique solution:

$$k_1 = \frac{c_2}{v_1+c_2} \quad \text{and} \quad k_2 = \frac{v_1}{v_1+c_2}$$

Example, n = 5 cycles

| **INIT VALUES** | F | C | V | PL | W | K |
|---|---|---|---|---|---|---|
| BRANCH 1 C | 125 | 200 | 90 | 0 | 315 | 415 |
| BRANCH 2 V | 100 | 80 | 120 | 0 | 220 | 300 |
| TOTAL | 125 | 280 | 210 | 0 | 535 | 715 |

**Table 2E**

| **VALUES** | F | C | V | PL | W | K | W-F/n |
|---|---|---|---|---|---|---|---|
| BRANCH 1 C | 118.7707641 | 190.0332226 | 85.51495017 | 0 | 299.3023256 | 394.318937 | 275.5481728 |
| BRANCH 2 V | 106.8936877 | 85.51495017 | 128.2724252 | 0 | 235.166113 | 320.681063 | 213.7873754 |
| TOTAL | 225.6644518 | 275.5481728 | 213.7873754 | 0 | 534.4684385 | 715 | |

**Table 2F**

$$k_1 = 0.55149501678321 \; k_2 = \; 0.44850498321678$$

As the comparison of Tables 2E and 2F illustrates, fixed capital appears as a joint value whatever its amount, since internal needs (in C and V) do not depend on it.

## 5. Conclusion

The two-branch system shows that compliance with fundamental equalities forces a unique allocation of capital between the two branches. When there is exploitation and consequently a profit made, this allocation of capitals leads to a specific rate of profit r*. The particular case of the system without fixed capital and without surplus-value is useful for showing the general sense of Marx's fundamental equalities: they express the adequacy between the production and the solvent social need. When taking these equalities into account in the system with non-zero fixed capital and surplus-value, the value of r* is determined both by the degree of exploitation of human labour and by the quantity of goods that can be absorbed by the market.

Note that we considered the possibility to set a difference Δr between the rate of profit of branches 1 and 2, the values r* and r*+ Δr are then found through the same interpolation method.

In general, the allocation of capital between branches is constrained by the adequacy of production to the solvent need. *The allocation of capital between branches is therefore inseparable from the question of the transformation of values into market production prices*.

## B- Transformation problem for a three-branch economy model

The approach is identical to the previous case with two branches. In the three-branch model, an additional unknown appears for the linear system in K which becomes underdetermined with an infinity of solutions.

### 1. Three-branch model with fixed capital

#### a) *Definitions*

In this model, production is divided into three branches. The first branch produces energy (E), the second, raw materials (C), and the third, goods that are consumed by the workers of the three branches (V). In each branch, the capital invested is distributed into four subtypes of capital.

We note that for branch i, the invested capital $K_i$ is subdivided into $F_i$, $E_i$ $C_i$ et $V_i$ :

$$\boldsymbol{K_i = F_i + E_i + C_i + V_i}$$

For each branch indexed i :

$F_i$ is the fixed capital, i.e. the capital invested in the purchase of infrastructure and machinery. This capital is amortized over a number n of cycles. We define $D_i = {F_i}/{n}$ the quantity of value transmitted by the fixed capital in each cycle.

$E_i$ is the capital needed for energy in each cycle.

$C_i$ is the capital needed for the purchase of raw materials in each cycle.

$V_i$ is the variable capital, defined as the capital needed to reproduce the workers' labour power in each cycle.

The proportion of non-variable capital in branch i is defined as the organic composition of branch i : $CO_i = \frac{F_i+E_i\ +C_i}{V_i}$

The total production of branch i produces a surplus value noted $PL_i$ depending on the exploitation rate of labour. This exploitation rate depends on the struggle between workers and employers and the resulting balance of power. Our initial assumption is that this balance of power equilibrates across branches and that labour is exploited on average by the same exploitation rate noted $e$. (The assumption of different exploitation rates for different branches does not change the general conclusions of the transformation).

Accordingly, ∀i, $\boldsymbol{PL_i = eV_i}$

The total production of branch i in each cycle is noted $W_i$ :

$$\boldsymbol{W_i = D_i + E_i + C_i + V_i + PL_i = D_i + E_i + C_i + (1 + e)V_i}$$

In Capital, Volume III, Marx postulates that commodities are exchanged at market production prices that differs from values. The transformation of value into price is resolved by a transformation coefficient specific to the type of commodity. This coefficient is denoted $x_i$ for branch i. This means that for branch i, the total production price is $\boldsymbol{x_i W_i}$.

In the three-branch model, coefficient $x_1$ applies to raw materials, coefficient $x_2$ to energy and $x_3$ to the commodities produced by branch 3. Coefficient $x_3$ also applies to the wages of workers who reproduce their labour power with consumption of commodities produced by branch 3.

Accordingly, the total price capital invested in industry i, denoted $K_{pi}$, is given by:

$$\boldsymbol{K_{pi}} = \boldsymbol{F_i} + x_1 \boldsymbol{C_i} + x_2 \boldsymbol{E_i} + x_3 \boldsymbol{V_i}$$

Note that fixed capital does not have a transformation coefficient because the capitalist obtains this fixed capital at the market price; this capital is then depreciated over n production cycles. This is in accordance with V. Laure van Bambeke's approach (13).

The profit (in prices) of branch i is noted $\boldsymbol{S_i}$ and defined by the following equation which gives the total production in price of branch i, at each cycle:

$$\boldsymbol{x_i W_i} = \boldsymbol{D_i} + \boldsymbol{x_1 C_i} + \boldsymbol{x_2 E_i} + \boldsymbol{x_3 V_i} + \boldsymbol{S_i}$$

Marx's fundamental equalities I and II as defined in paragraph A-1-b are generalizable to any number of branches of systems with or without fixed capital.

b) *Solving the transformation*

In what follows, we determine the coefficients $x_i$, while complying with the two fundamental equalities. The simplest but by no means obligatory case is to consider the same rate of profit for each branch. This rate of profit, noted r is such that, for each branch $\boldsymbol{S_i} = \boldsymbol{rK_{pi}}$.

The total capital $K_T$ is allocated into $K_1$ , $K_2$ et $K_3$ in branches 1, 2 and 3 respectively. This allocation is not a fixed figure, it depends on the transformation. On the other hand, the organic composition of the branches being defined by the quantities $\boldsymbol{CO_i} = \frac{(\boldsymbol{F_i}+\boldsymbol{E_i}+\boldsymbol{C_i})}{\boldsymbol{V_i}}$ along with the socio-technical coefficients $\boldsymbol{f_i} = \boldsymbol{F_i}/\boldsymbol{K_i}$, $\boldsymbol{e_i} = \boldsymbol{E_i}/\boldsymbol{K_i}$, $\boldsymbol{c_i} = \boldsymbol{C_i}/\boldsymbol{K_i}$ et $\boldsymbol{v_i} = \boldsymbol{V_i}/\boldsymbol{K_i}$ are set by the means of technology, the nature of the commodities under consideration, their compositions, and the degree of qualification of the workforce.

For each branch, $\boldsymbol{f_i} + \boldsymbol{e_i} + \boldsymbol{c_i} + \boldsymbol{v_i} = 1$ (1b)

For the following demonstration, we generalize this notation by normalizing the different terms involved to $K_i$.

$$\boldsymbol{w_i} = \boldsymbol{W_i}/\boldsymbol{K_i}, \ \boldsymbol{d_i} = \boldsymbol{D_i}/\boldsymbol{K_i}, \ \boldsymbol{pl_i} = \boldsymbol{PL_i}/\boldsymbol{K_i}, \ \boldsymbol{s_i} = \boldsymbol{S_i}/\boldsymbol{K_i}$$

Hence, the relations defined in the previous paragraph can be written:

$$pl_i = \boldsymbol{e} v_i \tag{2b}$$

$$d_i = {}^{f_i}/_{\boldsymbol{n}} \tag{3b}$$

$$w_i = d_i + e_i + c_i + v_i + pl_i = d_i + e_i + c_i + (1 + e)v_i \tag{4b}$$

$$x_i w_i = d_i + x_1 c_i + x_2 e_i + x_3 v_i + s_i \tag{5b}$$

$$s_i = r(nd_i + x_1 c_i + x_2 e_i + x_3 v_i) \tag{6b}$$

Having defined $k_1 = {}^{K_1}/_{K_T}$ , $k_2 = {}^{K_2}/_{K_T}$ et $k_3 = {}^{K_3}/_{K_T}$

fundamental equality I can be written:

$$k_1[x_1 w_1 - x_1 e_1 - x_2 c_1 - x_3 v_1 - d_1] + k_2[x_2 w_2 - x_1 e_2 - x_2 c_2 - x_3 v_2 - d_2] + k_3[x_3 w_3 - x_1 e_3 - x_2 c_3 - x_3 v_3 - d_3] - [k_1 pl_1 + k_2 pl_2 + k_3 pl_3] = 0$$

Or else:

$$k_1[x_1(w_1 - e_1) - x_2 c_1 - x_3 v_1 - (d_1 + pl_1)] + k_2[x_2(w_2 - c_2) - x_1 e_2 - x_3 v_2 - (d_2 + pl_2)] + k_3[x_3(w_3 - v_3) - x_1 e_3 - x_2 v_3 - (d_3 + pl_3)] = 0$$

For the resolution of the transformation problem, it will be useful to use the z-function defined as:

$$z = k_1[x_1(w_1 - e_1) - x_2 c_1 - x_3 v_1 - (d_1 + pl_1)] + k_2[x_2(w_2 - c_2) - x_1 e_2 - x_3 v_2 - (d_2 + pl_2)] + k_3[x_3(w_3 - v_3) - x_1 e_3 - x_2 v_3 - (d_3 + pl_3)] \tag{7b}$$

Having defined $t = 1 + r$, from equations (5b) and (6b) we derive the system of equations of the branches in price:

$$\left.\begin{aligned}
(e_1 t - w_1)x_1 + c_1 t x_2 + v_1 t x_3 &= -d_1(1 + nr) \\
e_2 t x_1 + (c_2 t - w_2)x_2 + v_2 t x_3 &= -d_2(1 + nr) \\
e_3 t x_1 + c_3 t x_2 + (v_3 t - w_3)x_3 &= -d_3(1 + nr)
\end{aligned}\right\} \tag{8b}$$

This system of equations, when its determinant is non-zero, allows to determine a unique triplet $x_1$, $x_2$, $x_3$ as a function of $r$

Fundamental equality II can be written:

$$k_1 w_1 + k_2 w_2 + k_3 w_3 = k_1 x_1 w_1 + k_2 x_2 w_2 + k_3 x_3 w_3$$

As with the two-branch model, a similar system of equations determines $k_i$ as a function of $x_i$. But this system of equations has an additional degree of freedom. Therefore, unlike the two-branch case, there are an infinity of solutions. By fixing one $k_i$, for example $k_3 = k_3{}^{fixed}$, we have:

$$k_1 + k_2 = 1 - k_3{}^{fixed}$$

and $$k_1 w_1(1 - x_1) + k_2 w_2(1 - x_2) = -k_3{}^{fixed} w_3(1 - x_3)$$

The determinant of this system is:

$$Det = \begin{vmatrix} 1 & 1 \\ w_1(1 - x_1) & w_2(1 - x_2) \end{vmatrix}$$

<u>Case $Det \neq 0$:</u>

The coefficient $k_3$ is fixed and we assume here different organic compositions of branch 1 and 2 (at identical exploitation rates).

The solutions are:

$$k_1 = \frac{\begin{vmatrix} 1 - k_3{}^{fixed} & 1 \\ -k_3{}^{fixed} w_3(1 - x_3) & w_2(1 - x_2) \end{vmatrix}}{\begin{vmatrix} 1 & 1 \\ w_1(1 - x_1) & w_2(1 - x_2) \end{vmatrix}}$$

$$k_2 = \frac{\begin{vmatrix} 1 & 1 - k_3{}^{fixed} 1 \\ w_1(1 - x_1) & -k_3{}^{fixed} w_3(1 - x_3) 0 \end{vmatrix}}{\begin{vmatrix} 1 & 1 \\ w_1(1 - x_1) & w_2(1 - x_2) \end{vmatrix}}$$

If at least one of the coefficients $k_i$ (for i=1 or i=2) is found negative for all values of $r$ considered this means that the fixed value $k_3{}^{fixed}$ is not acceptable and another one must be chosen. Among the alternatives that we consider acceptable are those for which the production of a type of commodity in one cycle would not be enough to fulfill consumption of this same commodity. The possibility of stocks built up in previous cycles is allowed.

<u>Case $Det = 0$:</u>

In case of identical exploitation rates and organic compositions for two branches, to avoid cancellation of the determinant, one of these two branches can be chosen as the one for which we fix the amount of capital to return to the Det≠0 case. In case the three branches have equal organic compositions, prices are identical to values and transformation coefficients are equal to 1.

Let $r^*$ be the value that corresponds to the first cancellation of z when z is decreasing (see chapter İ). Fundamental equality I is verified for $r = r^*$. Fundamental equality II is verified since it was used to determine $k_1$ and $k_2$ as a function of $x_1$ and $x_2$. The $r^*$ value can be approached as precisely as necessary using the algorithm described in the Appendix. The transformation coefficients are determined as a function of $r^*$ in a unique way.

c) *Simple reproduction*

In such a scheme, the production in value of all the commodities, excluding consumer goods, is exactly equal to the needs by the various branches, while the value produced which corresponds to consumer goods is equal to the total wages to which the total surplus value is added. In what follows we do not consider fixed capital and its depreciation. This does not alter the inferences made.

When the organic compositions of the branches are identical, the transformation coefficients $x_i$ are equal to unity and it is then possible, whatever the number of branches, to impose the condition of simple reproduction which is expressed in the case of three branches (E, C, V) by the following equalities.

$$k_1 + k_2 + k_3 = 1$$

$$k_1(w_1 - e_1) - k_2 e_2 - k_3 e_3 = 0$$

$$-k_1 c_1 + k_2(w_2 - c_2) - k_3 c_3 = 0$$

This system of three equations admits one and only one solution if the determinant is non-zero and implies the validity of the third equation below which is then redundant.

$$-k_1(v_1 + pl_1) - k_2(v_2 + pl_3) - k_3(w_3 - v_3 - pl_3) = 0$$

The production of branch 3 (V) is equal to the total surplus-value to which is added the total consumption of the productive class, while the productions in value of branches 1 and 2 correspond exactly to their total consumption by all branches.

When the organic compositions are different, for the system in K, one of Marx's two equalities (sum of prices=sum of values) adds an additional equation to the system (we may recall that Marx's first equality is already satisfied at the level of the system in x by the appropriate choice of the rate of profit):

$$k_1 w_1(1 - x_1) + \ k_2 w_2(1 - x_2) + \ k_3 w_3(1 - x_3) = 0$$

Considering $k_1 + k_2 + k_3 = 1$, only one constraint remains to determine the system. This means that the simple reproduction scheme in the sense of Marx is impossible in the general case with 3 branches of unequal organic composition.

Among an infinite number of possible constraints, we may choose arbitrarily in value:

$$-k_1(v_1 + pl_1) - k_2(v_2 + pl_3) - k_3(w_3 - v_3 - pl_3) = 0$$

Or in price:

$$-k_1(v_1 x_3 + pl_1) - k_2(v_2 x_3 + pl_3) - k_3(w_3 - v_3 - pl_3) = 0$$

d) *Examples*

Table 3A corresponds to the allocation of a capital of 1000 monetary units (m.u.) between the three branches E, C, V. This allocation is based on the arbitrary socio-technical coefficients used by V. Laure van Bambeke ((13), page 176). Tables 3B in value and 3C in price correspond to the solved transformation. The capital of branch 3 has been set at 300 m.u., thereby determining the only possible remaining allocation of the capital in the first two branches. When using the same values as V. Laure Van Bambeke for the capital of branch 3, our solution differs slightly from the "least bad solution" determined by the Moore-Penrose method that V. Laure van Bambeke uses (Table 3E). [5] Applying the above mentioned condition which would correspond to simple reproduction for a system with branches of identical organic composition is shown in Tables 3F and 3G (K1, K2, K3 are imposed by the condition).

| **INIT VALUES** | F/n | E | C | V | PL | W | K |
|---|---|---|---|---|---|---|---|
| BRANCH 1 E | 8.315044 | 19.401807 | 38.803467 | 24.94517 | 24.94517 | 116.410658 | 166.300884 |
| BRANCH 2 C | 1.196996 | 19.949964 | 39.899885 | 47.879993 | 47.879993 | 156.806831 | 119.699802 |
| BRANCH 3 V | 15.073325 | 116.355582 | 232.711164 | 214.2 | 214.2 | 792.540071 | 713.999996 |
| TOTAL | 24.585365 | 155.707353 | 311.414516 | 287.025163 | 287.025163 | 1065.75756 | 1000 |

**Table 3A**

| **VALUES** | F/n | E | C | V | PL | W | K |
|---|---|---|---|---|---|---|---|
| BRANCH 1 E | 16.63376468 | 38.81219294 | 77.62409181 | 49.90137006 | 49.90137006 | 232.8727896 | 332.67530 |
| BRANCH 2 C | 3.673247605 | 61.22088752 | 122.4416431 | 146.9303737 | 146.9303737 | 481.1965256 | 367.32538 |
| BRANCH 3 V | 6.333329867 | 48.88890027 | 97.77780055 | 90.0000005 | 90.0000005 | 333.0000317 | 300 |
| TOTAL | 26.64034215 | 148.9219807 | 297.8435355 | 286.8317443 | 286.8317443 | 1047.069347 | 1000 |

**Table 3B**

| **PRICES** | F/n | E | C | V | S | W | K |
|---|---|---|---|---|---|---|---|
| BRANCH 1 E | 16.63376468 | 46.46117676 | 70.84793374 | 49.31873652 | 95.50500825 | 278.76662 | 332.96549 |
| BRANCH 2 C | 3.673247605 | 73.28610577 | 111.7531583 | 145.2148584 | 105.2633344 | 439.1907044 | 366.9866 |
| BRANCH 3 V | 6.333329867 | 58.5237696 | 89.24233408 | 88.9491873 | 86.06340165 | 329.1120225 | 300.049 |
| TOTAL | 26.64034215 | 178.2710521 | 271.8434261 | 283.4827823 | 286.8317443 | 1047.069347 | 1000 |

**Table 3C**

[5] "Least bad solution" is the expression used by the author himself when using the Moore-Penrose method to obtain an approximated solution. The example displayed in Table 3E uses (for comparison) the exact same capital amount in Branch 3 as the one used in the V. Laure van Bambeke's example (367.9263). The precision of the solutions obtained by our method is limited only by the number of significant digits used (14 here).

|  | K | Kp | $x_i$ | r |
|---|---|---|---|---|
| BRANCH 1 E | 332.67530 | 332.96549 | 1.197076827 | 0.286831548657402 |
| BRANCH 2 C | 367.32538 | 366.9866 | 0.912705477 | 0.286831548657402 |
| BRANCH 3 V | 300 | 300.049 | 0.988324298 | 0.286831548657402 |

**Table 3D** For an allocation with $K_3 = 300$.

|  | K | Kp | $x_i$ | r |
|---|---|---|---|---|
| BRANCH 1 M | 305.5184211 | 305.7797645 | 1.19704086 | 0.286828108 |
| BRANCH 2 E | 326.5559609 | 326.2448384 | 0.912672947 | 0.286828108 |
| BRANCH 3 C | 367.9263 | 367.9760791 | 0.98829028 | 0.286828108 |

**Table 3E** For an allocation with $K_3 = 367.9263$.

Reproduction simple (K1, K2, K3 sont imposés par la condition)

| **VALUES** | F/n | E | C | V | PL | W | K |
|---|---|---|---|---|---|---|---|
| BRANCH 1 E | 12.23844056 | 28.55641674 | 57.11261712 | 36.71537761 | 36.71537761 | 171.3382296 | 244.7688170591 |
| BRANCH 2 C | 2.3535598 | 39.22605697 | 78.45202939 | 94.14269285 | 94.14269285 | 308.3170319 | 235.3563772164 |
| BRANCH 3 V | 10.97514318 | 84.72046958 | 169.4409392 | 155.9626472 | 155.9626472 | 577.0618463 | 519.8754877244 |
| TOTAL | 25.56714354 | 152.5029433 | 305.0055857 | 286.8207176 | 286.8207176 | 1056.717108 | 1000 |

**Table 3F**

| **PRICES** | F/n | E | C | V | S | W | K |
|---|---|---|---|---|---|---|---|
| BRANCH 1 E | 12.23844056 | 34.1809331 | 52.12104427 | 36.28269709 | 70.26215942 | 205.0852744 | 244.9690800373 |
| BRANCH 2 C | 2.3535598 | 46.9520823 | 71.59541801 | 93.03324738 | 67.43619301 | 281.3705005 | 235.1163456971 |
| BRANCH 3 V | 10.97514318 | 101.4071453 | 154.6320084 | 154.1246708 | 149.1223652 | 570.2613329 | 519.9152562655 |
| TOTAL | 25.56714354 | 182.5401607 | 278.3484707 | 283.4406152 | 286.8207176 | 1056.717108 | 1000 |

**Table 3G**

|  | K | Kp | $x_i$ | r |
|---|---|---|---|---|
| BRANCH 1 M | 244.7688170591 | 244.9690800373 | 1.196961558795741 | 0.286820522033078 |
| BRANCH 2 E | 235.3563772164 | 235.1163456971 | 0.9126012235097054 | 0.286820522033078 |
| BRANCH 3 C | 519.8754877244 | 519.9152562655 | 0.9882152780008686 | 0.286820522033078 |

**Table 3H**

### e) *Geometrical interpretation*

The two planes that correspond in K-space to fundamental equality II and conservation of total capital are given by the following two equations:

$$K_1 w_1 (1 - x_1) + K_2 w_2 (1 - x_1) + K_3 w_3 (1 - x_3) = 0$$

$$K_1 + K_2 + K_3 = K_T$$

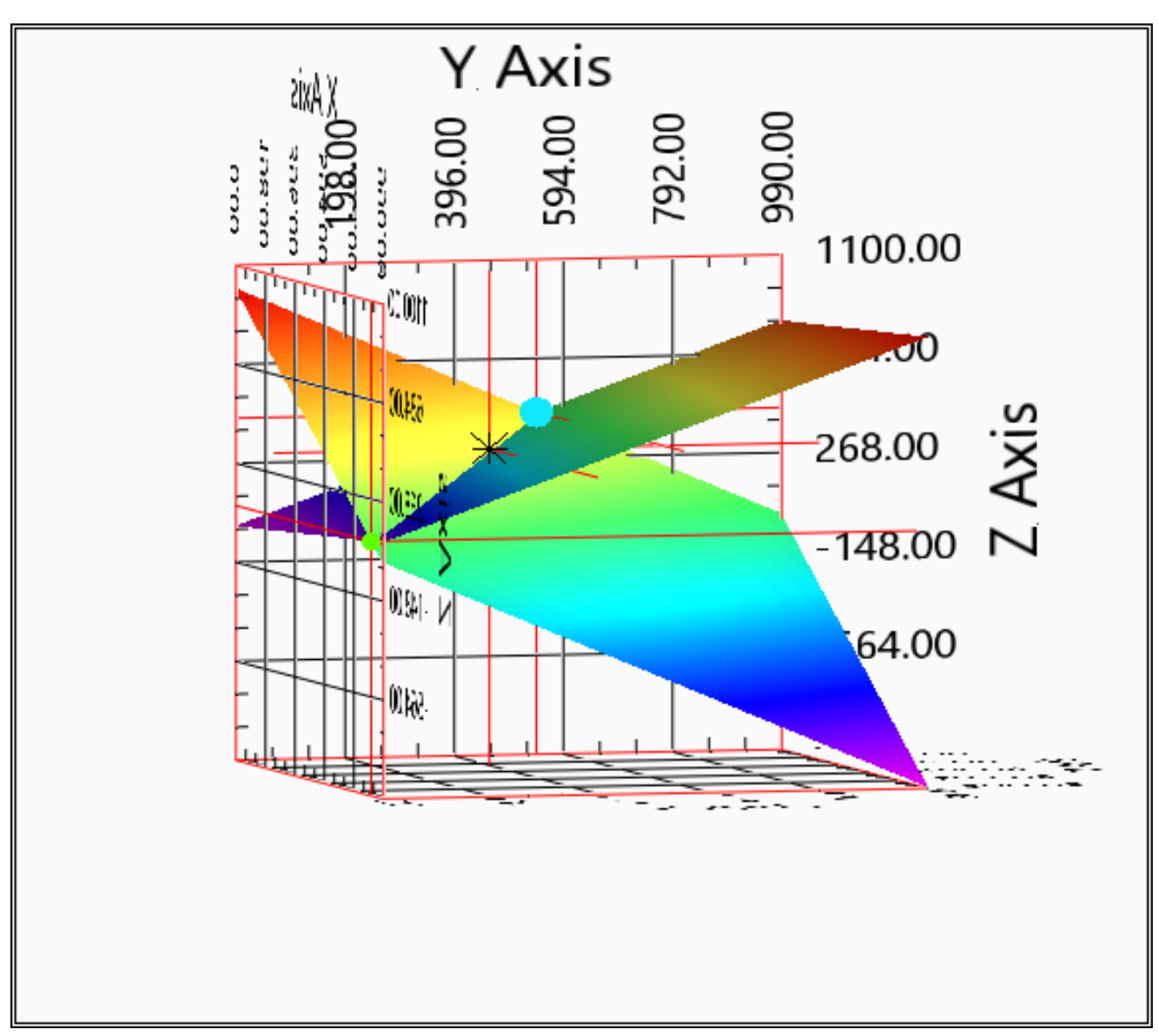

**Figure 2A** The asterisk indicates one of the solutions that lies on the intersection of the two planes. The green and blue dots indicate the approximate boundaries of the solution set.

At the intersection of these two planes is a straight line. The solution $(K_1, K_2, K_3)$ lies on this line (Figure 2A). If we choose a different value of $K_3$, the rate of profit $r$ will also be different as well as $x_1$, $x_2$ and $x_3$ the coefficients of the transformation. The quantities $w_1$, $w_2$ and $w_3$ being constants it follows that the first equation above is modified and now corresponds to a slightly different plane. The second equation expressing the constancy of the total capital does not change. We can deduce that the intersection line of the two planes changes for each value of $K_3$ chosen. The set of solutions is thus positioned along a curved line in the 3 dimensional space $K_1$, $K_2$, $K_3$. (Figure 2B).[6] On Figure 2A the limits of the set of solutions indicated by the blue and green spherical points can only be approximate because of this curvature.

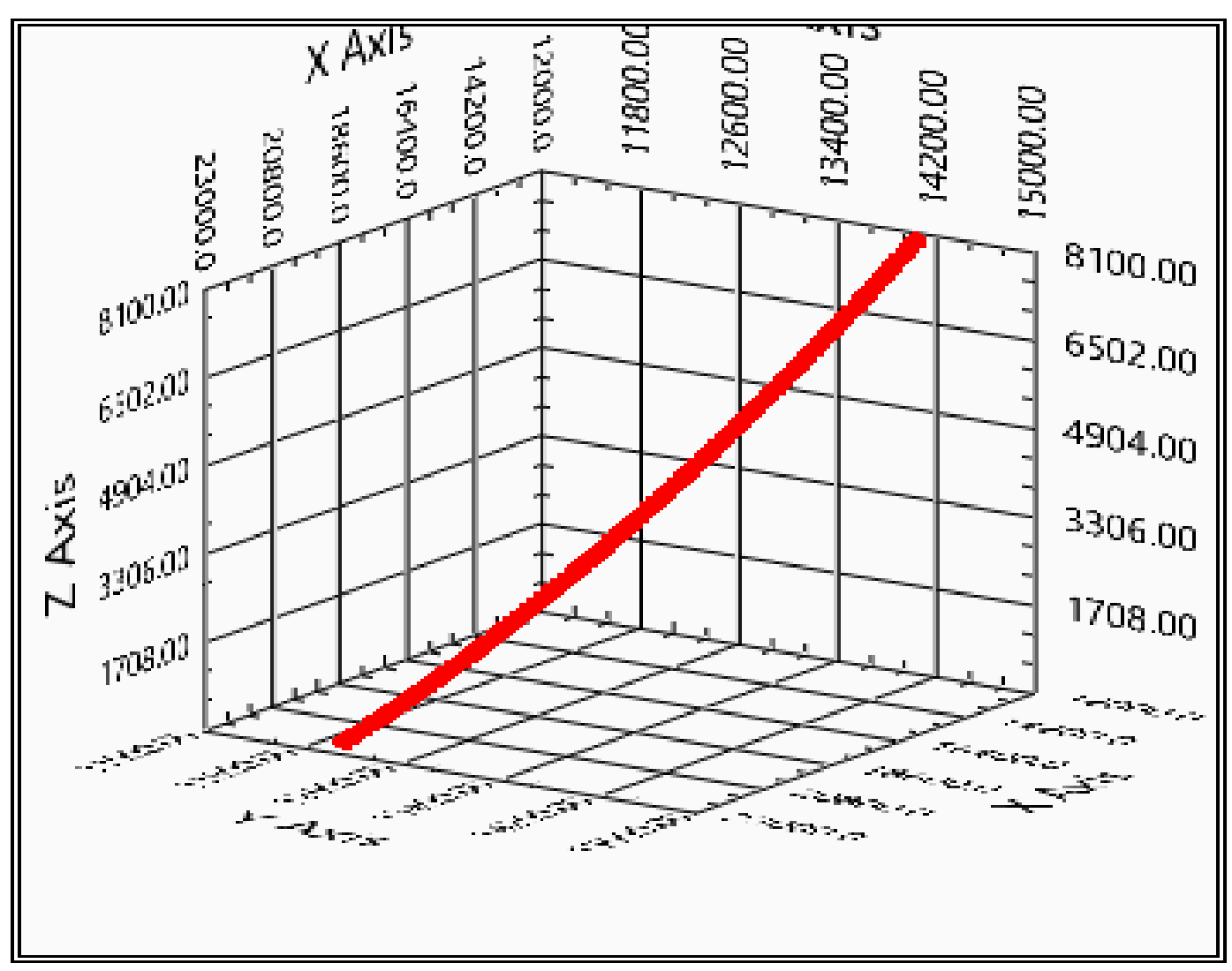

[6] not on a straight line as stated by V. Laure van Bambeke ((13) page 181)

**Figure 2B** The set of solutions is not on a straight segment but on a curved segment (red line). This example uses parameters chosen to obtain a marked curvature.

By an analogous reasoning, we would establish that for a four-branch model, the set of solutions in the space $K_1$, $K_2$, $K_3$, $K_4$ would be distributed over a slightly curved portion of the surface.

### f) *Conclusion*

Whatever the socio-technical coefficients of the three branches, a value $r^*$ of the rate of profit can be found while both fundamental equalities are complied with. Since, in this model, surplus value is the only profit-creating mechanism, at identical operating rates, the internal rate of profit (calculated in value) is greater for the least capital-intensive branch (low organic composition). But in prices, a comparable rate of profit is observed between branches (let us say uniform). Marx described this process by expressing that the "brother" capitalists share the total surplus value, and this sharing is translated into a uniform rate of profit (in prices) (14). In other words, in prices, a highly capitalistic branch can have the same rate of profit as a branch with a low organic composition, even if in value terms it is the latter branch that generates proportionally more surplus-value. *It is notable that, unlike the simplest case with two branches, the conformed allocations are here infinite*.

The *surplus in value* of branch i is defined as the positive difference between the production (in value) of commodity i and the total consumption of this commodity by all branches.[7] There is no profit possible without surplus in value (16). This surplus can be monopolized in shares, in luxury goods or in money by the owners of the means of production. Workers who produce more than they consume cannot absorb all the overproduction. But they can direct their spending towards one product rather than another. This choice will drive the competition between capitalist producers to make their goods desired over others.  If, for example, an increased urge for cell phones is generated, it will require an increase in the extraction of certain minerals, altering the pattern of global production. Such competition between capitalists, while satisfying a uniform rate of profit in all branches, is only possible if there is more than one possible allocation for capital between branches. We show that, as soon as surplus value is non-zero (and therefore surplus in value and profit are non-zero), there is an infinite number of possible capital allocations in conformity with the transformation and complying with fundamental equalities. We may say that the occurrence of surplus brings, in both senses of the word, *play* into the economy.

In the "real" world, if one takes a momentary view, it seems that only one of an infinite number of possible capital allocations is being settled. This equilibrium is determined by the reality of the world, the attractiveness of certain products rather than others, the relative success of branches of production, or other parameters that are not relevant to the transformation problem (e.g.

[7] "Surplus *in* value" is not to mistaken with the "surplus value" which has been defined in A-1-a.

demography). The transformation merely consists in finding the possible allocations, i.e. compatible with the laws of conservation (fundamental equalities) and the establishment of a uniform rate of profit. Besides, these allocations change all the time due to the effects of technology, advertising, i.e. ultimately competition between the different commodities. If the transformation were so rigid as to allow only one solution, then it would clearly contradict the reality of the world where capitalists are competing with each other.

## 2. Three-branch model with zero fixed capital and zero profit

### a) *Solving the transformation*

In the case of zero fixed capital and zero profit (i.e. zero surplus value), equations (5) and (6) for the system of equations of the branches in prices becomes:

$$(e_1 - w_1)x_1 + c_1x_2 + v_1x_3 = 0$$

$$e_2x_1 + (c_2 - w_2)x_2 + v_2x_3 = 0$$

$$e_3x_1 + c_3x_2 + (v_3 - w_3)x_3 = 0$$

Or:

$$w_1x_1 = e_1x_1 + c_1x_2 + v_1x_3$$

$$w_2x_2 = e_2x_1 + c_2x_2 + v_2x_3$$

$$w_3x_3 = e_3x_1 + c_3x_2 + v_3x_3$$

Or eventually:

$$[e_1 - w_1]x_1 + c_1x_2 + v_1x_3 = 0$$

$$e_2x_1 + [c_2 - w_2]x_2 + v_2x_3 = 0$$

$$e_3x_1 + c_3x_2 + [v_3 - w_3]x_3 = 0$$

Given that $w_i = c_i + e_i + v_i \quad \forall i,$ we deduce that $x_i = x_j \ \forall i, j$

As explained below, the compliance with fundamental equality II is only achieved for $x_1 = x_2 = x_3 = 1.$

Indeed, fundamental equality II is

$$k_1w_1 + k_2w_2 + k_3w_3 = k_1x_1w_1 + k_2x_2w_2 + k_3x_3w_3$$

$x_1 = x_2 = x_3$ implies $x_1 = x_2 = x_3 = 1$

Fundamental equality I is now:

$$k_1(x_1(w_1 - e_1) - x_2c_1 - x_3v_1) + k_2(x_2(w_2 - c_2) - x_1e_2 - x_3v_2) + k_3(x_3(w_3 - v_3) - x_1e_1 - x_2c_2) = 0$$

which can be written:

$[k_1 w_1 - (k_1 e_1 + k_2 e_2 + k_3 e_1)]x_1 + [k_2 w_2 - (k_1 c_1 + k_2 c_2 + k_3 c_3)]x_2 + [k_3 w_3 - (k_1 v_1 + k_2 v_2 + k_3 v_3)]x_3 = 0$

Condition to fulfill the solvent need:

Now, the quantity of commodities produced in each of the three branches must be greater than or equal to the total quantity of that commodity consumed in the three branches.[8] These inequalities apply to prices, which means:

$$k_1 w_1 x_1 \geq (k_1 e_1 + k_2 e_2 + k_3 e_3)x_1$$

$$k_2 w_2 x_2 \geq (k_1 c_1 + k_2 c_2 + k_3 c_3)x_2$$

$$k_3 w_3 x_3 \geq (k_1 v_1 + k_2 v_2 + k_3 v_3)x_3$$

Therefore, fundamental equality I is only possible when the following three equalities are satisfied simultaneously:

$$k_1 w_1 = k_1 e_1 + k_2 e_2 + k_3 e_3$$

$$k_2 w_2 = k_1 c_1 + k_2 c_2 + k_3 c_3$$

$$k_3 w_3 = k_1 v_1 + k_2 v_2 + k_3 v_3$$

*This model with no fixed capital, nor surplus value, highlights the significance of fundamental equalities: they express the adequacy between production and the solvent need.*

These equalities can also be written:

$$k_1(w_1 - e_1) - k_2 e_2 - k_3 e_3 = 0$$

$$-k_1 c_1 + k_2(w_2 - c_2) - k_3 c_3 = 0$$

$$-k_1 c_1 + k_2(w_2 - c_2) - k_3 c_3 = 0$$

The determinant of this system is zero, so we shall retain only two independent equations.

Taking into account the equality $k_1 + k_2 + k_3 = 1$, as the third independent equation of the system, we have :

$$k_1[(w_1 - e_1)] - k_2 e_2 - k_3 e_3 = 0$$

$$-k_1[(c_1)] + k_2(w_2 - c_2) - k_3 c_3 = 0$$

$$k_1 + k_2 + k_3 = 1$$

---

[8] These inequalities may be transiently violated when considering buildup of stocks in previous cycles.

$$D = \begin{vmatrix} (w_1 - e_1) & -e_2 & -e_3 \\ -c_1 & (w_2 - c_2) & -c_3 \\ 1 & 1 & 1 \end{vmatrix}$$

$D \neq 0$, then there is a unique solution for $(k_1, k_2, k_3)$ :

$$k_1 = \frac{\begin{vmatrix} 0 & -e_2 & -e_3 \\ 0 & (w_2 - c_2) & -c_3 \\ K_T & 1 & 1 \end{vmatrix}}{D}$$

$$k_2 = \frac{\begin{vmatrix} (w_1 - e_1) & 0 & -e_3 \\ -c_1 & 0 & -c_3 \\ 1 & 1 & 1 \end{vmatrix}}{D}$$

$$k_3 = \frac{\begin{vmatrix} (w_1 - e_1) & -e_2 & 0 \\ -c_1 & (w_2 - c_2) & 0 \\ 1 & 1 & 1 \end{vmatrix}}{D}$$

With, $\forall i$, $K_i = k_i\, K_T$

### b) *Examples*

Table 4A replicates the example of Tables 3, but with no surplus value and no fixed capital. The transformation then yields only one possible allocation (Table 4B).

| **INIT VALUES** | F/n | E | C | V | PL | W | K |
|---|---|---|---|---|---|---|---|
| BRANCH 1 E | 0 | 19.401807 | 38.803467 | 24.94517 | 0 | 83.150444 | 83.150444 |
| BRANCH 2 C | 0 | 19.949964 | 39.899885 | 47.879993 | 0 | 107.729842 | 107.729842 |
| BRANCH 3 V | 0 | 116.355582 | 232.711164 | 214.2 | 0 | 563.266746 | 563.266746 |
| TOTAL | 0 | 155.707353 | 311.414516 | 287.025163 | 0 | 754.147032 | 754.147032 |

**Table 4A**

| **VALUES or PRICES** | F/n | E | C | V | PL | W | K |
|---|---|---|---|---|---|---|---|
| BRANCH 1 E | 0 | 35.77724779 | 71.55422451 | 45.99929935 | 0 | 153.3307716 | 153.33077165 |
| BRANCH 2 C | 0 | 56.78908263 | 113.5780429 | 136.2940243 | 0 | 306.6611498 | 306.66114982 |
| BRANCH 3 V | 0 | 60.76444123 | 121.5288825 | 111.8617868 | 0 | 294.1551105 | 294.15511053 |
| TOTAL | 0 | 153.3307716 | 306.6611498 | 294.1551105 | 0 | 754.147032 | 754.147032 |

**Table 4B**

Table 4C replicates the example of Table 3, but with an absence of surplus value only. The transformation then gives only one possible allocation (Table 4D).

| INIT VALUES | F/n | E | C | V | PL | W | K |
|---|---|---|---|---|---|---|---|
| BRANCH 1 E | 8.315044 | 19.401807 | 38.803467 | 24.94517 | 0 | 91.465488 | 166.300884 |
| BRANCH 2 C | 1.196996 | 19.949964 | 39.899885 | 47.879993 | 0 | 108.926838 | 119.699802 |
| BRANCH 3 V | 15.073325 | 116.355582 | 232.711164 | 214.2 | 0 | 578.340071 | 713.999996 |
| TOTAL | 24.585365 | 155.707353 | 311.414516 | 287.025163 | 0 | 778.732397 | 1000 |

**Table 4C**

| VALUES or PRICES | F/n | E | C | V | PL | W – F/n | K |
|---|---|---|---|---|---|---|---|
| BRANCH 1 E | 15.0284791 | 35.0665193 | 70.13277292 | 45.08550597 | 0 | 150.284798 | 300.5695891 |
| BRANCH 2 C | 3.339651635 | 55.66094614 | 111.3217723 | 133.5864923 | 0 | 300.569211 | 333.9657271 |
| BRANCH 3 V | 7.715375723 | 59.55733274 | 119.1146655 | 109.6396104 | 0 | 288.311609 | 365.4653658 |
| TOTAL | 26.08350645 | 150.2847982 | 300.5692107 | 288.3116086 | 0 | 739.165617 | 1000 |

**Table 4D**

## 3. Conclusion

The three-branch system with zero fixed capital and zero surplus value highlights the meaning of fundamental equalities. They express the adequacy between the production and the solvent need. In this simplified model, the fundamental equalities force a unique allocation of capital between the three branches that ensures this adequacy.

To generalize, whatever the number of branches, the economic model without fixed capital (or the one in which its production is endogenous -paragraph F), nor surplus-value, leads to a unique allocation of capital such that commodities are produced in quantities that correspond exactly to their consumption (solvent social need).

The solving approach presented here and the associated algorithm (Appendix) is generalizable to any number of branches. For an economic system with N branches, the number of unknowns to be determined is 2N + 1 and the number of *independent* equations in the system to solve is N + 3. [9,10] When N = 2 (two branches), 2N + 1 = N +3, the system is determined and yields a single solution. When N > 2 then 2N +1 > N+3, the system is underdetermined and yields an infinity of solutions. Contrary to what has been stated by V. Laure van Bambeke, whatever the branch number, the system is never overdetermined. [11]

---

[9] 2N + 1 is the number of the following unknowns: $x_1, x_2, \dots, x_N, r, k_1, k_2, \dots, k_N$

[10] The N + 3 equations are N equations in $x_i$, two equations in $k_i$ (Marx's fundamental equalities) and one equation in $k_i$ for the constancy of the total capital committed.

[11] « Pour une économie formée de deux branches, le système complet est compose de quatre équations pour deux inconnues (*x*1 et *x*2) et trois paramètres (*r, K*1 et *K*2) . Il semble surdéterminé. » (4) This view that the equation system to be solved is overdetermined is also the core of the method used by V. Laure van Bambeke in later works (13).

## C- Particularity of the cases for which the fixed capital is zero

### 1. Zero fixed capital and non-zero uniform profit

We deal with this particular case because it provides a means of responding to the criticisms that have been made on the role of fixed capital in the approach of V. Laure van Bambeke (11). We show that the transformation problem is solved in the same way whether or not fixed capital is considered. In the case of zero fixed capital and uniform profit and when the number of branches is higher than two, we show a noticeable property: considering a given total amount of capital, although there is an infinity of possible allocations of capital between branches complying with both fundamental equalities and a uniform rate of profit, all these allocations yield the same the total surplus value.

#### a) *The rate of profit derived from the eigenvalue of the matrix of socio-technical coefficients*

The system without fixed capital is closed, i.e. each commodity is produced from commodities also produced by the system. The system of equations for prices (8b) reads:

$$w_1x_1 = (1+r)e_1x_1 + (1+r)c_1x_2 + (1+r)v_1x_3$$

$$w_2x_2 = (1+r)e_2x_1 + (1+r)c_2x_2 + (1+r)v_2x_3$$

$$w_3x_3 = (1+r)e_3x_1 + (1+r)c_3x_2 + (1+r)v_3x_3$$

This system of homogeneous equations has a zero determinant, hence an infinite number of solutions.

Let $\boldsymbol{x_u}(x_1{}^*, x_2{}^*, x_3{}^*)$, the particular solution whose modulus is unity (unit vector). All the collinear vectors $q.\boldsymbol{x_u}(x_1{}^*, x_2{}^*, x_3{}^*)$ (q real $> 0$) are also solutions. Thus, there exists a consistent system of relative prices, with transformation coefficients:

$$x_i = q \times x_i{}^*, \qquad \forall i = 1,2,3$$

Note that, amongst this infinity of solutions, a particular one can be found, which complies with both fundamental equalities (see part C-1-c). For this particular solution, $q$ is denoted $q^*$.

From the above system of equations, we obtain:

$$(1+r) = \frac{w_i x_i{}^*}{e_i x_1{}^* + c_i x_2{}^* + v_i x_3{}^*}, \forall i = 1,2,3 \qquad \textbf{(9)}$$

Thus, the $r$-value is unambiguously defined and can be calculated from the unit vector $\boldsymbol{x_u}$. In fact, $\frac{1}{(1+r)}$ is the eigenvalue and $\boldsymbol{x_u}$ is the unit eigenvector of the linear transformation of equation 9 equivalent to matrix equation:

$$\begin{bmatrix} \mathbf{e_1}/w_1 & \mathbf{c_1}/w_1 & \mathbf{v_1}/w_1 \\ \mathbf{e_2}/w_2 & \mathbf{c_2}/w_2 & \mathbf{v_2}/w_2 \\ \mathbf{e_3}/w_3 & \mathbf{c_3}/w_3 & \mathbf{v_3}/w_3 \end{bmatrix} \begin{pmatrix} x_1 \\ x_2 \\ x_3 \end{pmatrix} = \frac{1}{(1+r)} \begin{pmatrix} x_1 \\ x_2 \\ x_3 \end{pmatrix}$$

The above matrix, that we define as the *matrix of socio-technical coefficients*, accounts for the requirements in each commodity that goes into the production of a given commodity.

The rate of profit $r$ considered above is therefore calculated independently of the capital employed. This also means that $r$ does not vary with capital allocation: $dr = 0$.

b) *Invariance of total surplus value with capital allocation*

Note that we define three distinct quantities:

$$r_i = \frac{S_i}{K_{pi}} \qquad pl_i = \frac{PL_i}{K_i} \qquad s_i = \frac{S_i}{K_i}$$

$r_i$ represents the real rate of profit, the one appreciated in practice by the capitalist. $pl_i$ is the internal rate of profit, which is inaccessible but plays a pivotal role "behind the curtain". $s_i$ is the profit in price per unit of committed capital in value. Unlike $s_i$, $r_i$ and $pl_i$ are ratios of "homogeneous" quantities (in prices and values, respectively).

The total profit $S$ can be written as the scalar product:

$$S = \boldsymbol{r}.\mathbf{K_p} = \sum_i r_i.K_{pi} \qquad \textbf{(10)}$$

(Bold characters are used for vector notation.)

Infinitesimal change in total profit amount can be written:

$$dS = \boldsymbol{r}.\mathbf{dK_p} + \mathbf{K_p}.\boldsymbol{dr} \qquad \textbf{(11)}$$

In $\mathrm{K_p}$-space, vector $\mathbf{dK_p}$ lies within the plane defined by the equation $\sum_i.K_{pi} = \mathrm{K_T}$ (where $\mathrm{K_T}$ denotes the total capital committed). Since the profit $\boldsymbol{r}$ is uniform, its components are identical and its vector is perpendicular to the $\mathrm{K_p}$-space. Therefore, the scalar product of the first member in the above equation is zero. Since $dr = 0$, the scalar product of the second member is zero too. Therefore:

$$dS = 0$$

Compliance with fundamental equality I (the sum of the profits is equal to the sum of the surplus values) implies that total surplus value $PL$ remains constant.

$$dPL = \boldsymbol{pl} \,.\, \boldsymbol{dK} = \sum_i pl_i \; dK_i = 0$$

It follows that the only capital flows allowed are in the plane defined by the equality of the sum of the capitals and perpendicular to vector $\boldsymbol{pl}$ $(pl_1, pl_2, pl_3)$ in the K-space.

c) *Rate of profit determination from committed capital amounts*

The most straight forward approach to calculate the average rate of profit is from the capital committed to each branch. This way, the internal rate of profit (in value, named $r_v$) and the rate of profit (in price) are expressed as follows.

$$r_v = \frac{\sum_i K_i[w_i - (e_i + c_i + v_i)]}{K_T} = \frac{\sum_i K_i[w_i - 1]}{K_T}$$

$$r = \frac{q^* \sum_i K_i[w_i x_i^* - (e_i x_1^* + c_i x_2^* + v_i x_3^*)]}{K_T}$$

With arbitrary allocations of capital, not only do these two average rates of profit (in value and price) not coincide, but they generally vary from one allocation to another.

As illustrated in Tables 5A and B, when the total capital $K_T$ is allocated arbitrarily between branches without complying with the fundamental equalities, the internal rate of profit $r_v$ as well as the total production differ from one allocation to another. In the absence fixed capital, this may seem rather counter intuitive.

| **INIT VALUES** | F/n | E | C | V | PL | W | K |
|---|---|---|---|---|---|---|---|
| BRANCH 1 E | 0 | 19.401807 | 38.803467 | 24.94517 | 24.94517 | 108.095614 | 83.150444 |
| BRANCH 2 C | 0 | 19.949964 | 39.899885 | 47.879993 | 47.879993 | 155.609835 | 107.729842 |
| BRANCH 3 V | 0 | 116.355582 | 232.711164 | 214.2 | 214.2 | 777.466746 | 563.266746 |
| TOTAL | 0 | 155.707353 | 311.414516 | 287.025163 | 287.025163 | 1041.172195 | 754.14703200000 |

**Table 5A** $r_v$ = 0.380595760.

| **INIT VALUES** | F/n | E | C | V | PL | W | K |
|---|---|---|---|---|---|---|---|
| BRANCH 1 E | 0 | 58.655991 | 117.311538 | 75.414814 | 75.414814 | 326.7971584 | 251.382344 |
| BRANCH 2 C | 0 | 46.552270 | 93.1044400 | 111.72563 | 111.725633 | 363.107977 | 251.382344 |
| BRANCH 3 V | 0 | 51.928751 | 103.857503 | 95.5960891 | 95.5960891 | 346.978433 | 251.382344 |
| TOTAL | 0 | 157.137013 | 314.273481 | 282.736537 | 282.736537 | 1036.883569 | 754.14703200000 |

**Table 5B** Other allocation with the total capital divided into three equal parts. $r_v$ = 0.374909036.

$q^*$ is determined from the compliance with fundamental equality II:

$$q^* = \frac{\sum_i K_i w_i}{\sum_i K_i w_i x_i^*}$$

Now, for $q^*$, which corresponds to capital allocation complying with both fundamental equalities, since $K_T = q^* \sum_i K_i \; (e_i x_1^* + c_i x_2^* + v_i x_3^*)$, it can be verified that:

$$r_v = r.$$

Compliance with both fundamental equalities implies the equality between the global rates of profit calculated in value or in price and their concordance with that obtained from matrix of socio-technical coefficients.

d) *Algorithm to determine $q^*$*

As a reminder, fundamental equality II for the solution $q\,\boldsymbol{x_u}$ reads:

$$K_1\, w_1(1 - q x_1^*) + K_2\, w_1(1 - q x_2^*) + K_3\, w_1(1 - q x_3^*) = 0$$

By linear approximation and iterative bracketing (for $K_3$ fixed), the couple $(K_1, K_2)$ is solution of the system of equations:

$$K_1\, w_1(1 - qx_1{}^*) + K_2\, w_1(1 - qx_2{}^*) + K_3\, w_1(1 - qx_3{}^*) = 0$$

$$K_1 + K_2 + K_3 = K_T$$

As in the situation with fixed capital, the solution must also comply with fundamental equality I which corresponds to the cancellation of the z-function:

$$z = K_1[qx_1{}^*(w_1 - e_1) - qx_2{}^*c_1 - qx_3{}^*v_1 - pl_1] + K_2[qx_2{}^*(w_2 - c_2) - qx_1{}^*e_2 - qx_3{}^*v_2 - pl_2] + K_3[qx_3{}^*(w_3 - v_3) - qx_1{}^*e_3 - qx_2{}^*v_3 - pl_3]$$

Following the bracketing by a pair (z1, z2) such that z1<0 and z2>0, we determine by interpolation the particular value $q^*$ which cancels the z-function. This method is very similar to the one used with fixed capital, except that only the modulus $q^*$ of the price transformation vector is to be found (see Appendix).

e) *The various capital allocations complying with both fundamental equalities*

Considering the transformation and the compliance with both fundamental equalities, the average rates of profit in value ($r_v$) and in price ($r$) are equal. Furthermore, since their value is calculated from the eigenvalue of the matrix of socio-technical coefficients, independently of capital allocation, the rate of profit does not vary with capital allocation amongst the infinite possibility as soon as there are three or more branches.

Tables 6A (value) 6B (price) and 7A (value) 7B (price) are constructed from the random allocation in Table 5A and correspond to a choice of capital in branch 3 of 230 and 300 monetary units respectively. They illustrate the constancy of PL and S (and thus $r_v$ and $r$) for different capital allocations.

| **VALUES** | F/n | E | C | V | PL | W | K |
|---|---|---|---|---|---|---|---|
| BRANCH 1 E | 0 | 44.98323645 | 89.96613207 | 57.83556554 | 57.83556554 | 250.6204996 | 192.7849340 |
| BRANCH 2 C | 0 | 61.36333074 | 122.7265292 | 147.272238 | 147.272238 | 478.6343359 | 331.3620979 |
| BRANCH 3 V | 0 | 47.5117412 | 95.02348239 | 87.46477641 | 87.46477641 | 317.4647764 | 230 |
| TOTAL | 0 | 153.8583084 | 307.7161437 | 292.5725799 | 292.5725799 | 1046.719612 | 754.147032 |

**Table 6A**

| **PRICES** | F/n | E | C | V | S | W | K |
|---|---|---|---|---|---|---|---|
| BRANCH 1 E | 0 | 47.98869505 | 86.4912726 | 58.15294787 | 74.73225603 | 267.3651716 | 192.63292 |
| BRANCH 2 C | 0 | 65.46319026 | 117.9863294 | 148.080419 | 128.6175843 | 460.147523 | 331.52994 |
| BRANCH 3 V | 0 | 50.68613643 | 91.3532874 | 87.94475398 | 89.22273959 | 319.2069174 | 229.98418 |
| TOTAL | 0 | 164.1380218 | 295.8308894 | 294.1781208 | 292.5725799 | 1046.719612 | 754.147032 |

**Table 6B**

| | r | $x_i$ |
|---|---|---|
| BRANCH 1 E | 0.38795164275602 | 1.066812858 |
| BRANCH 2 C | 0.38795164275602 | 0.961375916 |
| BRANCH 3 V | 0.38795164275602 | 1.005487667 |

**Table 6C**

| **VALUES** | F/n | E | C | V | PL | W | K |
|---|---|---|---|---|---|---|---|
| BRANCH 1 E | 0 | 37.72783071 | 75.45537557 | 48.50719063 | 48.50719063 | 210.1975875 | 161.6903969 |
| BRANCH 2 C | 0 | 54.15861783 | 108.3171189 | 129.9808983 | 129.9808983 | 422.4375334 | 292.4566350 |
| BRANCH 3 V | 0 | 61.97183634 | 123.9436727 | 114.084491 | 114.084491 | 414.084491 | 300 |
| TOTAL | 0 | 153.8582849 | 307.7161672 | 292.5725799 | 292.5725799 | 1046.719612 | 754.147032 |

**Table 7A**

| **PRICES** | F/n | E | C | V | S | W | K |
|---|---|---|---|---|---|---|---|
| BRANCH 1 E | 0 | 40.24853505 | 72.54098102 | 48.7733821 | 62.67859176 | 224.2414899 | 161.5629 |
| BRANCH 2 C | 0 | 57.77711008 | 104.1334697 | 130.6941907 | 113.5165014 | 406.1212719 | 292.60477 |
| BRANCH 3 V | 0 | 66.11235209 | 119.1564622 | 114.7105491 | 116.3774868 | 416.3568502 | 299.97936 |
| TOTAL | 0 | 164.1379972 | 295.830913 | 294.1781218 | 292.5725799 | 1046.719612 | 754.147032 |

**Table 7B**

| | r | $x_i$ |
|---|---|---|
| BRANCH 1 M | 0.38795164275602 | 1.066812862 |
| BRANCH 2 E | 0.38795164275602 | 0.961375919 |
| BRANCH 3 C | 0.38795164275602 | 1.00548767 |

**Table 7C**

f) *Matrix of socio-technical coefficients*

The matrix of socio-technical coefficients takes into account the requirements in each commodity which goes into the production of a given commodity. It ignores the conditions of satisfaction of solvent social needs, but it contains the information on the profit that will be realized for each unit produced and sold; it is a profit per unit of capital, and thus a potential rate of profit. We have shown that, in the case of a uniform rate of profit, the same rate is generated by any capital allocation complying with both Marx's fundamental equalities. This finding further supports the fundamental importance of these equalities and the coherence of Marx's theory of value. Only the compliance with the two fundamental equalities will make it possible to know the adequate allocation of the capital making this rate of profit effective.

g) *Permitted values of capital flows between branches in the case of a closed system (without fixed capital) and with a uniform rate of profit, an example*

We have shown that the total surplus value is constant for all permitted allocations of capital, i.e.:

$$\Delta PL = \boldsymbol{pl}\,.\,\Delta\boldsymbol{K} = \sum_i pl_i\,\Delta K_i = 0$$

The possible transfer vectors $\Delta\boldsymbol{K}$ lie in the $\sum_i K_i = K_T$ plane and are orthogonal to the vector $\boldsymbol{pl}$.

Let consider the case for which only branch 1 produces surplus value. For total $PL$ to remain unchanged, it is necessary that $K_1$ also remains unchanged. Therefore, only $\Delta\boldsymbol{K}$ transfers orthogonal to the $K_1$ axis are possible. This means that in this case, only transfers between branches 2 and 3 are possible and there is only one solution for $K_1$.

This case is illustrated by Tables 8A-C. Note that the value of the amount of capital in branch 1 (121.90 m.u.; Table 8B) is lower than that found when there was no surplus value at all (153.33 m.u. ; Table 4B). This example illustrates that with surplus value produced, branch 1 can satisfy the solvent social need with a smaller fraction of the total capital. In other words, there is an additional flexibility to meet the solvent social need. Therefore, the capital allocated to this branch can be reduced without lacking the value to meet the needs.

| **VALUES** | F/n | E | C | V | PL | W | K |
|---|---|---|---|---|---|---|---|
| BRANCH 1 E | 0 | 28.66090532 | 57.32159349 | 36.84972001 | 36.84972001 | 159.6819388 | 122.8322188 |
| BRANCH 2 C | 0 | 61.3545743 | 122.7090164 | 147.2512225 | 0 | 331.3148132 | 331.3148132 |
| BRANCH 3 V | 0 | 61.97183634 | 123.9436727 | 114.084491 | 0 | 300 | 300 |
| TOTAL | 0 | 151.987316 | 303.9742825 | 298.1854335 | 36.84972001 | 790.996752 | 754.147032 |

**Table 8A**

| **PRICES** | F/n | E | C | V | S | W | Kp |
|---|---|---|---|---|---|---|---|
| BRANCH 1 E | 0 | 22.95038931 | 60.35473264 | 38.60430098 | 5.956833231 | 127.8662562 | 121.90942 |
| BRANCH 2 C | 0 | 49.13003795 | 129.2020934 | 154.2625158 | 16.25149885 | 348.846146 | 332.59465 |
| BRANCH 3 V | 0 | 49.62431418 | 130.5020808 | 119.5165669 | 14.64138793 | 314.2843499 | 299.64296 |
| TOTAL | 0 | 121.7047414 | 320.0589069 | 312.3833837 | 36.84972001 | 790.996752 | 754.147032 |

**Table 8B**

| | r | $x_i$ |
|---|---|---|
| BRANCH 1 E | 0.048862779328106044 | 0.80075591 |
| BRANCH 2 E | 0.048862779328107064 | 1.052914425 |
| BRANCH 3 C | 0.048862779328106419 | 1.0476145 |

**Table 8C**

## 2. Zero fixed capital and different rates of profit between branches

In contrast to the previous situation, when rates of profit are no longer identical in all branches, total profit as well as total production change with capital allocation. Consider the following equality:

$$PL = \sum_i K_i \, pl_i = K_1 \, pl_1 + K_2 \, pl_2 + K_3 \, pl_3 = K_T . r$$

In this equation, we cannot simply replace $r$ by its average value $< r >$ when the capital allocation between branches that allows us to calculate $< r >$ is not known. $< r >$ can no longer be obtained by means of the eigenvalue of a matrix, even though we are dealing with an equation without a second member.

Note that in this configuration of different rates of profit, the particular value of $q$ ($q^*$) which cancels the z-function, changes with the allocations, so the modulus of the price vector is also modified and the set of solutions is a curved segment when there are three branches as in the case with fixed capital.

The idea of different rates of profit for different branches is quite conceivable, for example by imagining a situation of monopoly for one of the branches. Even if the situation without fixed capital looks artificial, its consideration is nevertheless of theoretical interest since it provides a way to challenge the consistency of the concepts. Even when the rate of exploitation remains similar in all branches, a monopoly situation leads to an unequal share of surplus value between capitalists. In this case the different rates of profit $r_i$ remain determined independently of the capital employed (equation 9 always applies provided that the small index i is added to the rate of profit concerning branch i) while the average rate of profit $< r >$ varies according to the distribution of capital among the N branches.

$$< r > = \frac{1}{\mathrm{K_T}} \sum_{\mathrm{i}}^{\mathrm{N}} \mathrm{K_i} \, \mathrm{r_i}$$

When rates of profit differ between branches then the vector $\boldsymbol{r}$ is no longer perpendicular to the capital flow $\boldsymbol{dK}_{pi}$. The first scalar product of equation 11 no longer cancels out, on the other hand the modulus of $\boldsymbol{r}$ can no longer be a factor in the sum of equation 10 and the total amount of profit is therefore no longer invariable and depends on capital allocation.

## 3. Difference from the case with fixed capital amortization

We have shown that, in a system devoid of fixed capital and for uniform rate of profit, the sum of profit or surplus-value is constant whatever the capital mobility across branches, provided Marx's fundamental equality are complied with. In a system including fixed capital, the situation is quite different.

In this case the expression for the profit of branch i contains a term $d_i$ corresponding to the depreciation of this fixed capital.

$$S_i = K_{ip} r_i = K_i s_i = K_i (w_i x_i - e_i x_1 - c_i x_2 - v_i x_3 - d_i)$$

Since the change in prices, a function of the capital allocation, does not affect the supplementary coefficient, the production and costs of a branch no longer vary proportionally with each other. The change with capital allocation of the price vector $\boldsymbol{x}$ is no longer a change of the module alone, the ratio of commodity prices changes and total production as well as the rate of profit can vary with the capital allocation. The rate of profit in this case can no longer be derived from the calculation of the eigenvalue of a matrix.

## 4. Conclusion

The consideration of the case with zero fixed capital and a uniform rate of profit, for which the exact calculation of the rate of profit is possible without having to consider the amounts of capital allocated to the different branches, may have misled economic theories into thinking that the allocation of capital between branches is not relevant (12, 15, 17). However, we show that even in these cases, capital allocation as determined through the transformation is essential. Indeed, with an arbitrary capital allocation (Tables 5) the rate of profit in value does not coincide with that obtained from the eigenvalue of the matrix. The coincidence of these two rates occurs when fundamental equalities are complied with, and thus when the solvent social need are satisfied. In this case, the total profit is invariant, which explains why it is possible to determine the rate of profit from the socio-technical coefficient matrix (independently of capital allocation) (15). In the case of non-uniform rates of profit, rates of profit of each branch are invariant for all *conformed* allocations (still determined using equation 9) but capital allocation becomes indispensable to the calculation of the average rate of profit. [12]

---

[12] By *conformed* allocations, we mean allocations complying with both fundamental equalities

## D- Convergence criteria in a "real" process

In the above, we have developed a method for the direct determination of the allocation of capital that corresponds to the final state of equilibrium of the system with the end point of the process being known (uniform rates of profit or rates differing by a given deviation) and the allocation of capital initially entered defining the socio-technical coefficients for each branch, our method makes it possible to deduce what capital transfers have been required for the transformation problem to be solved. In what we will call here (by linguistic abuse) a "real process", the rates of profit of the different branches gradually reach their equilibrium values through successive exchanges of capital that are systematically transferred to the most profitable branches. In a simulation, convergence to a uniform rate of profit could be achieved through a process of successive transfers adjusted according to rate of profit gradients between branches.

In order to simulate this "real process" of convergence, we set initial rates of profit different between branches and followed the capital flows leading, step by step, to an intermediate rate of profit. For a three-branch model without fixed capital, with the rate of profit of branch 2 as a reference, the starting rates of profit $r_1$ and $r_3$ are such that $r_1 = r_2 + \Delta r$ and $r_3 = r_2 - \Delta r$. The simulation consists of gradually decrementing the differential $\Delta r$ to zero (and then negative) through successive iterations (see Appendix 2). The capital amounts $K_1$ and $K_3$ (in value) are determined by the program at each iteration to comply with both fundamental equalities.

Figure 3 shows an example of such convergence where the higher rate of profit branch attracts capital from the lower rate of profit branch.

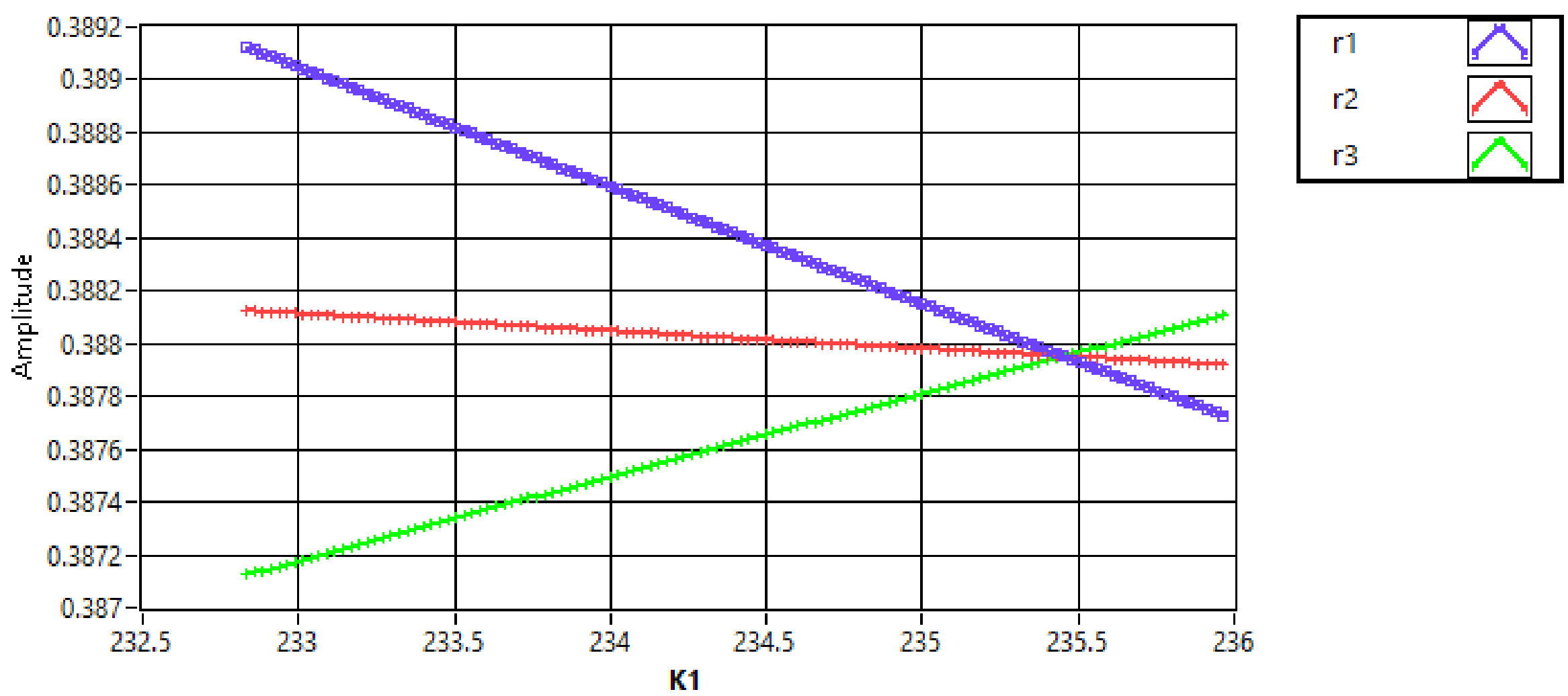

**Figure 3** The rates of profit (y-axis) of the three branches are plotted as a function of the capital K1 (in value) committed to branch 1. Capital K2 is held fixed and transfers only take place between 1 and 3 in either direction. For each value of r1, r2, r3 the equilibrium capital allocation is calculated according to our algorithm (see Appendix) so that both fundamental equalities (in addition to the 3 branch equalities) are complied with.

In other cases, such as the one illustrated in Figure 4, convergence requires a transfer to the least profitable branch. Indeed, whether one is on one side or the other of the point of intersection, the transfer to the most profitable branch leads to a move away from the intersection point, which thus is not an equilibrium point.

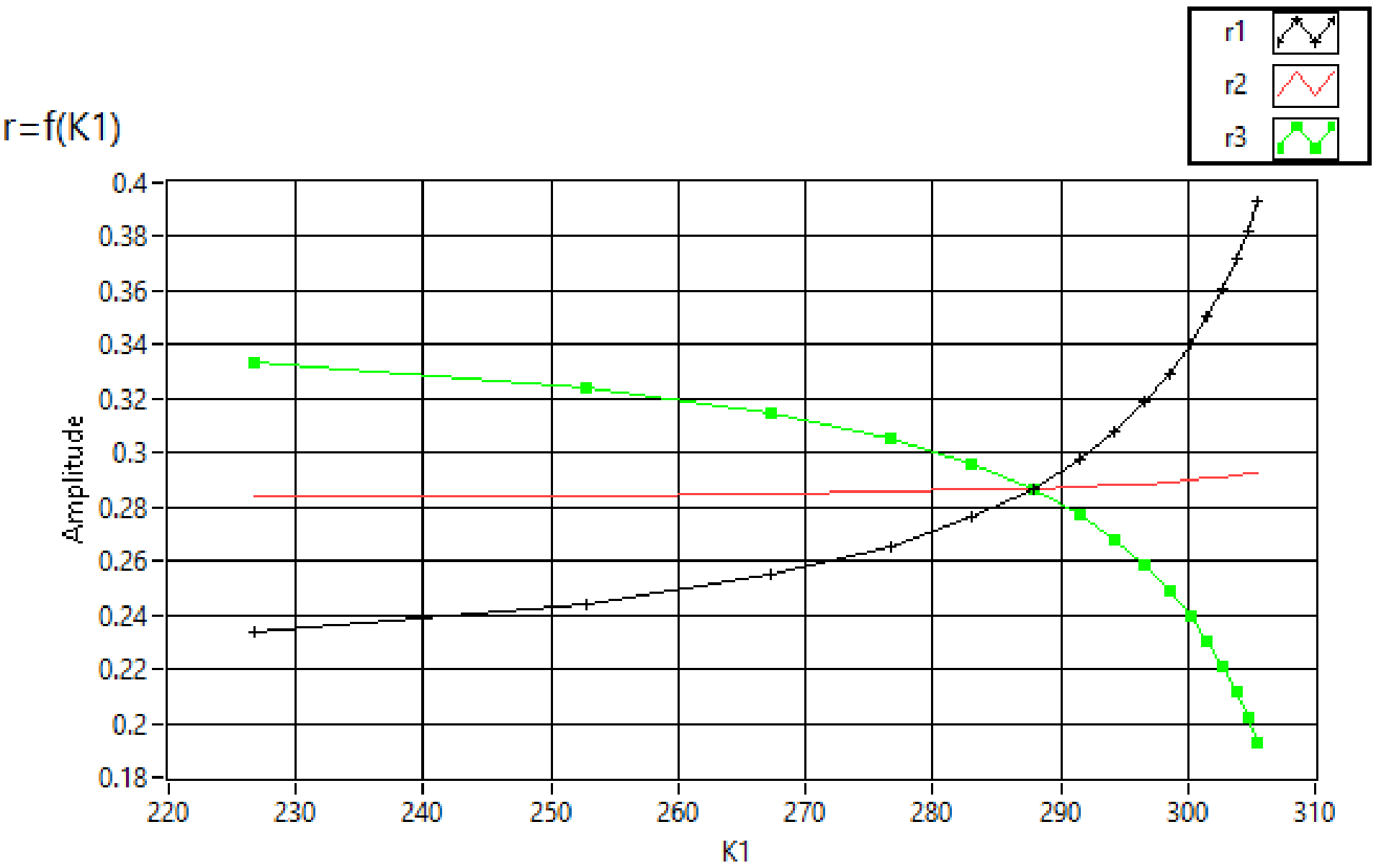

**Figure 4** Example of divergent courses of rates of profit. Here it can be seen that convergence only occur if the transfer of capital is made to the least profitable branch.

The examples in Figures 3 and 4 illustrate this obvious point: if the index $i$ denotes the most profitable branch, and $K_i$ the amount of capital committed to this branch, convergence can only occur if

$$\frac{dr_i}{dK_i} < 0$$

The sign of the variation of $r$ being identical to that of $s$, we can examine the condition

$$\frac{ds_i}{dK_i} < 0$$

Reminding that $S_i$ is the profit made in this branch and that

$$s_i = \frac{S_i}{K_i} = \boldsymbol{w_i x_i - d_i - (e_i x_1 + c_i x_2 + v_i x_3)}$$

We infer:

$$\frac{ds_i}{dK_i} = \frac{dS_i}{dK_i}\frac{K_i}{{K_i}^2} - \frac{S_i}{K_i^2} = \frac{1}{K_i}\frac{dS_i}{dK_i} - \frac{S_i}{K_i^2}$$

The condition becomes

$$\frac{1}{K_i}\frac{dS_i}{dK_i} < \frac{S_i}{K_i^2}$$

i.e.

$$\frac{dS_i}{dK_i} < \frac{S_i}{K_i}$$

Conclusion

Convergence to a uniform rate of profit is possible only if, for the most profitable branch of the economy, the rate of increase in profit associated with the increase in capital (in value) is lower than the ratio of profit to committed capital (in value). In other words, in the most profitable branch, the greater the capital employed, the more difficult it must be to increase the profit of this branch. Increasingly large capital injections would be required to increase the profit in the same proportions as before.

This condition of convergence seems to be a "naturally" complied with condition in accordance with the limits generally imposed by nature. It is directly related to the general law of diminishing returns, but it is not an essential physical limit that always applies in the real world. It would seem that the non-compliance with the condition established above, for example in the information sector, should lead to a speculative bubble phenomenon.

## E- Invalidity of the consequences attributed to Okishio's theorem

The Okishio's theorem (18) states that increasing a capitalist's profits by reducing their production costs increases the general rate of profit in society. This theorem has been used as an argument against the tendency of the average rate of profit to fall (TRPF) predicted by Marx (19-21). On the opposite of this mistaken reasoning, we show that not only the Okishio's theorem is not incompatible with the TRPF, but it contributes to it, as soon as capital mobility across branches and its resulting possible reallocations are taken into account.

### 1. The Okishio theorem does not prevent the TRPF

In this section, starting from a three-branch model with different rates of profit for each branch, we present simulations along the lines of the previous chapter showing a convergence to a uniform rate of profit that is accompanied by a TRPF.

We note the overall organic composition $CO$, calculated as follows: $CO = \frac{F+E+C}{V}$

F, E, C, V, designate $\sum_i F_i$, $\sum_i E_i$, $\sum_i C_i$, $\sum_i V_i$, respectively.

The organic composition in price is: $CO_p = \frac{F+x_1E+x_2C}{x_3V}$

The costs are: $D + x_1E + x_2C + x_3V$

with $D = F/n$
n being the number of cycles for the amortization of the fixed capital (equal to 10 in the following examples).

#### a) *Case with zero fixed capital (F = 0)*

The parameters of this simulation are given in Appendix, section 2.

We start with the allocation displayed in Tables 9 A-B (in value and price) for which we have assigned a higher rate of profit for branch 1 (E), an intermediate one for branch 2 (C) and a lower one for branch 3 (V). We apply the capital transfer process described in chapter D to equalize the rates of profit. For these simulations, the amount of capital $K_2$, set by the program at each iteration, is also gradually decreased (see Appendix, section 2).

| **VALUES** | F/n | E | C | V | PL | W |
|---|---|---|---|---|---|---|
| BRANCH 1 E | 0 | 54.32456192 | 108.6487122 | 69.84583613 | 69.84583613 | 302.6649464 |
| BRANCH 2 C | 0 | 71.29627221 | 142.5923907 | 171.1113371 | 171.1113371 | 556.1113371 |
| BRANCH 3 V | 0 | 28.16163884 | 56.32327769 | 51.84300518 | 51.84300518 | 188.1709269 |
| TOTAL | 0 | 153.782473 | 307.5643807 | 292.8001784 | 292.8001784 | 1046.94721 |

| PRICES | F/n | E | C | V | S | W | Taux de profit |
|---|---|---|---|---|---|---|---|
| BRANCH 1 E | 0 | 58.00818831 | 104.4593118 | 70.18737752 | 90.53309824 | 323.1879759 | 0.389130454 |
| BRANCH 2 C | 0 | 76.13071211 | 137.0941514 | 171.9480599 | 149.4952412 | 534.6681645 | 0.388125001 |
| BRANCH 3 V | 0 | 30.07121625 | 54.15150078 | 52.09651396 | 52.77183896 | 189.0910699 | 0.387119547 |
| TOTAL | 0 | 164.2101167 | 295.704964 | 294.2319513 | 292.8001784 | 1046.94721 | |

**Tables 9 A, B** Values and prices at the start of the simulation.

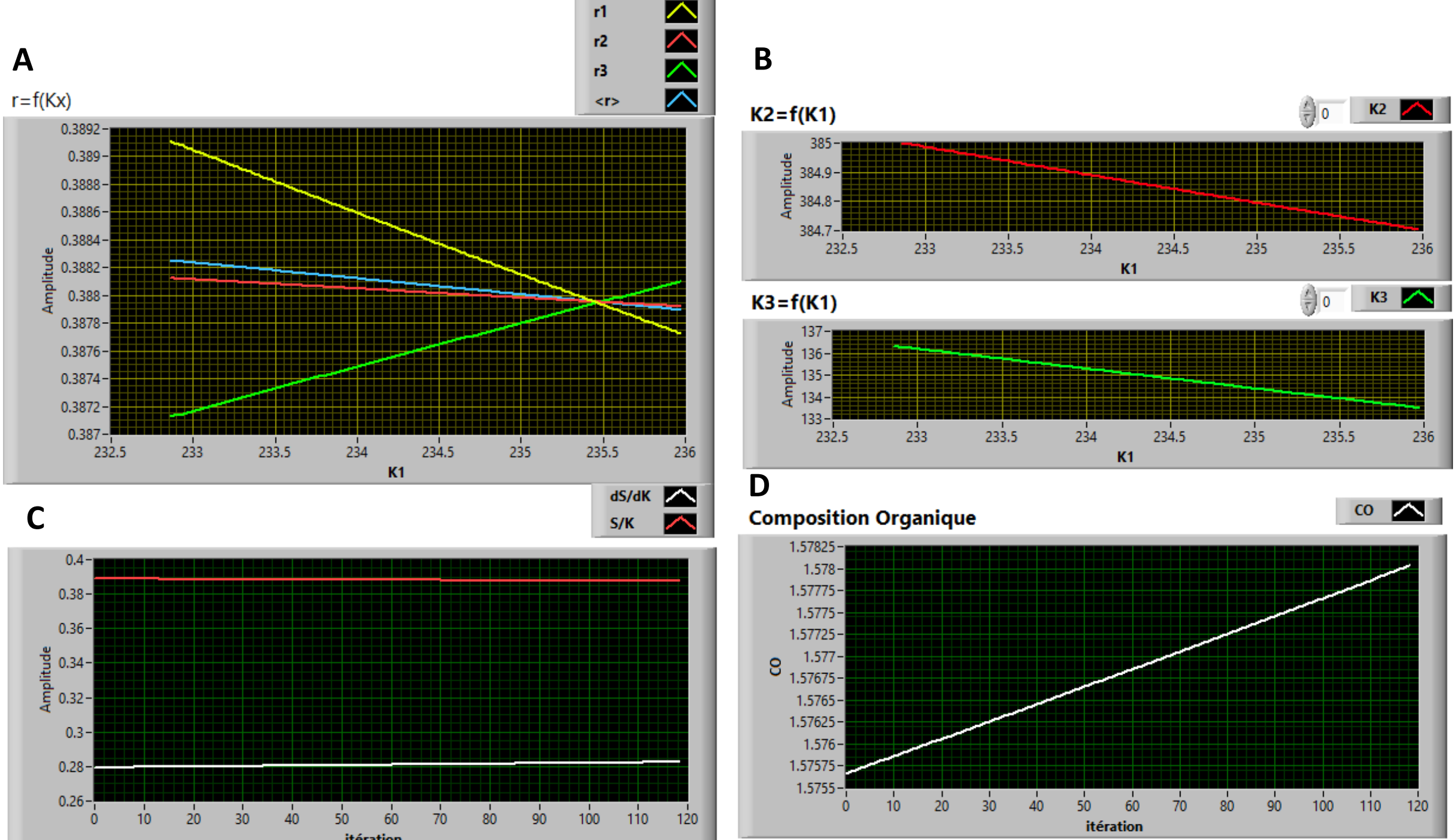

**Figure 4** Simulations from initial values in Table 9. (The capital of branch 2 with initial value 385 m.u. is decremented by 0.0025 m.u. at each iteration, see Appendix.) **A.** Convergence to uniform rate of profit occurs here when capital is transferred to the most profitable branch. The overall average rate of profit (blue plot) falls as convergence occurs. **B.** Evolution of capital as a function of K1 the capital of branch E. **C.** dS1/dK1 (white plot) and S/K (red plot) showing compliance with the convergence criterion $\frac{dS_1}{dK_1} < \frac{S_1}{K_1}$ (see chapter D). **D.** Evolution of the organic composition of total capital.

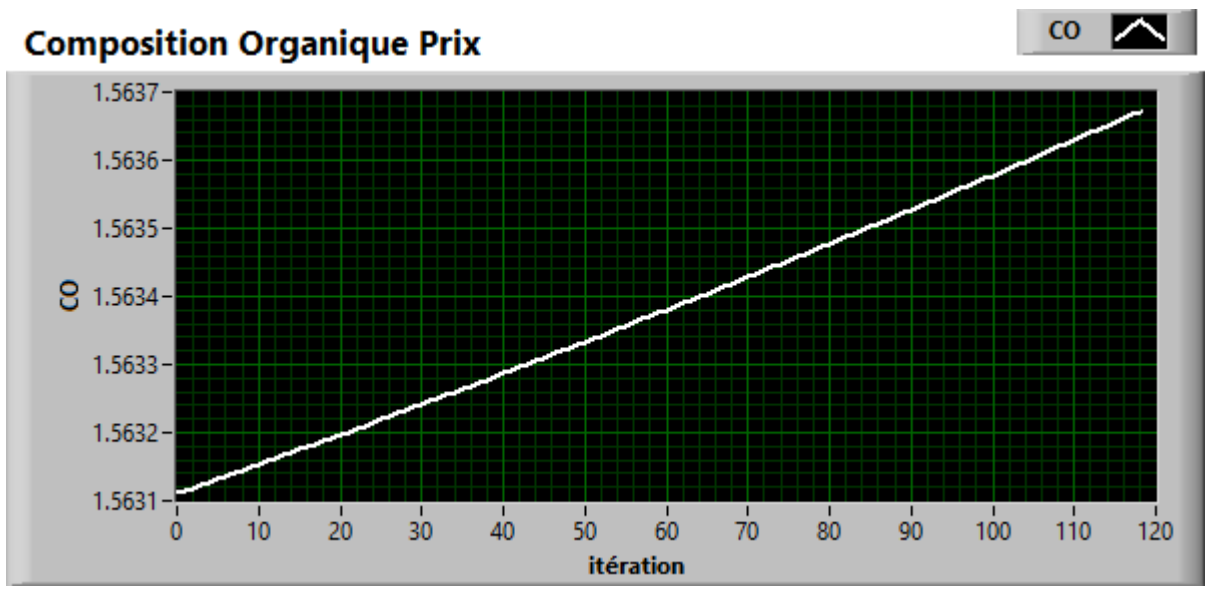

**Figure 4E** Change in the organic composition of total capital.

For the example in Figure 4, the organic composition calculated in terms of price changes along the same direction as the organic composition in terms of value. However, it should be noted that the organic composition in price is not necessarily a monotonic function of the rate of profit. For example, starting from a larger rate of profit differential (Figure 5), although the average rate of profit decreases monotonically (blue curve), the organic composition in price (white curve) does not, instead it exhibits an parabolic-like shape.

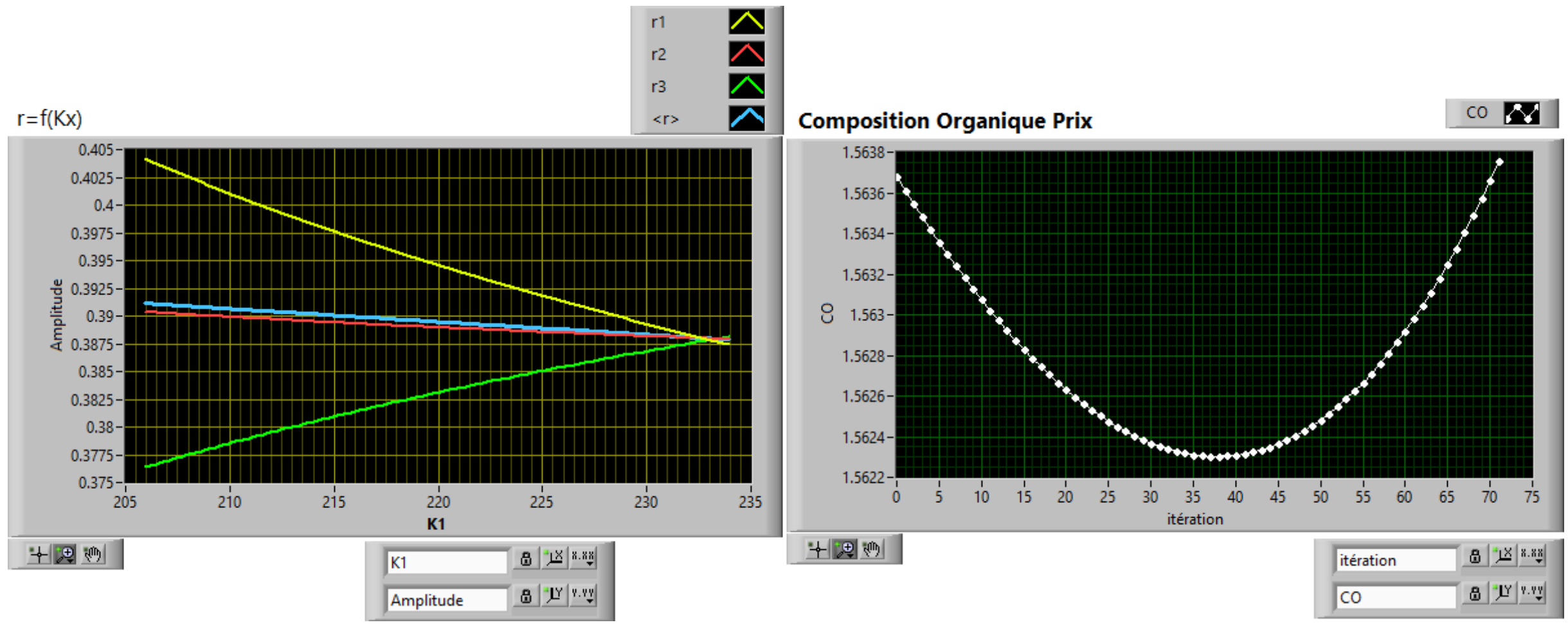

**Figure 5** Example illustrating a case for which the organic composition in price is not a monotonic function of the rate of profit. (The capital of branch 2 with initial value 385 m.u. is decremented by 0.0025 m.u. at each iteration, see Appendix.)

In the zero fixed capital case, total production costs remain constant (equal to total committed capital). So they are not shown in Figures 4 and 5. This is in contrast with the non-zero fixed capital case (see below, Figure 6C).

These examples illustrate the observation of a TRPF, in the absence of fixed capital.

### b) *Case with non-zero fixed capital*

Tables 10 and 11 display values and prices at the beginning and at the end of the simulation (at iteration #101), respectively.

| **VALUES** | F/n | E | C | V | PL | W |
|---|---|---|---|---|---|---|
| BRANCH 1 E | 32.036068 | 33.93699782 | 67.87373852 | 43.63326468 | 43.63326468 | 221.1133337 |
| BRANCH 2 C | 3.849993503 | 64.16665702 | 128.3331757 | 154.0002322 | 154.0002322 | 504.3502907 |
| BRANCH 3 V | 5.260801586 | 40.60972813 | 81.21945627 | 74.75880071 | 74.75880071 | 276.6075874 |
| TOTAL | 41.14686309 | 138.713383 | 277.4263705 | 272.3922976 | 272.3922976 | 1002.071212 |

**Table 10A** Values at the start of the simulation.

| PRICES | F/n | E | C | V | S | W |
|---|---|---|---|---|---|---|
| BRANCH 1 E | 32.036068 | 45.20922369 | 59.7470034 | 41.5738036 | 115.9903679 | 294.5564666 |
| BRANCH 2 C | 3.849993503 | 85.47971055 | 112.967443 | 146.7315237 | 94.93418134 | 443.962852 |
| BRANCH 3 V | 5.260801586 | 54.09831161 | 71.49479659 | 71.23023502 | 61.46774839 | 263.5518932 |
| TOTAL | 41.14686309 | 184.7872458 | 244.209243 | 259.5355623 | 272.3922976 | 1002.071212 |

**Tableau 10B** Prices at the start of the simulation.

| | K | Kp | $x_i$ | r |
|---|---|---|---|---|
| BRANCH 1 E | 465.804681 | 466.8907107 | 1.332151533 | 0.248431518 |
| BRANCH 2 C | 385 | 383.6786122 | 0.880266871 | 0.247431518 |
| BRANCH 3 V | 249.196001 | 249.4313591 | 0.952800665 | 0.246431518 |
| TOTAL | 1100 | 1100 | | |

**Tableau 10C** Capitals, transformation coefficients and rates of profit at the start of the simulation.

| VALUES | F/n | E | C | V | PL | W |
|---|---|---|---|---|---|---|
| BRANCH 1 E | 32.07991603 | 33.98344767 | 67.96663786 | 43.69298588 | 43.69298588 | 221.4159733 |
| BRANCH 2 C | 3.848993505 | 64.14999035 | 128.2998424 | 153.9602322 | 153.9602322 | 504.2192906 |
| BRANCH 3 V | 5.249453301 | 40.52212727 | 81.04425453 | 74.59753551 | 74.59753551 | 276.0109061 |
| TOTAL | 41.17836283 | 138.6555653 | 277.3107348 | 272.2507536 | 272.2507536 | 1001.64617 |

**Tableau 11A** Values at the end of the simulation.

| PRICES | F/n | E | C | V | S | W |
|---|---|---|---|---|---|---|
| BRANCH 1 E | 32.07991603 | 45.20008407 | 59.83667647 | 41.67182581 | 115.7084158 | 294.4969182 |
| BRANCH 2 C | 3.848993505 | 85.3234488 | 112.9530076 | 146.8383048 | 94.94236624 | 443.9061209 |
| BRANCH 3 V | 5.249453301 | 53.89693174 | 71.34998863 | 71.14678577 | 61.5999715 | 263.2431309 |
| TOTAL | 41.17836283 | 184.4204646 | 244.1396727 | 259.6569164 | 272.2507536 | 1001.64617 |

**Tableau 11B** Prices at the end of the simulation.

| | K | Kp | $x_i$ | r |
|---|---|---|---|---|
| BRANCH 1 E | 466.4422317 | 467.5077466 | 1.330061756 | 0.247500532 |
| BRANCH 2 C | 384.9 | 383.6046962 | 0.880383058 | 0.247500532 |
| BRANCH 3 V | 248.6584503 | 248.8882391 | 0.953741773 | 0.247500532 |
| TOTAL | 1100 | 1100 | | |

**Tableau 11C** Capitals, transformation coefficients and rates of profit at the end of the simulation.

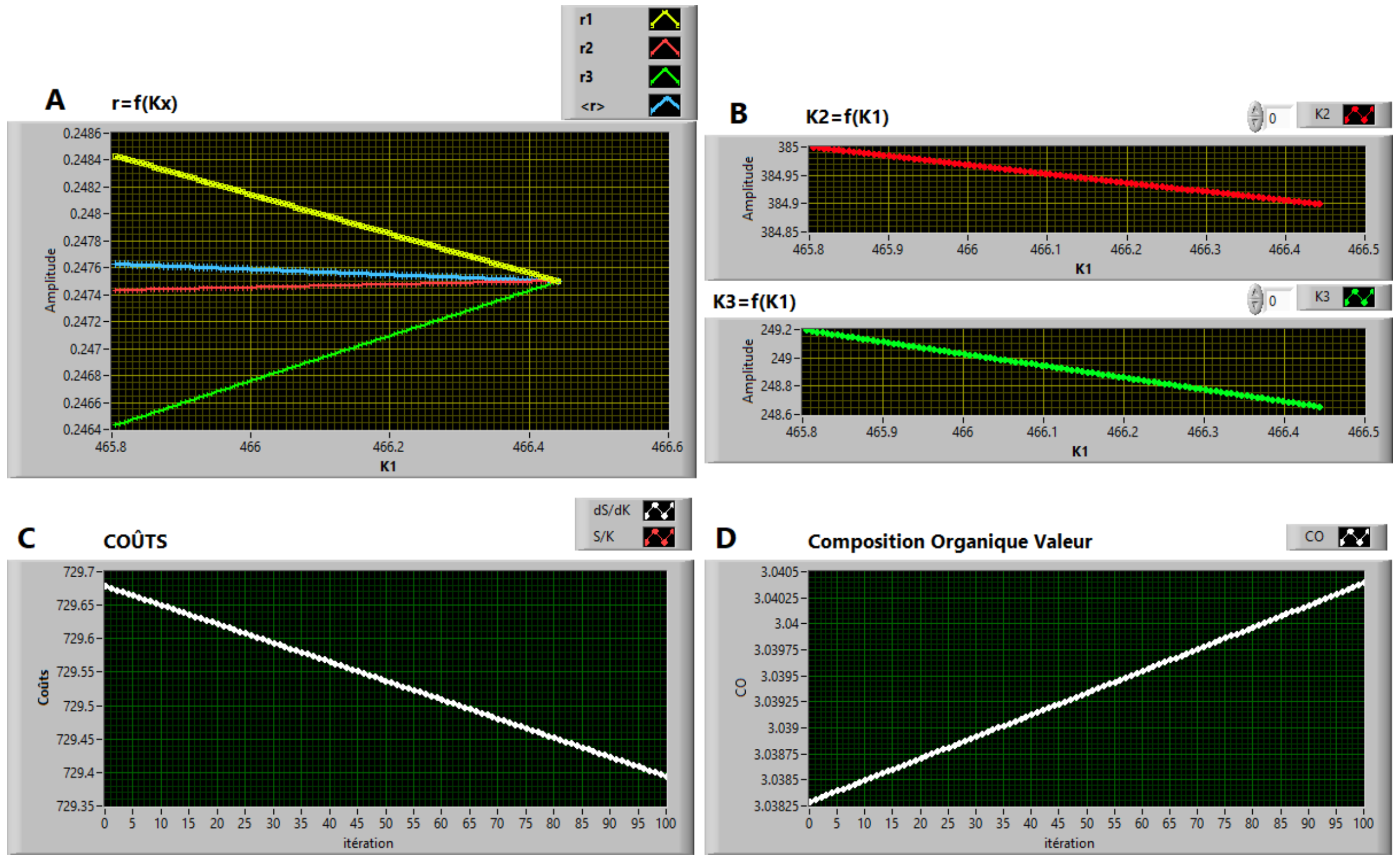

**Figure 6** Simulations based on the starting figures displayed in Table 10. **A.** Convergence to the same rate of profit when capital is transferred to the most profitable branch. The overall average rate of profit (blue plot) falls as convergence occurs. **B.** Evolution of capital as a function of K1 (the capital of branch E). **C.** Evolution of production costs (in prices): a monotonic decrease is observed. **D.** Evolution of the organic composition of total capital *in value*, showing a monotonic increase.

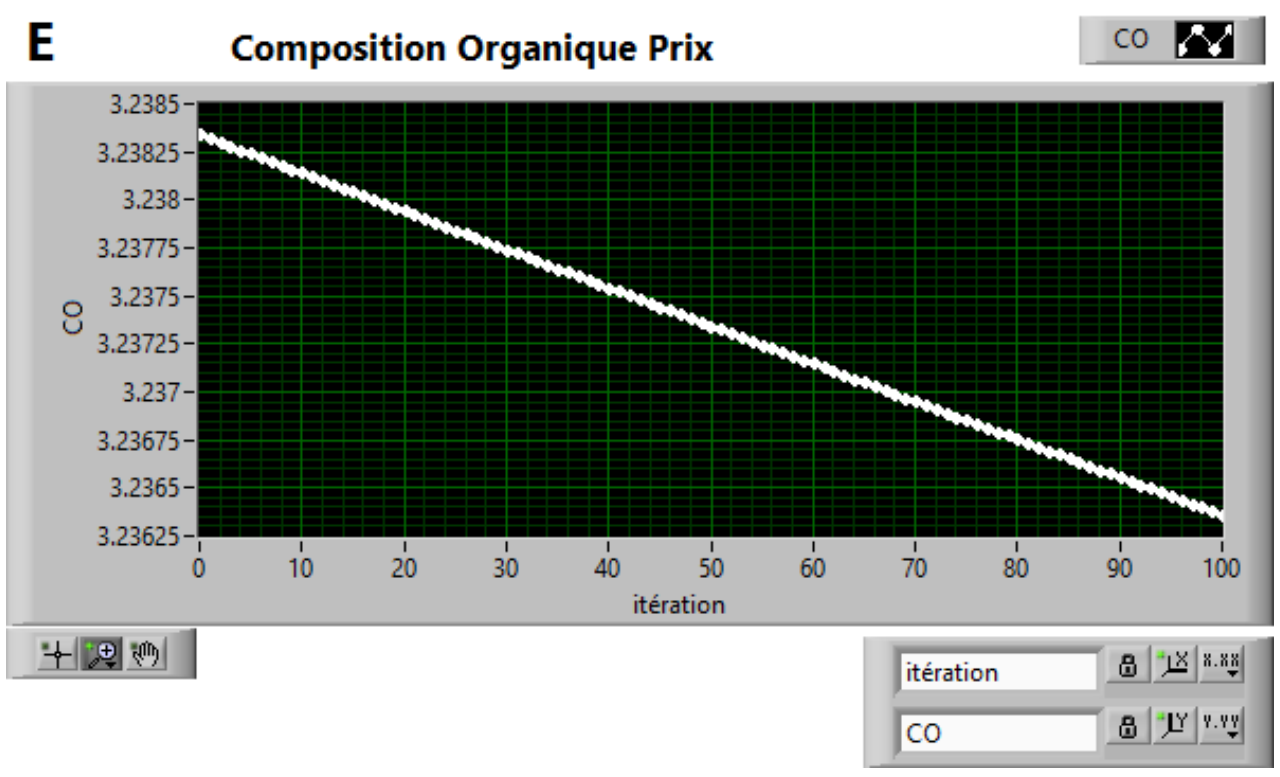

**Figure 6E** Evolution of the organic composition of total capital *in price*, showing a monotonic decrease.

This example shows that capital reallocation driven by the convergence of the rates of profit in accompanied by a decrease of the apparent organic composition, i.e., expressed in price (Figure 6E). In contrast, the organic composition in value increases (Figure 6D). This example illustrates that even in the case of a decreasing price organic composition, the TRPF is associated with an increasing organic composition in value.

This model enables to observe that a TRPF can occur when capital flows to the more profitable (and also higher organic composition) branch until the equalization of rates of profit. This fall in the rate of profit that occurs, although production costs fall (in the case of fixed capital), is a logical consequence of the increase in organic composition in value. The organic composition can also increase if there is no fixed capital and the fall in the average rate of profit is also observed in that case. It is interesting to note that in this last example with non-zero fixed capital, the organic composition calculated in price decreases (Figure 6E). This last example illustrates that what actually governs the TRPF is the increase in the organic composition *in value*.

## 2. The Okishio theorem favors the TRPF

In the previous section we started directly from a situation with different rates of profit for each branch. Here we are interested in the genesis of a higher rate of profit that may appear in a sector. We start with the following allocation (Table 12), which is used to define social-technical coefficients. (This example comprises five branches M, E, C, V and L, the definition of which is detailed later, chapters F and G.)

| | F/n | E | C | V | PL | W | K |
|---|---|---|---|---|---|---|---|
| BRANCH 1 M | 10 | 30 | 60 | 40 | 40 | 180 | 230 |
| BRANCH 2 E | 15 | 20 | 50 | 30 | 30 | 145 | 250 |
| BRANCH 3 C | 5 | 35 | 65 | 80 | 80 | 265 | 230 |
| BRANCH 4 V | 3 | 32 | 70 | 60 | 60 | 225 | 192 |
| BRANCH 5 L | 2 | 0.5 | 1 | 20 | 20 | 33.5 | 31.5 |
| TOTAL | 35 | 117.5 | 246 | 220 | 230 | 844.7 | 933.5 |

**Table 12** Amortization period: n=10 cycles.

We are interested (for example) in the energy branch E. Its capital is (150 + 20 + 50 + 30) =250 m.u.. We deduce that for this branch: $f_E/n$ = 15/250 = 0.06 and that $v_E$= 30/250 = 0.120.

Let us imagine that capitalists in this branch (E) start using a technical improvement that allows them to save production costs, especially labour, at the cost of an increase in the proportion of fixed capital in value from 150 m.u. to 152 m.u. (for K2 = 250 m.u.) at the expense of labour power, which falls from 30 to 28 m.u. (for K2 = 250). We assume that the capital invested in this branch is constant. The new socio-technical coefficients become: $f_E'/n$=152/250 = 0.0608 and $v_E'$=28/250=0.112 (Table 13).

| | F/n | E | C | V | PL | W | K |
|---|---|---|---|---|---|---|---|
| BRANCH 1 M | 10 | 30 | 60 | 40 | 40 | 180 | 230 |
| BRANCH 2 E | 15.2 | 20 | 50 | 28 | 30 | 145 | 250 |
| BRANCH 3 C | 5 | 35 | 65 | 80 | 80 | 265 | 230 |
| BRANCH 4 V | 3 | 32 | 70 | 60 | 60 | 225 | 192 |
| BRANCH 5 L | 2 | 0.5 | 1 | 20 | 20 | 33.5 | 31.5 |
| TOTAL | 35.2 | 117.5 | 246 | 218 | 230 | 848.5 | 933.5 |

**Table 13** Amortization period: n=10 cycles.

From these socio-technical coefficients we can construct a conformed allocation, i.e. one that complies with both fundamental equalities. We choose to set the amount of capital for the last three branches at 242, 358 and 5 m.u., respectively. This choice corresponds to an arbitrary (but conformed) allocation among those that fulfill the needs of each branch without resorting to the use of stocks. From the socio-technical coefficients derived in Table 12 we obtain the transformation solution displayed in Table 14. The rate of profit is r = 0.2614634020.

| **VALUES** | F/n | E | C | V | PL | W |
|---|---|---|---|---|---|---|
| BRANCH 1 M | 4.381104518 | 13.14331355 | 26.28662711 | 17.52441807 | 17.52441807 | 78.85988132 |
| BRANCH 2 E | 13.66407577 | 18.21876769 | 45.54691922 | 27.32815153 | 27.32815153 | 132.0860657 |
| BRANCH 3 C | 5.260869565 | 36.82608696 | 68.39130435 | 84.17391304 | 84.17391304 | 278.826087 |
| BRANCH 4 V | 5.59375 | 59.66666667 | 130.5208333 | 111.875 | 111.875 | 419.53125 |
| BRANCH 5 L | 0.317460317 | 0.079365079 | 0.158730159 | 1.587301587 | 3.174603175 | 5.317460317 |
| TOTAL | 29.21726017 | 127.9341999 | 270.9044142 | 242.4887842 | 244.0760858 | 914.6207443 |

**Table 14A**

| **PRICES** | F/n | E | C | V | S | W |
|---|---|---|---|---|---|---|
| BRANCH 1 M | 4.38110452 | 16.32383 | 24.3255872 | 16.7476419 | 26.4622152 | 88.2403788 |
| BRANCH 2 E | 13.6640758 | 22.6274801 | 42.1490194 | 26.1168213 | 59.4918342 | 164.049231 |
| BRANCH 3 C | 5.26086957 | 45.7375364 | 63.2891634 | 80.4428737 | 63.2946078 | 258.025051 |
| BRANCH 4 V | 5.59375 | 74.1052488 | 120.783694 | 106.9161 | 93.5365822 | 400.935374 |
| BRANCH 5 L | 0.31746032 | 0.09857043 | 0.14688854 | 1.51694387 | 1.29084639 | 3.37070954 |
| TOTAL | 29.2172602 | 158.892666 | 250.694352 | 231.74038 | 244.076086 | 914.620744 |

**Table 14B**

| | K (Value) | Kp | $x_i$ | $r$ |
|---|---|---|---|---|
| BRANCH 1 M | 100.7654039 | 101.2081042 | 1.118951453 | 0.261463402 |
| BRANCH 2 E | 227.7345961 | 227.5340784 | 1.24198741 | 0.261463402 |
| BRANCH 3 C | 242 | 242.0782691 | 0.925397812 | 0.261463402 |
| BRANCH 4 V | 358 | 357.7425423 | 0.955674636 | 0.261463402 |
| BRANCH 5 L | 5 | 4.937006012 | 0.63389463 | 0.261463402 |
| TOTAL | 933.5 | 933.5 | | |

**Table 14C**

This equilibrium is the one from which we start. The value produced by branch E is :

$$W_E = D_E + E_E + C_E + V_E + PL_E$$

Now we are just after the technical improvement has taken place, the socio-technical coefficients in Table 13 now apply, the primed letters indicate the new values, we now have :

$$W'_E = D'_E + E_E + C_E + V'_E + PL'_E$$

The socially necessary labour (corresponding to the average of the considered industry) to produce commodity E has not yet changed and the two quantities $W_E$ and $W'_E$ are equal. If $V'_E + D'_E < V_E + D_E$ , an extra surplus value is achieved thanks to the new technique.

Let's consider the moment *immediately* following this technical change, when the amounts of capital allocated to the different branches have not started to change yet. We can calculate:

$$D'_E = \frac{f'_E}{n} K_E = 0{,}0608 * 227{,}734 = 13{,}846$$

$$V'_E = v'_E K_E = 0{,}112 * 227{,}734 = 25{,}506$$

$$D_E = 13{,}664 < D'_E = \ et\ V_E = 27{,}328 > V'_E$$

The productions in prices before and after are:

$$x_E W_E = D_E + x_E E_E + x_C C_E + x_v V_E + S_E$$

$$x'_E W_E = D'_E + x_E E_E + x_C C_E + x_v V'_E + S'_E$$

Assuming that the prices for this quantity have not changed (yet):

$$x'_E W_E = x_E W_E$$

The cost saving relative to the standard production process previously in use is source of extra profit made $(S'_E - S_E)$ thanks to the modernizing capitalists. Its expression is:

$$S'_E - S_E = x_v(V_E - V'_E) - (D'_E - D_E)$$

$$r'_E = \frac{S'_E}{F'_E + x_E E_E + x_C C_E + x_v V'_E}$$

From Table 14B (prices) we read:

$$S_E = 59.49183422278\ \ V_E = 26.1168212538384878\ \ D_E = 13.6640757654306455$$

$$x_v = 0{,}9556746355$$

From the above calculated $D'_E$ et $V'_E$, we find :

$$S'_E = 61.0508\ \ et\ \ r'_E = 0.26822$$

$$r'_E = 0.26822 \quad > r_E = 0.26146340205694413$$

The *transient rate of profit* of branch E having become higher than that of the other sectors. Assuming capitalists of other sectors have not yet modernized their production, capital will flow in branch E until an equalization of rates of profit is reached. The new equilibrium state in value and price is then given by Tables 15, it was obtained by keeping the capital of the last three branches fixed. The transfer from branch 1 to branch 2, *which is also the most capitalistic*, is the solution indicated by the algorithm for the compliance with Marx's two fundamental equalities. The new amounts of capital, the rates of profit and the transformation coefficients are shown in Table 16.

| **VALUES** | F/n | E | C | V | PL | W |
|---|---|---|---|---|---|---|
| BRANCH 1 M | 3.828900457 | 11.48670137 | 22.97340274 | 15.31560183 | 15.31560183 | 68.92020823 |
| BRANCH 2 E | 14.6184656 | 19.23482316 | 48.0870579 | 26.92875242 | 28.85223474 | 137.7213338 |
| BRANCH 3 C | 5.260869565 | 36.82608696 | 68.39130435 | 84.17391304 | 84.17391304 | 278.826087 |
| BRANCH 4 V | 5.59375 | 59.66666667 | 130.5208333 | 111.875 | 111.875 | 419.53125 |
| BRANCH 5 L | 0.317460317 | 0.079365079 | 0.158730159 | 1.587301587 | 3.174603175 | 5.317460317 |
| TOTAL | 29.61944594 | 127.2936432 | 270.1313285 | 239.8805689 | 243.3913528 | 910.3163393 |

**Table 15A**

| **PRICES** | F/n | E | C | V | S | W |
|---|---|---|---|---|---|---|
| BRANCH 1 M | 3.828900457 | 14.29538511 | 21.24403114 | 14.62665421 | 23.06288203 | 77.05785294 |
| BRANCH 2 E | 14.6184656 | 23.93804763 | 44.46720264 | 25.71740597 | 62.65529921 | 171.396421 |
| BRANCH 3 C | 5.260869565 | 45.83065913 | 63.24300388 | 80.38748545 | 63.11484395 | 257.836862 |
| BRANCH 4 V | 5.59375 | 74.25612895 | 120.6956008 | 106.8424837 | 93.27135049 | 400.659314 |
| BRANCH 5 L | 0.317460317 | 0.098771121 | 0.146781409 | 1.515899388 | 1.286977104 | 3.36588934 |
| TOTAL | 29.61944594 | 158.4189919 | 249.7966199 | 229.0899287 | 243.3913528 | 910.3163393 |

**Table 15B**

| | K (Value) | Kp | $x_i$ | r |
|---|---|---|---|---|
| BRANCH 1 M | 88.06471052 | 88.45507503 | 1.11807342 | 0.260729891 |
| BRANCH 2 E | 240.4352895 | 240.3073122 | 1.244516128 | 0.260729891 |
| BRANCH 3 C | 242 | 242.0698441 | 0.924722879 | 0.260729891 |
| BRANCH 4 V | 358 | 357.7317135 | 0.955016614 | 0.260729891 |
| BRANCH 5 L | 5 | 4.936055093 | 0.632988145 | 0.260729891 |
| TOTAL | 933.5 | 933.5 | | |

**Table 16**

The transfer of capital to the most profitable branch (and also the most capitalistic in our example) leads to a new equilibrium for which the rate of profit has decreased from its initial $r_E$ value:

$$r''_E = 0.2607298905038033$$

To sum up the sudden saving in production costs of branch E is associated with the expected sudden transient increase in its rate of profit. But this increase is transient. The rate of profit eventually decreases to stabilize at a value below the initial one:

$$r'_E > r_E > r''_E$$

The phenomenon of the tendency of the average rate of profit to fall (TRPF), as stated by Marx, is therefore verified for the model described here, as soon as non-instantaneous capital flows are taken into account. [13] There would be no interest in innovation if the final result was reached immediately. Okishio's theorem accounts for a transitory phenomenon. It describes the transitory effect of the initial increase in the rate of profit in a given sector. This effect, followed by capital transfers until the rates of profit are equalized, is a process that realizes the TRPF. Note that the TRPF can occur even in the absence of fixed capital. It is the increase in organic composition in value, even if it falls when expressed in prices (Figure 7), that governs the fall in the average rate of profit. One might think that permanent technical innovation, occurring before a new state of equilibrium is reached, is capable of counteracting this TRPF. It is undeniable that innovation is one factor (among others) that changes the trajectory of the average rate of profit by giving it now and then positive impulses. In that sense, the word *tendency* chosen by Marx is adequate. Marx's theory of value predicts that, even if the system is drawn into a permanent disequilibrium, punctuated by changes in the conditions of production associated with transient increases in rate of profit, in the end, since the mass of surplus-value to be shared tends to diminish in the course of time (all other things being equal), then the TRPF inexorably occurs.

[13] Here we see that M. Husson's argument against V. Laure van Bambeke's proposal, claiming that one should consider the time it takes for capital flow to occur between branch, does not hold (Ref. 9). Rather, considering this time further supports our conclusions.

## F- Addition of a luxury branch

Simple reproduction is defined by Marx using a two-sector system I(C) and II(V), so that the consumption of capitalists (PL1 + PL2) is met by the commodities of branch II, the total production of which is V1 + V2 + PL1 + PL2. More generally, simple reproduction may be defined as a scheme in which the total production meets the total demand.

In line with the models proposed by Ladislaus von Bortkiewicz (7), we consider adding a luxury branch tailored in order to obtain a simple reproduction scenario for which, whatever i is, the price of the total production of the commodity of branch i is exactly equal to the price of the consumption of this commodity by all branches. We address whether such a model is coherent with Marx's conception (compliance with fundamental equalities).

Next, in order to model simple reproduction closer to Marx's conception, we consider the addition of a luxury branch (branch 4) such that the total production of branch 3 (goods mostly consumed by the working class) added to the total production of branch 4 (goods mostly consumed by the capitalists'class) meets de total demand ($\sum_i V_i + \sum_i PL_i$). This model is named Marx-type simple reproduction model.

### 1. Bortkiewicz-type simple reproduction model

#### a) *Model with zero fixed capital*

If we start from a system of total capital $K_{1,2,3}$ with three branches 1 (E; energy), 2 (C; raw materials), 3 (V; current consumer goods) and zero fixed capital, we have seen in Chapter C that total surplus value is constant whatever the capital allocation complying with both fundamental equalities, therefore:

$$\sum_{i=1}^{3} K_i pl_i = PL_{1,2,3} = constant \ \forall \ K_i \ such \ as \sum_{i=1}^{3} K_i = K_{1,2,3}$$

From chapter C, we have seen that in this case:

$$\sum_{i=1}^{3} S_i = S_{1,2,3} = PL_{1,2,3}$$

Let $E_{1,2,3}, C_{1,2,3}, V_{1,2,3}$ be the total consumption for these three branches in commodities of the designated type, and $E_L, C_L, V_L = E_4, C_4, V_4$ those for the luxury branch. We want the sum of the prices of a given commodity to be equal to the price of its total production.

This condition of total reproduction can be written:

$$\cancel{x_1} \sum_{i=1}^{4} E_i = \cancel{x_1} W_1 \quad ; \quad \cancel{x_2} \sum_{i=1}^{4} C_i = \cancel{x_2} W_2 \quad ; \quad \cancel{x_3} \sum_{i=1}^{4} V_i = \cancel{x_3} W_3 \qquad \textbf{(12)}$$

We feed the luxury branch with the surplus in value $E_L, C_L, V_L$ produced in the first three branches so that:

$$\left.\begin{aligned} E_L &= W_1 - (E_1 + E_2 + E_3) \\ C_L &= W_2 - (C_1 + C_2 + C_3) \\ V_L &= W_3 - (V_1 + V_2 + V_3) \end{aligned}\right\}$$

The total surplus in value being $PL_1 + PL_2 + PL_3$, we have:

$$E_L + C_L + V_L = PL_1 + PL_2 + PL_3$$

$$S_L = x_L W_L - (x_1 E_L + x_2 C_L + x_3 V_L)$$

Now, taking into account the L branch, the total surplus value and the total profit are:

$$PL_T = PL_{1,2,3} + PL_L = S_T = S_{1,2,3} + SL_L$$

We recall here that while the quantities $E_i, C_i, V_i$ have become $E'_i, C'_i, V'_i$ due to capital reallocation following the addition of the luxury branch, on the other hand, the sums $PL_{1,2,3}$ as well as $S_{1,2,3}$ have remained constant and equal to each other, which implies:

$$PL_L = S_L = rK_{pL}$$

Therefore, we have:

$$x_L = 1$$

We posit that the exploitation rate $e_L$ for the luxury branch L is equal to 1, as for the other branches ($e_L = 1$), i.e. $PL_L = V_L$

The committed capital of branch L is $K_L = E_L + C_L + V_L$, and equal to the (invariant) surplus $PL_{1,2,3}$ of the first three branches. Thus, we have:

$$rK_L = r(E_L + C_L + V_L) = r.PL_{1,2,3} = S_L = V_L \tag{13}$$

The total reproduction condition which translates into equations 12 imposes:

$$\cancel{x_3}W_3 - \cancel{x_3}(V_1 + V_2 + V_3) = \cancel{x_3}PL_L = \cancel{x_3}V_L$$

Since we have assumed that all exploitation rates are unity, the equation translate into:

$$W_3 = V_L + PL_{1,2,3}$$

In accordance with equation 13 and introducing the production per unit of capital $w_3 = W_3/K_3$, the above equality becomes:

$$K_3 w_3 = r.PL_{1,2,3} + PL_{1,2,3} = PL_{1,2,3}(1 + r)$$

$$K_3 = \frac{PL_{1,2,3}(1 + r)}{w_3}$$

This equality shows that the capital in branch 3 is determined *directly*: the quantities used to calculate it are independent of the capital allocation between branches. We recall that we are dealing here with all allocations that comply with both fundamental equalities. It is already known that, for a three-branch system, when the capital of one branch (in this case, branch 3) is *fixed*, then there remains only one possible solution. The *four-branch total reproduction system with luxury branch*, therefore has one-and-only one possible capital allocation. $K_3$ is fixed and the main algorithm provides the obligatory values for $K_1$ and $K_2$.

An example is given below after normalization of the capital to 1000 monetary units (m.u.).

| **VALUES** | F/n | E | C | V | PL | W |
|---|---|---|---|---|---|---|
| BRANCH 1 E | 0 | 36.6183671 | 73.2364567 | 47.0807379 | 47.0807379 | 204.0163 |
| BRANCH 2 C | 0 | 52.3117634 | 104.623414 | 125.54844 | 125.54844 | 408.032057 |
| BRANCH 3 V | 0 | 58.0607973 | 116.121595 | 106.884625 | 106.884625 | 387.951642 |
| BRANCH 4 L | 0 | 57.0253718 | 114.050592 | 108.437839 | 108.437839 | 387.951642 |
| TOTAL | 0 | 204.0163 | 408.032057 | 387.951642 | 387.951642 | 1387.951641 |

| **PRICES** | F/n | E | C | V | S | W |
|---|---|---|---|---|---|---|
| BRANCH 1 E | 0 | 39.0649449 | 70.4077658 | 47.3391015 | 60.8354001 | 217.647212 |
| BRANCH 2 C | 0 | 55.806862 | 100.582431 | 126.237409 | 109.645493 | 392.272194 |
| BRANCH 3 V | 0 | 61.9400053 | 111.636505 | 107.471173 | 109.03291 | 390.080592 |
| BRANCH 4 L | 0 | 60.8354001 | 109.645493 | 109.03291 | 108.437839 | 387.951642 |
| TOTAL | 0 | 217.647212 | 392.272194 | 390.080592 | 387.951642 | 1387.951641 |

| | K | Kp | xi | r |
|---|---|---|---|---|
| BRANCH 1 E | 156.9355618 | 156.8118124 | 1.066812861 | 0.387951642 |
| BRANCH 2 C | 282.4836177 | 282.6267015 | 0.961375919 | 0.387951642 |
| BRANCH 3 V | 281.0670172 | 281.0476828 | 1.00548767 | 0.387951642 |
| BRANCH 4 L | 279.5138033 | 279.5138033 | 1 | 0.387951642 |
| TOTAL | 1000 | 1000 | | |

**Tables 17 A, B, C**

### b) *Model with fixed capital*

Equations 12 still apply. But this time their vertical sums implies:

$$E_L + C_L + V_L = PL_1 + PL_2 + PL_3 + d_1 + d_2 + d_3$$

The exploitation rate being equal to unity, it can also be written:

$$E_L + C_L + V_L = d_1 + d_2 + d_3 + V_1 + V_2 + V_3$$

For any branch, we have

$$W_i = D_i + V_i + E_i + C_i + PL_i \quad i = 1,2,3$$

$$W_i - (V_i + E_i + C_i) = D_i + PL_i = D_i + V_i$$

Now, the surplus value of the L branch being still equal to its profit ($S_L = V_L$) ($x_L = 1$), we have

$$V_L = rK_L = r[n.D_L + E_L + C_L + V_L] = r[n.D_L + D_1 + D_2 + D_3 + V_1 + V_2 + V_3]$$

and:

$$W_3 = V_L + PL_{1,2,3} = \ r[n.D_L + dD_1 + D_2 + D_3] + (r+1)PL_{1,2,3}$$

Finally:

$$K_3 = \frac{PL_{1,2,3}(1+r) + r[n.d_L + d_1 + d_2 + d_3]}{w_3}$$

The fixed capital $n.d_L$ of the luxury branch can be chosen arbitrarily.

Note that here, unlike in the previous case, the surplus value and the profit depend on capital allocation. Each allocation corresponds to a different quantity of total surplus value. By feeding back the obtained value of $K_3$ several times into the input of the main algorithm in order to refresh $PL_{1,2,3}$, we execute a loop that converges rapidly and asymptotically towards $K_3$.

### c) *Conclusion on the Borkiewicz-type simple reproduction models*

It is possible to artificially construct a four-branch system of total reproduction with or without fixed capital and with identical rates of profit in all branches, this system complying with both Marx's fundamental equalities. The addition of the luxury branch interacts with the first three branches and modifies capital allocation. When there is no fixed capital the rate of profit is not modified.

An example is given below with a total capital to 1000 monetary units (m.u.) and an amortizing period of n = 10 cycles.

| **VALUES** | F/n | E | C | V | PL | W |
|---|---|---|---|---|---|---|
| BRANCH 1 E | 11.68238094 | 27.25894178 | 54.51767702 | 35.0471962 | 35.0471962 | 163.5533921 |
| BRANCH 2 C | 2.524161581 | 42.06942435 | 84.13875802 | 100.9667859 | 100.9667859 | 330.6659157 |
| BRANCH 3 V | 5.455205741 | 42.11039296 | 84.22078591 | 77.5213876 | 77.5213876 | 286.8291598 |
| BRANCH 4 L | 2.233404974 | 52.11463306 | 107.7886947 | 73.29379013 | 73.29379013 | 308.724313 |
| TOTAL | 21.89515323 | 163.5533921 | 330.6659157 | 286.8291598 | 286.8291598 | 1089.772781 |

| PRICES | F/n | E | C | V | S | W |
|---|---|---|---|---|---|---|
| BRANCH 1 E | 11.68238094 | 32.63036677 | 49.757351 | 34.63716777 | 67.07462469 | 195.7818912 |
| BRANCH 2 C | 2.524161581 | 50.35928238 | 76.7920048 | 99.78554292 | 72.33213176 | 301.7931234 |
| BRANCH 3 V | 5.455205741 | 50.40832393 | 76.86687025 | 76.61443991 | 74.12861323 | 283.4734531 |
| BRANCH 4 L | 2.233404974 | 62.38391809 | 98.37689739 | 72.43630246 | 73.29379013 | 308.724313 |
| TOTAL | 21.89515323 | 195.7818912 | 301.7931234 | 283.4734531 | 286.8291598 | 1089.772781 |

**Tables 18 A, B**

| | K | Kp | xi | r |
|---|---|---|---|---|
| BRANCH 1 E | 233.6476244 | 233.8486949 | 1.197051853 | 0.286813354 |
| BRANCH 2 C | 252.4165841 | 252.1784459 | 0.91268289 | 0.286813354 |
| BRANCH 3 V | 258.4046239 | 258.4416915 | 0.988300678 | 0.286813354 |
| BRANCH 4 L | 255.5311677 | 255.5311677 | 1 | 0.286813354 |
| TOTAL | 1000 | 1000 | | |

**Table 19**

The luxury branch has a transformation coefficient equal to unity and therefore has an organic composition of its capital equal to the average of the organic compositions of all the branches, consistent with the idea that this luxury branch could serve as a monetary standard (7)

Note that the socio-technical coefficients of the luxury branch have been constrained to obtain a reproduction system. If this constrain is ignored and the luxury branch simply added with random socio-technical coefficients, the supplied algorithm leads to a transformation coefficient $x_L$ that has no reason to be unity as is illustrated in the example below (Tables 20).

| VALUES | F/n | E | C | V | PL | W |
|---|---|---|---|---|---|---|
| BRANCH 1 E | 16.25105639 | 40.8813405 | 77.31916682 | 46.53165992 | 46.53165992 | 227.5148835 |
| BRANCH 2 C | 2.871510536 | 45.45965176 | 83.72240806 | 114.8607886 | 114.8607886 | 361.7751475 |
| BRANCH 3 V | 6.118836912 | 50.58938706 | 101.1787662 | 87.0434776 | 87.0434776 | 331.9739454 |
| BRANCH 4 L | 4.081632657 | 4.081632657 | 18.36734695 | 36.73469383 | 36.73469383 | 99.99999992 |
| TOTAL | 29.3230365 | 141.012012 | 280.587688 | 285.1706199 | 285.1706199 | 1021.263976 |

| PRICES | F/n | E | C | V | S | W |
|---|---|---|---|---|---|---|
| BRANCH 1 E | 16.25105639 | 49.41898169 | 69.55260122 | 46.32534963 | 93.48100295 | 275.0289919 |
| BRANCH 2 C | 2.871510536 | 54.95342547 | 75.31264887 | 114.3515232 | 77.94640294 | 325.435511 |
| BRANCH 3 V | 6.118836912 | 61.15445243 | 91.01554852 | 86.65754757 | 85.55566662 | 330.5020521 |
| BRANCH 4 L | 4.081632657 | 4.934039028 | 16.52238132 | 36.57182096 | 28.18754739 | 90.29742135 |
| TOTAL | 29.3230365 | 170.4608986 | 252.4031799 | 283.9062414 | 285.1706199 | 1021.263976 |

| | K | Kp | xi | r |
|---|---|---|---|---|
| BRANCH 1 E | 327.2427311 | 327.8074964 | 1.208839561 | 0.285170425 |
| BRANCH 2 C | 272.7579537 | 273.3327029 | 0.899551872 | 0.285170425 |
| BRANCH 3 V | 300 | 300.0159176 | 0.995566238 | 0.285170425 |
| BRANCH 4 L | 100 | 98.84456787 | 0.902974214 | 0.285170425 |
| TOTAL | 1000 | 1000 | | |

**Tables 20 A, B, C**

However, as illustrated in the example below, in a situation with zero surplus value, the solution is unique.

| **PRICES = VALUES** | F/n | E | C | V | PL | W |
|---|---|---|---|---|---|---|
| BRANCH 1 E | 13.54259217 | 34.06789741 | 64.43285399 | 38.77651264 | 0 | 150.8198562 |
| BRANCH 2 C | 3.414165106 | 54.05056148 | 99.54416694 | 136.5670407 | 0 | 293.5759342 |
| BRANCH 3 V | 7.380786296 | 61.02294604 | 122.0458825 | 104.9953309 | 0 | 295.4449458 |
| BRANCH 4 L | 1.678451284 | 1.678451284 | 7.553030773 | 15.10606152 | 0 | 26.01599486 |
| TOTAL | 26.01599486 | 149.1414049 | 286.0229035 | 280.3388843 | 0 | 765.8567311 |

| | K |
|---|---|
| BRANCH 1 E | 272.7031858 |
| BRANCH 2 C | 324.3034202 |
| BRANCH 3 V | 361.8720225 |
| BRANCH 4 L | 41.12205641 |
| TOTAL | 1000 |

## 2. Marx-type simple reproduction model with zero fixed capital

With the addition of a luxury branch L the commodities of which are not used by the other branches an additional constraint is allowed for the complete determination of the system:

$$k_1 + k_2 + k_3 + k_L = 1$$

$$k_1(w_1 - e_1) - k_2e_2 - k_3e_3 - k_Le_L = 0$$

$$-k_1c_1 + k_2(w_2 - c_2) - k_3c_3 - k_Lc_L = 0$$

$$k_1w_1(1 - x_1) + k_2w_2(1 - x_2) + k_3w_3(1 - x_3) + k_Lw_L(1 - x_L) = 0$$

This system is completely determined and also implies the equality below:

$$k_3(w_3 - v_3 - pl_3) = k_1(v_1 + pl_1) + k_2(v_2 + pl_2) - k_L w_L$$

All the systems whose number of branches exceeds 4, including a luxury branch, will enable Marx-type simple reproduction (see Chapter G for the five-branch model).

Example

| **INIT VALUES** | F/n | E | C | V | PL | W |
|---|---|---|---|---|---|---|
| BRANCH 2 E | 0 | 19.401807 | 38.803467 | 15 | 15 | 88.205274 |
| BRANCH 3 C | 0 | 19.949964 | 39.899885 | 47.879993 | 47.879993 | 155.609835 |
| BRANCH 4 V | 0 | 116.355582 | 232.711164 | 215 | 215 | 779.066746 |
| BRANCH 5 L | 0 | 10 | 15 | 8 | 8 | 41 |
| TOTAL | 0 | 165.707353 | 326.414516 | 285.879993 | 285.879993 | 1063.881855 |

**Table 20D**

| **VALUES** | F/n | E | C | V | PL | W |
|---|---|---|---|---|---|---|
| BRANCH 2 E | 0 | 36.53209782 | 73.06391885 | 28.24383663 | 28.24383663 | 166.0836899 |
| BRANCH 3 C | 0 | 42.1242002 | 84.24830961 | 101.0982481 | 101.0982481 | 328.5690061 |
| BRANCH 4 V | 0 | 80.2313795 | 160.462759 | 148.2502712 | 148.2502712 | 537.194681 |
| BRANCH 5 L | 0 | 7.196012424 | 10.79401864 | 5.756809939 | 5.756809939 | 29.50365094 |
| TOTAL | 0 | 166.0836899 | 328.5690061 | 283.349166 | 283.349166 | 1061.351028 |

**Table 20E**

| **VALUES** | F/n | E | C | V | PL | W |
|---|---|---|---|---|---|---|
| BRANCH 2 E | 0 | 41.48030033 | 68.87465802 | 27.87935414 | 50.34509435 | 188.5794068 |
| BRANCH 3 C | 0 | 47.82984224 | 79.41777015 | 99.79359035 | 82.68866512 | 309.7298679 |
| BRANCH 4 V | 0 | 91.09856579 | 151.2623169 | 146.3371236 | 141.5642575 | 530.2622639 |
| BRANCH 5 L | 0 | 8.17069849 | 10.17512274 | 5.682519167 | 8.751148993 | 32.77948939 |
| TOTAL | 0 | 188.5794068 | 309.7298679 | 279.6925873 | 283.349166 | 1061.351028 |

**Table 20F**

## G- The addition of a branch that produces fixed capital

### 1. General case

We consider the following four basic branches in a model with n production cycles:

$\boldsymbol{K_i = F_i + E_i + C_i + V_i}$

$F_i$: capital in machine tools, buildings...

Branch 1 produces a commodity, the consumption of which by branch i is equal to the amortization $D_i = K_i d_i = {}^{F_i}/_{n}$.

As for the previous models, $E_i$ , $C_i$ and $V_i$ are respectively the capital needed for purchase of energy in each cycle, the capital needed for purchase of raw materials in each cycle and the variable capital, defined as the capital needed to reproduce the workers' labour force in each cycle. $\boldsymbol{e}$ is the exploitation rate.

Similarly to the description in Chapter B, we have:

$$w_i = d_i + e_i + c_i + v_i + pl_i = d_i + e_i + c_i + (1 + e)v_i \quad (4b)$$

$$x_i w_i = d_i + x_2 e_i + x_3 c_i + x_4 v_i + s_i \quad (5c)$$

Note that the coefficient $x_1$, which applies for each cycle to the production of branch 1, does not apply to $d_i$ since the latter corresponds to the amortization of a single purchase at the beginning of the n production cycles. If one imagines the next production cycle, the purchase of the fixed capital by branch i will then cost $x_1 F_i$, and will be amortized in $x_1 D_i = {}^{x_1 F_i}/_{n}$ equal parts.

$$s_i = r(nd_i + x_2 e_i + x_3 c_i + x_4 v_i)$$

Fundamental equality II reads:

$$k_1 w_1 + k_2 w_2 + k_3 w_3 + k_4 w_4 = k_1 x_1 w_1 + k_2 x_2 w_2 + k_3 x_3 w_3 + k_4 x_4 w_4$$

The z-function reads:

$$\begin{aligned} z = k_1[x_1 w_1 - x_2 e_1 - x_3 c_1 - x_4 v_1 - (d_1 + pl_1)] \\ + k_2[x_2(w_2 - e_2) - x_3 c_2 - x_4 v_2 - (d_2 + pl_2)] \\ + k_3[x_3(w_3 - c_3) - x_2 e_3 - x_4 v_3 - (d_3 + pl_3)] \\ + k_4[x_4(w_4 - v_4) - x_2 e_4 - x_3 c_3 - (d_4 + pl_4)] \end{aligned}$$

or: $z = k_1 w_1 x_1 - (k_1 d_1 + k_2 d_2 + k_3 d_3 + k_4 d_4) - [k_2 w_2 - (k_1 e_1 + k_2 e_2 + k_3 e_3 + k_4 e_4)]x_2 + [k_3 w_3 - (k_1 c_1 + k_2 c_2 + k_3 c_3 + k_4 c_4)]x_3 + [k_4 w_4 - (k_1 v_1 + k_2 v_2 + k_3 v_3 + k_4 v_4)]x_4 - \sum_1^4 k_i pl_i$

Marx's fundamental equality I is complied with when $z = 0$

The condition of the fulfillment of the solvent need in each cycle can be expressed by the following inequalities:

$$k_1 w_1 x_1 \geq k_1 d_1 + k_2 d_2 + k_3 d_3 + k_4 d_4$$

$$k_2 w_2 \geq k_1 e_1 + k_2 e_2 + k_3 e_3 + k_4 e_4$$

$$k_3 w_3 \geq k_1 c_1 + k_2 c_2 + k_3 c_3 + k_4 c_4$$

$$k_4 w_4 \geq k_1 v_1 + k_2 v_2 + k_3 v_3 + k_4 v_4$$

Having defined $t = 1 + r$, in the general case with surplus value we have the price system:

$$(-w_1)x_1 + e_1 t x_2 + c_1 t_1 x_3 + v_1 t x_4 = -d_1(1 + nr)$$

$$0x_1 + (e_2 t - w_2)x_2 + c_2 t x_3 + v_3 t x_4 = -d_2(1 + nr)$$

$$0x_1 + e_3 t x_2 + (c_3 t - w_3)x_3 + v_3 t x_4 = -d_3(1 + nr)$$

$$0x_1 + e_4 t x_2 + c_4 t x_3 + (v_4 t - w_4)x_4 = -d_4(1 + nr)$$

A quadruplet of transformation coefficients $(x_1, x_2, x_3, x_4)$ can be found for a given value of $r$ (non-zero determinant of the system). As before, there exists a value $r^*$ which cancels the variable $z$ whose derivative with respect to r is negative for $r^*$. We note that we have to choose two values among the fractions of capital $k_i$ to know univocally the two other fractions. These two chosen capitals can lead to an empty set of solutions and even when the set of solutions is not empty, these solutions may not ensure the fulfillment of the solvent social need (the production of a branch of goods being lower than the consumption of this good). Nevertheless, we allow these solutions as they make sense if we assume the existence of stocks resulting from the surpluses of previous cycles. We will be guided in the choice of fractions $k_i$ by starting with their values given from the case when the surplus value is zero (e=0), which we discuss in the next paragraph. In this case, we have one and only one possible solution, which gives us a starting point in the space of solutions. Then, we just have to progressively increase the exploitation rate e up to the wished value by adjusting the fractions $k_i$ if necessary.

For the next series of n cycles, the fixed capital in branch i, which was $nD_i$ will become $nD_i x_1$. If $x_1 > 1$, the branch that produces fixed capital will have sucked in value from the other branches' surplus value over the previous series of n cycles,. If $x_1 < 1$, on the opposite, part of the surplus value of the fixed capital branch will have been given up to the other branches.

The amortization of the fixed capital purchased at the beginning of a period of n cycles appears in the left-hand column (Table 21A) without a transformation coefficient and is "homogeneous" to a quantity of value. But this fixed capital is renewed during the next period of n cycles at a price produced by the coefficient $x_1$. The spaces of values and prices communicate and transform one into the other: *at the interface between two periods of n cycles, what is a price at the end of the period becomes also a value at the beginning of the next period*. This duality price/value reflects the competition between the different branches to hoover up surplus-value. At the interface between two periods of n cycles: The reproduction price of fixed capital is equal to its value increased or decreased by the fraction of surplus-value respectively taken from or given up to the other branches (depending on whether $x_1$>1 or <1) during the n preceding temporal steps.

## 2. Cases with no surplus value

The transformation coefficients are all unity, so the expression for z-function becomes:

$$z = [k_1 w_1 - (k_1 d_1 + k_2 d_2 + k_3 d_3 + k_4 d_4)] + [k_2 w_2 - (k_1 e_1 + k_2 e_2 + k_3 e_3 + k_4 e_4)] + [k_3 w_3 - (k_1 c_1 + k_2 c_2 + k_3 c_3 + k_4 c_4)] + [k_4 w_4 - (k_1 v_1 + k_2 v_2 + k_3 v_3 + k_4 v_4)]$$

Taking into account the inequalities imposed by the satisfaction of needs, the expression of z can only be cancelled if:

$$k_1 w_1 = k_1 d_1 + k_2 d_2 + k_3 d_3 + k_4 d_4$$

$$k_2 w_2 = k_1 e_1 + k_2 e_2 + k_3 e_3 + k_4 e_4$$

$$k_3 w_3 = k_1 c_1 + k_2 c_2 + k_3 c_3 + k_4 c_4$$

$$k_4 w_4 = k_1 v_1 + k_2 v_2 + k_3 v_3 + k_4 v_4$$

These four equations are mutually dependent (zero determinant).

On the other hand, the conservation of the total capital reads:

$$k_1 + k_2 + k_3 + k_4 = 1$$

Associated with three of the first four equations, we obtain a system that has a unique solution (non-zero determinant). In the case of a closed system, if there is no surplus value, each branch produces strictly as needs require.

## 3. Adding a luxury branch

Rather than a simple reproduction system as in chapter F, which dictates the luxury branch organic composition to match production surplus, it may seem more realistic to consider characteristic socio-technical coefficients for this branch. Besides, as in the case illustrated by V. Laure van Bambeke, the consumer goods industry must also produce for the non-productive ("middle") classes.[14] Furthermore, each commodity branch has a "top of the range" for the ruling class, but whose products can also be bought occasionally by the surplus-value producers.

$$k_L w_L = k_L (d_L + e_L + c_L + v_L)$$

---

[14] The case illustrated by V. Laure van Bambeke is Table 10.7 page 202 of reference (4), which also corresponds to the second table in Chapter H below.

System of price equations:

$$(-w_1)x_1 + e_1tx_2 + c_1t_1x_3 + v_1tx_4 = -d_1(1 + nr)$$

$$0x_1 + (e_2t - w_2)x_2 + c_2tx_3 + v_3tx_4 = -d_2(1 + nr)$$

$$0x_1 + e_3tx_2 + (c_3t - w_3)x_3 + v_3tx_4 = -d_3(1 + nr)$$

$$0x_1 + e_4tx_2 + c_4tx_3 + (v_4t - w_4)x_4 = -d_4(1 + nr)$$

$$(-w_L)x_L + e_Ltx_2 + c_Ltx_3 + v_Ltx_4 = -d_L(1 + nr)$$

In a similar reasoning as in the previous case, the condition of fulfillment of needs is translated into the following system of equalities:

$$k_1w_1 = k_1d_1 + k_2d_2 + k_3d_3 + k_4d_4 + k_Ld_L$$

$$k_2w_2 = k_1e_1 + k_2e_2 + k_3e_3 + k_4e_4 + k_Le_L$$

$$k_3w_3 = k_1c_1 + k_2c_2 + k_3c_3 + k_4c_4 + k_Lc_L$$

$$k_4w_4 = k_1v_1 + k_2v_2 + k_3v_3 + k_4v_4 + k_Lv_L$$

$$k_1 + k_2 + k_3 + k_4 + k_L = 1$$

When its determinant is non-zero this system of five independent equations with five unknowns has *one and only one solution*. Now since the branch L does not produce any fundamental goods, $k_L = 0$ is a solution and therefore it is *the only one* possible.

In the model where fixed capital is produced within the system, we conclude that *an absence of profit means no luxury branch can exist* (the luxury branch being defined simply as a branch whose production is not required for the operation of the other branches).

## 4. Solution with constraint for simple reproduction

Marx's two equalities dictate the amount of capital allocated to two branches when the other three have been chosen. We can add three other conditions to obtain an exactly determined system. For example, we can impose that the production in value of the first three sectors (M,E,C) is equal to their consumption. In such a scheme, the total surplus value is used for the consumption of goods from branches V and L by the non-productive classes and the capitalists. Therefore, this is a simple reproduction scheme. The three constraints lead to the following additional equations:

$$k_1w_1 = k_1d_1 + k_2d_2 + k_3d_3 + k_4d_4 + k_Ld_L$$

$$k_2w_2 = k_1e_1 + k_2e_2 + k_3e_3 + k_4e_4 + k_Le_L$$

$$k_3w_3 = k_1c_1 + k_2c_2 + k_3c_3 + k_4c_4 + k_Lc_L$$

We already had the two equations:

$$k_1 w_1(1-x_1) + k_2 w_2(1-x_2) + k_3 w_3(1-x_3) + k_4 w_4(1-x_4) + k_L w_4(1-x_L)$$

$$k_1 + k_2 + k_3 + k_4 + k_L = 1$$

This system generally has one and only one solution (nonzero determinant). The example below illustrates the effect of these constraints applied to values (Tables 21A-C) or to prices (Tables 21 D-F). Note that when the constraints are applied to values, since the transformation coefficient $x_1$ (that applies to the "fixed capital" sector (machines, buildings,…) is greater than 1, the price of the renewed fixed capital is more expensive than the price it costed at the onset of the series of cycles. The system is non-stationary.

| **VALUES** | F/n | E | C | V | PL | W |
|---|---|---|---|---|---|---|
| BRANCH 1 M | 0.942281451 | 4.400705989 | 8.795475011 | 5.654254129 | 5.654254129 | 25.44697071 |
| BRANCH 2 E | 10.92332862 | 25.48849501 | 50.97480536 | 32.76998587 | 32.76998587 | 152.9266007 |
| BRANCH 3 C | 2.334280629 | 38.90545524 | 77.8085762 | 93.37122515 | 93.37122515 | 305.7907624 |
| BRANCH 4 V | 10.89494353 | 84.12486001 | 168.1981097 | 154.8310307 | 154.8310307 | 572.8799746 |
| BRANCH 5 L | 0.352136478 | 0.007084491 | 0.013796114 | 0.186433967 | 0.186433967 | 0.745885017 |
| TOTAL | 25.44697071 | 152.9266007 | 305.7907624 | 286.8129298 | 286.8129298 | 1057.790193 |

**Table 21 A**

| **PRICES** | F/n | E | C | V | S | W |
|---|---|---|---|---|---|---|
| BRANCH 1 M | 0.942281451 | 5.267352601 | 8.026570932 | 5.587541245 | 8.118033268 | 27.9417795 |
| BRANCH 2 E | 10.92332862 | 30.50803459 | 46.5185667 | 32.38334243 | 62.70970537 | 183.0429777 |
| BRANCH 3 C | 2.334280629 | 46.5672443 | 71.00651815 | 92.26956549 | 66.88079835 | 279.0584069 |
| BRANCH 4 V | 10.89494353 | 100.6918666 | 153.4941611 | 153.0042248 | 148.035536 | 566.120732 |
| BRANCH 5 L | 0.352136478 | 0.008479665 | 0.012590052 | 0.184234288 | 1.068856792 | 1.626297273 |
| TOTAL | 25.44697071 | 183.0429777 | 279.0584069 | 283.4289083 | 286.8129298 | 1057.790193 |

**Table 21 B**

| | K (Value) | Kp | xi | Rate of profit |
|---|---|---|---|---|
| BRANCH 1 M | 28.27324963 | 28.30427928 | 1.09803952 | 0.28681293 |
| BRANCH 2 E | 218.4665725 | 218.64323 | 1.196933541 | 0.28681293 |
| BRANCH 3 C | 233.4280629 | 233.1861342 | 0.912579585 | 0.28681293 |
| BRANCH 4 V | 516.1034357 | 516.1396877 | 0.988201294 | 0.28681293 |
| BRANCH 5 L | 3.728679349 | 3.726668782 | 2.180359219 | 0.28681293 |
| TOTAL | 1000 | 1000 | | |

**Table 21 C**

| VALUES | F/n | E | C | V | PL | W |
|---|---|---|---|---|---|---|
| BRANCH 1 M | 0.859545387 | 4.014306478 | 8.023197278 | 5.157788099 | 5.157788099 | 23.21262534 |
| BRANCH 2 E | 10.9134292 | 25.4653957 | 50.92860872 | 32.74028761 | 32.74028761 | 152.7880089 |
| BRANCH 3 C | 2.332166525 | 38.87021947 | 77.73810678 | 93.286661 | 93.286661 | 305.5138148 |
| BRANCH 4 V | 10.93433944 | 84.42905391 | 168.8063109 | 155.3908968 | 155.3908968 | 574.9514977 |
| BRANCH 5 L | 0.44900193 | 0.009033287 | 0.017591139 | 0.237718091 | 0.237718091 | 0.951062539 |
| TOTAL | 25.48848248 | 152.7880089 | 305.5138148 | 286.8133516 | 286.8133516 | 1057.417009 |

**Table 21 D**

| PRICES | F/n | E | C | V | S | W |
|---|---|---|---|---|---|---|
| BRANCH 1 M | 0.859545387 | 4.804875763 | 7.321838029 | 5.096954377 | 7.405268926 | 25.48848248 |
| BRANCH 2 E | 10.9134292 | 30.48049851 | 46.47661165 | 32.35413108 | 62.65309553 | 182.877766 |
| BRANCH 3 C | 2.332166525 | 46.52524078 | 70.94251914 | 92.18638803 | 66.82057382 | 278.8068883 |
| BRANCH 4 V | 10.93433944 | 101.0563386 | 154.0498661 | 153.5581331 | 148.5715338 | 568.1702111 |
| BRANCH 5 L | 0.44900193 | 0.010812285 | 0.016053384 | 0.234914316 | 1.36287948 | 2.073661395 |
| TOTAL | 25.48848248 | 182.877766 | 278.8068883 | 283.4305209 | 286.8133516 | 1057.417009 |

**Table 21 E**

| | K (Value) | Kp | xi | Rate of profit |
|---|---|---|---|---|
| BRANCH 1 M | 25.79074572 | 25.81912204 | 1.098043935 | 0.286813352 |
| BRANCH 2 E | 218.2685841 | 218.4455333 | 1.196937949 | 0.286813352 |
| BRANCH 3 C | 233.2166525 | 232.9758132 | 0.912583572 | 0.286813352 |
| BRANCH 4 V | 517.9696559 | 518.0077322 | 0.988205463 | 0.286813352 |
| BRANCH 5 L | 4.754361821 | 4.751799289 | 2.180362816 | 0.286813352 |
| TOTAL | 1000 | 1000 | | |

**Table 21 F**

Now, if the surplus value is zeroed, there is a reorganization of capital away from the luxury branch and into the core industries, as illustrated in Table 22.

| PRICES | F/n | E | C | V | S | W |
|---|---|---|---|---|---|---|
| BRANCH 1 M | 1.228824113 | 5.738936736 | 11.47013111 | 7.373682046 | 0 | 25.811574 |
| BRANCH 2 E | 13.74645109 | 32.07596898 | 64.14918867 | 41.23935328 | 0 | 151.210962 |
| BRANCH 3 C | 3.322774916 | 55.38068952 | 110.7580563 | 132.9109966 | 0 | 302.3725173 |
| BRANCH 4 V | 7.513523881 | 58.01536678 | 115.9951413 | 106.7767487 | 0 | 288.3007806 |
| BRANCH 5 L | 0 | 0 | 0 | 0 | 0 | 0 |
| TOTAL | 25.811574 | 151.210962 | 302.3725173 | 288.3007806 | 5.68E-14 | 767.695834 |

**Table 22**

These examples shed light on the link between surplus value and the luxury branch in a system where no constrain is applied to force its reproducibility. As is demonstrated in chapter G, in a system that produces everything it consumes, *there can be no luxury industry without profit*.

## H- Transformation's neutral element and the non-equivalence between exploitation and productivity

Let's consider again the example of Table 21 with five branches and n=10 amortization cycles. Before running the transformation algorithm, the initially entered values were:

| **INIT VALUES** | F/n | E | C | V | PL | W |
|---|---|---|---|---|---|---|
| BRANCH 1 M | 6.665533413 | 31.12982091 | 62.21764476 | 39.9972002 | 39.9972002 | 180.0073995 |
| BRANCH 2 E | 10 | 23.334 | 46.666 | 30 | 30 | 140 |
| BRANCH 3 C | 2 | 33.334 | 66.666 | 80 | 80 | 262 |
| BRANCH 4 V | 4.222 | 32.6 | 65.18 | 60 | 60 | 222.002 |
| BRANCH 5 L | 18.888 | 0.38 | 0.74 | 10 | 10 | 40.008 |
| TOTAL | 41.77553341 | 120.7778209 | 241.4696448 | 219.9972002 | 219.9972002 | 844.0173995 |

We have generally considered a uniform exploitation rate of 100% for all branches. For this reason, the values of PL and V were chosen to be identical in the initial layout. A capital allocation that preserves the 100% exploitation rate, complies with both fundamental equalities and a uniform rate of profit amongst branches is obtained from the algorithm (see Appendix) and leads to the following tables (in values and in price).

| **VALUES** | F/n | E | C | V | PL | W |
|---|---|---|---|---|---|---|
| BRANCH 1 M | 0.859573858 | 4.014439445 | 8.023463034 | 5.157958943 | 5.157958943 | 23.21339422 |
| BRANCH 2 E | 10.9135 | 25.4655609 | 50.9289391 | 32.7405 | 32.7405 | 152.789 |
| BRANCH 3 C | 2.3323067 | 38.87255577 | 77.74277923 | 93.292268 | 93.292268 | 305.5321777 |
| BRANCH 4 V | 10.93390257 | 84.42568071 | 168.7995665 | 155.3846884 | 155.3846884 | 574.9285266 |
| BRANCH 5 L | 0.449418107 | 0.00904166 | 0.017607444 | 0.23793843 | 0.23793843 | 0.951944073 |
| TOTAL | 25.48870124 | 152.7872785 | 305.5123553 | 286.8133538 | 286.8133538 | 1057.415043 |

| **PRICES** | F/n | E | C | V | S | W |
|---|---|---|---|---|---|---|
| BRANCH 1 M | 0.859573858 | 4.80503501 | 7.322080722 | 5.097123319 | 7.405514379 | 25.48932729 |
| BRANCH 2 E | 10.9135 | 30.48069683 | 46.47691422 | 32.35434168 | 62.65350314 | 182.8789559 |
| BRANCH 3 C | 2.3323067 | 46.52803809 | 70.94678478 | 92.19193095 | 66.82459191 | 278.8236524 |
| BRANCH 4 V | 10.93390257 | 101.0523031 | 154.0437149 | 153.5520014 | 148.5656016 | 568.1475236 |
| BRANCH 5 L | 0.449418107 | 0.010822307 | 0.016068264 | 0.235132062 | 1.364142737 | 2.075583477 |
| TOTAL | 25.48870124 | 182.8768953 | 278.8055629 | 283.4305294 | 286.8133538 | 1057.415043 |

Now, this table in prices could conceivably be interpreted as a table in values for which the exploitation rates would be different from one branch to another. In that case these exploitation rates might be better defined as *productivity rates* (see below). In turns out that *this table would then be a neutral element of the transformation*.

It could be argued that the worker's time is "better" exploited when it is associated with more sophisticated machines and leads to a greater quantity or quality of goods produced over the same working time. So one could say that the increased value results from higher productivity rather than higher exploitation. We shall call this conception the "productivity-only-conception" (keeping in mind that it differs from the Marxist conception, in which productivity is rather about the quantity of commodities).[15] In the productivity-only-conception, the difference between values and prices would not be relevant: the price table could be interpreted as an economy with specific productivity rates for each sector, making it unnecessary to transform values into prices, and useless the coexistence of a value system and a price system.

However, once at the equilibrium, with a same rate of profit in all branches, the productivity-only-conception and Marx's conception have contrasting implications. In the productivity-only-conception, rates of profit are not impacted by capital fluxes across branches. In contrast, in Marx's conception, the average (uniform) rate of profit is affected by capital fluxes across branches. Therefore, in the productivity-only-conception improvement in productivity, all other things being equal, can only lead to an increase in the average rate of profit. In Marx's conception the improvement in productivity, all other things being equal, leads to an opposite effect: a tendency for the rate of profit to fall (TRPF). Note that for both conceptions, the direction of change in the average profit rate is a trend that may not materializes as other factors (rate of surplus value, GDP growth) may thwart it.

The major issue with the ability of the productivity-only-conception to explain capitalism is that it would mean that productivity creates value even when there is no human labour. Such a conception would make capitalist-compatible a world with no working class, i.e a world with no capitalism.

[15] For example, Marx writes at the end of chapter 3, in Capital, volume II « Increased productivity can increase only the substance of capital but not its value; but therewith it creates additional material for the self-expansion of that value. »

## I- Variations of the z-function and determination of $\boldsymbol{r}^*$

Considering the expression of the z-function given by equation 7a (or 7b) as a function of the coefficients $k_i$ et $x_i$, we showed that it could be expressed as a function of $r$. Its form would then be the ratio of two polynomials whose degrees increase with the number of branches considered. This polynomial form implies that there can be several values of $r$ which cancel the z-function, i.e., for which fundamental equality I is verified. However, amongst these possible $r$ values, only the ones for which capital values and transformation coefficient values are positive and for which fundamental equality II is complied with make economic sense. The smallest of these valid $r$ values is named $r^*$. Here, we show how $r^*$ can be unambiguously identified amongst all the $r$ values which cancel the z-function.

We have defined the z-function as:

$$z = \sum_i S_i - \sum_i PL_i = \sum_i (S_i - PL_i)$$

In its first step our algorithm calculates the two unfixed capitals in such a way as to comply with fundamental equality II, i.e.:

$$\sum_i W_i = \sum_i W_i\, x_i$$

We consider a value of $r$ close enough to $r^*$ so that the provisional solutions for the $k_i$ are all positive. We consider the case of three branches for clarity and without impairing generalization. By detailing each side of the fundamental equality II we have:

$$\sum_i (d_i + E_i + C_i + V_i + PL_i) = \sum_i (d_i + E_i x_1 + C_i x_2 + V_i x_3 + S_i)$$

When $r < r^*$ prices are globally lower than when $r = r^*$, so:

$$\sum_i (d_i + E_i + C_i + V_i) > \sum_i (d_i + E_i x_1 + C_i x_2 + V_i x_3)$$

Therefore, when $r < r^*$, fundamental equality II can be complied with only for $S_i > PL_i$, i.e. for $z > 0$.

We conclude that in the vicinity of the solution $r^*$ and for $r < r^*, z > 0$

Therefore, as illustrated in Figure 8, we define $r^*$ as the first value that both cancels z and is such that

$$\left[\frac{dz}{dr}\right]_{r*} < 0$$

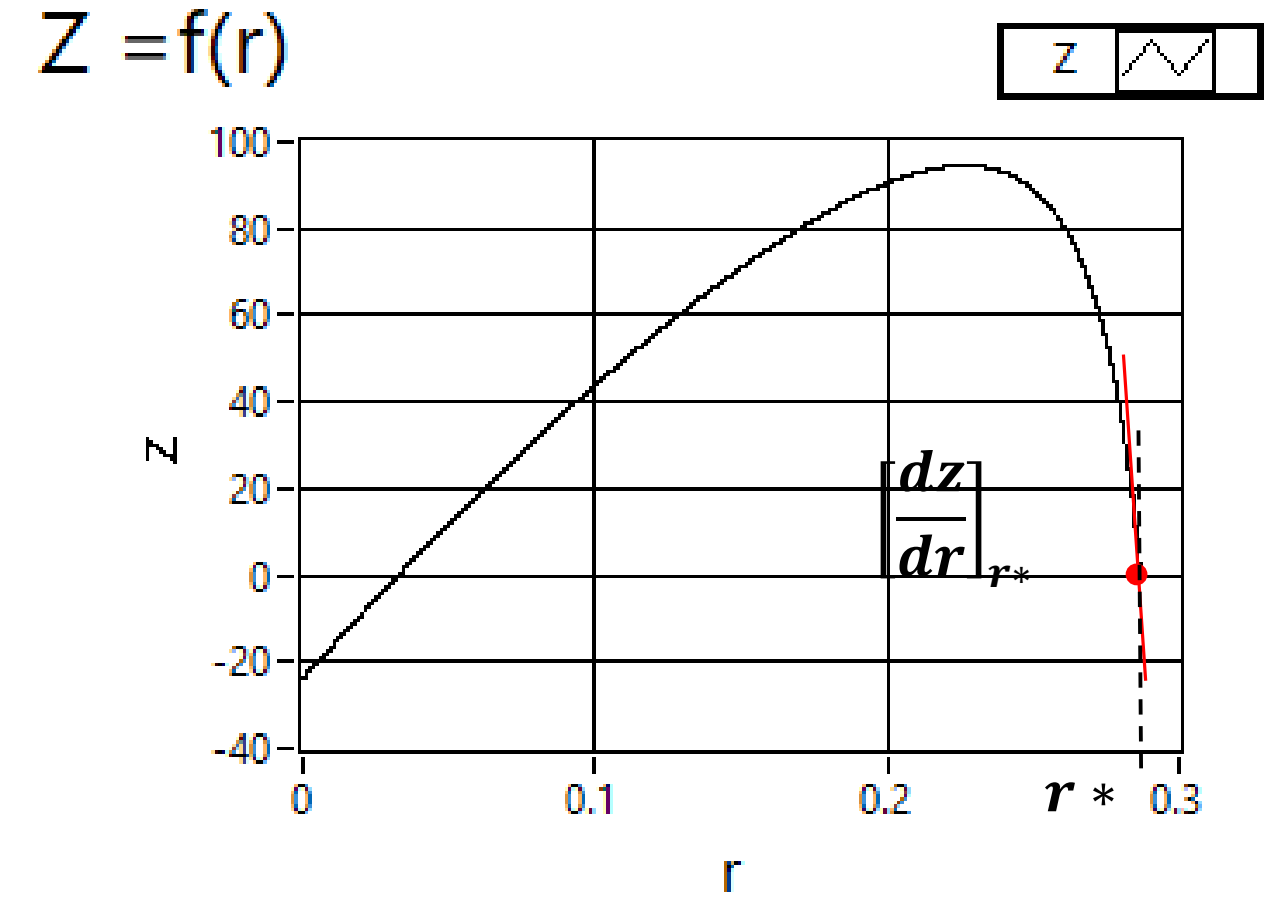

**Figure 8**

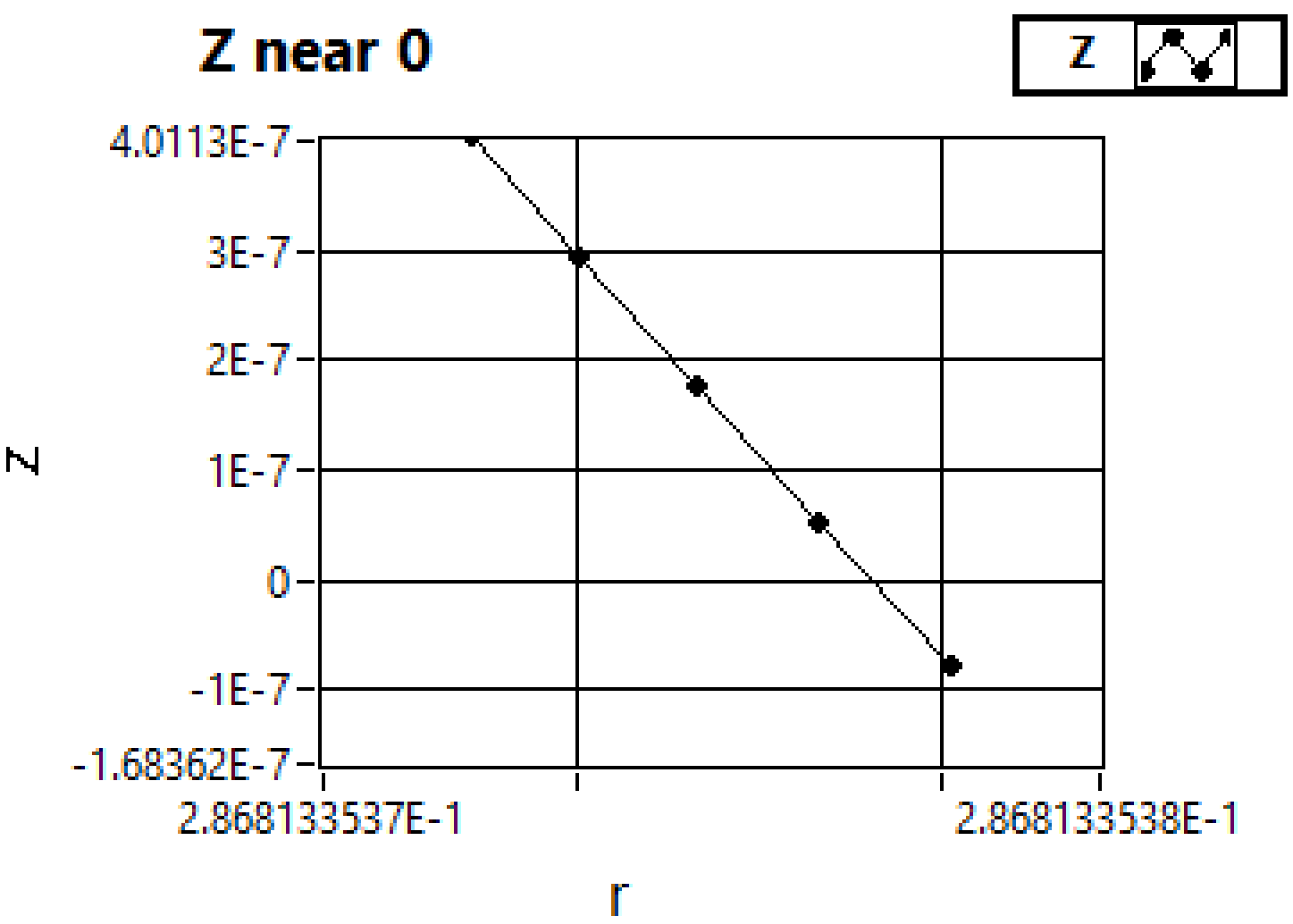

**Figure 9**

In other words, $r^*$ is the first value found that cancels the z-function when its graph crosses the r-axis from top to bottom. In the vicinity of $r^*$ the z-function is continuous, implying there is an exact solution. We have not attempted to determine an analytical form of this solution, rather, we provide an algorithm (see Appendix) enabling an approximation by means of interpolation of $r^*$ with accuracy (Figure 9). The transformation coefficients are determined in terms of $r^*$ in a unique way (see Chapters A and B).

Note that for certain initial configurations, in particular for certain choices of the amounts of committed capital that we fix, there are no solutions and the z function never crosses the ordinate axis with a negative derivative.

## GENERAL CONCLUSION

To use the expression formulated by the philosopher of science Etienne Klein, «le monde s'oppose à ce qu'il nous montre», it is common in the history of science for an essential law to be discovered against the appearance of the facts. [16] The example given by E. Klein is that Galileo postulated, in spite of common observation, that bodies subject to attraction must fall at the same speed, regardless of their weight. We want to recall that the author whose sentence "Facts are stubborn" became famous also drawn up an indictment without appeal against the positivist trend and its vulgar empiricism often mistaken with the scientific method.[17,18] Likewise, the law of value conceptualized by K. Marx may seem to contradict the facts. This apparent contradiction has brought attacks on his work, as illustrated by von Böhm-Bawerk's criticism (23):

> The law of value maintains that quantity of labor alone determines the exchange relations; facts show that it is *not* only the quantity of labor, or the factors in harmony with it, which determine the exchange relations. These two propositions bear the same relation to each other as Yes to No—as affirmation to contradiction.[19]

Or more:

> Marx has not deduced from facts the fundamental principles of his system, either by means of a sound empiricism or a solid economico-psychological analysis; he founds it on no firmer ground than a formal dialectic.

Marx wrote Volume III of Capital before Volume I (24). It is in Volume III that the method of transforming values into prices is presented, and he "is as well aware of this fact as any orthodox economist" according to Lexis (25), and as well as Galileo in his statement on the fall of objects, that this law of value is in apparent contradiction with the facts. Behind the appearances, Marx seeks to reveal the essence of the principles and, as Lexis acknowledges, "profit and rate of profit are simply the surface of the phenomenon, while surplus value the rate of surplus value are the hidden but essential objects of investigation" (25).

Marx never completed the full writing of Volume III. For the transformation of values into market production prices, only a general orientation is given. We may argue that Marx did not dispose of the appropriate mathematical tools to strictly demonstrate the validity of his theory of value. At this stage, Marx authorized himself approximations. As recalled in our introduction, Marx himself had warned about the approximation of using values instead of prices (Table 1) to calculate the market price of production (Volume III, Chapter 9). The meaning of this passage in the book has been controversial. M. Husson argues instead that Marx, using prices to calculate the cost of production, warns against double counting profit (26). On the contrary, others, like Bortkiewicz, argue that Marx uses values for inputs (6). Although we do not share M. Husson's conclusions, we recognize the difficulty of interpreting this passage from Volume III.

---

[16] Literally translation of the quote of E. Klein: "the world opposes to what it shows us" from (22)

[17] "Facts are stubborn", from Lenin's *Collected Works*, Progress Publishers, Moscow, Volume 27, 1972, pages 79-84

[18] See "Materialism and Empirio-criticism": V. I. Lenine Collected Works Volume 14 1908, Progress Publishers, Moscow

[19] Note that this criticism does not acknowledge the meaning which Marx gives to "socially necessary labour time", which reflects exchange relationships.

The fundamental principle underlying the theory of value and exploitation is that all value of a commodity is rooted in the human labour socially necessary to its production. This principle breaks down into two postulates which can be summarized in a simplified way as follows.

1) Any value is derived from work.
2) Any profit is derived from surplus labour.

Marx proposed that these postulates apply to the capitalist economy, provided that we admit a mismatch between values and prices. The problem of transformation is to unveil the rules underlying this mismatch for the different branches of a global economy. Its solving by complying with the two postulates would appear to us a very arduous task if the flow of capitals towards an adequate allocation was not taken into account. There has certainly been no lack of ingenuity and brilliant minds, some of which well-equipped mathematically, to tackle it. However, we believe that the problem was wrongly formulated, even if some compromise solutions were considered satisfactory by some (27-31).

As far as we are aware, V. Laure van Bambeke has been the first to propose a solution of the transformation problem by including in it, as an inseparable component of the problem, the determination of the allocation of the capital (between branches) which complies with Marx's fundamental equalities. In the introduction, we have pointed out by an analogy with electrostatics that this way of seeing things sounds quite natural. However, this analogy has limitations, especially when it comes to consider time. The distribution of electrical charges on the surface of a conductor is almost instantaneous, whereas investments and the modernization of production tools take time. Precisely, even if capital flows are faster than in Marx's time, this temporal shift allows for the existence of a transitional regime during which capitalists in a sector gain an advantage by lowering their production costs. This advantage leads to a temporarily higher rate of profit in the sector. This rate differential then drives inflows of capital into that industry. The subsequent equalization of rates is later accompanied by a TRPF (with exploitation rates and production conditions being elsewhere unchanged).

In V. Laure van Bambeke's first model (4, 13) in the case of two or three branches, the imported fixed capital is not transformed and therefore has a price equal to its value. But this fixed capital has itself been constituted by socially necessary labour within the framework of a production with a given organic composition.  In his four-branch model (13) Laure van Bambeke addresses this issue by adding a branch that produces fixed capital within the economic system with an associated transformation coefficient. In order to show the consistency of the law of value with the way profit is shared, we have gone further by solving the cases for which fixed capital is zero. This thought experiment provides the possibility of conceiving an economic system in which no prior value exists. Moreover, as M. Husson points out in his critique of V. Laure van Bambeke's work (11), one does not see why this simplified problem would be more difficult to solve than the obviously more complex one that incorporates fixed capital. Above all, by taking our thought experiment even further, by considering models without fixed capital and without profit, we unveil the link between Marx's fundamental equalities and the satisfaction of the solvent social need. This approach provides evidence that the allocation of capital between branches cannot be considered as an exogenous and fixed feature of the problem of transformation. It must be determined together with the transformation coefficients.

In the context of models without fixed capital, we show that, for a uniform rate of profit, the latter is invariant whatever the capital allocation is that complies with both fundamental equalities. This property may explain the success of the classical theory which calculates the rate of profit from the eigenvalue of the socio-technical matrix, independently of the amounts of capital and their allocation. The widespread use of this approach may explain the reluctance to admit that the transformation problem includes in its solution the determination of an adequate capital allocation between branches.

We argue that the solving approach presented here is more appropriate than the one presented by Laure van Bambeke.

In his 2018 article, V. Laure Van Bambeke writes:

> 2. But taking into account Marx's constraints within the model of price determination from values leads to an overdetermined analytical system that has no solution in the classical sense.
>
> 3. However, such a system always has an approximate solution in the sense of the least squares method. After a brief introduction to the mathematical method of Moore and Penrose, we apply it to a complete model of price determination from values.
>
> 4. We then admit that the prices thus calculated are the effective prices that can cause an unbalanced situation characterized by profitability gaps between branches and capital transfers to the most profitable activities. An equilibrium point is reached after many iterations when the allocation of capital becomes efficient and the two equalities, supposedly irreconcilable and contradictory, are simultaneously complied with. [20]

Yet Laure van Bambeke in his book (12) has correctly characterized the general problem as bilinear, in $x_i$ coefficient of transformation of value into price for commodity i) and in $K_i$ (amount of capital in value attributed to branch i). Its detour through the Moore-Penrose method to obtain approximate values is unnecessary and the system actually has a set (or an infinity of sets) of exact solutions. Marx's postulates 1) and 2) mentioned above form the basis of the fundamental conception and must operate even when the rates of profit of the industries are different. They cannot be considered as approximations. In the Moore-Penrose method, the state of equality of rates of profit is approached progressively by successive transfers, which eventually lead to the correct solution, but there is no theoretical reason why Marx's equalities should not be complied with during this "approach phase". Yet, with this method, since the differences (the differences between the rates and the two differences between surplus values and profits as well as between prices and values) are adjusted according to the least squares method, the fundamental equalities will only be (approximately) complied with once the equality of the rates has been (approximately) reached.

[20] Translation of the original text, which is in French (4).

In our algorithm, the rate of profit is used as an adjustment variable, but it is also directly tied to exploitation since it depends primarily on the rate of surplus value. However, for the profit to be effectively realized, the produced commodity must also be sold and the absorption capacity of the market has to be taken into account. Our algorithm achieves this by strictly complying with both fundamental equalities. Now, any unsold goods are lost value and this amount of lost value is down to the tendency of the capitalist system to expand production capacities faster than the markets can expand, thereby leading to overproduction crises. Note that in our models, the surplus value either takes the form of luxury or "high-end" commodities or is used by the capitalist class to develop its capital and resist competition.

In a monopoly situation, one branch may benefit from a higher rate of profit than other branches. A clever "marketing" operation may also provide a commodity with a unique characteristic, such that the firm that produces it finds itself in a monopoly situation. For example, the company that manufactures Nike sports shoes may make extra profit than other branches not because it is the only one that can make sports shoes but because it produces sports shoes bearing the seek after Nike logo. [21] The argument that a commodity can be sold at a price that has more to do with the urge consumers have for it than with the amount of labour it contains or the amount of capital invested by the capitalist has often been used to challenge the relevance of the concept of objective value in Marxist theory. However, subjective urge has to meet the objective reality of the means of production. In a global economy, an extra profit may eventually be achievable only thanks to the transfer of the surplus-value produced by wage earners from all branches. In the framework of Marxist theory, a non-desirable commodity has no value no matter how much time is spent on its production. This is not to say that a desired commodity contains some value as if value was a substance. Value, however objective it may be, arises from complex interactions that involve the whole social field from production to consumption. The number of hours socially necessary and the efforts that are put into producing the commodity remain indispensable for understanding the dynamics of capitalism, but they result from struggles between workers and their employers that fluctuate constantly according to time and place.

[21] Example used by Fréderic Lordon (32).

While value is carried along with the commodity, it is not a substance contained in it. [22] These considerations, which endow the commodity with a social and relational dimension, clearly differentiate the concept of value in Marxism from its classical Ricardian meaning, as stated by Marx and other authors (8).

We categorically reject the "non-stationary" theory that M. Husson has put forward to explain the failures of previous attempts to provide a resolution of the problem of transformation that complies with fundamental equalities: the idea that the inputs and outputs that make up commodities change their value in the course of the production process rending the capitalist system intrinsically out of balance (26). Of course, this theory seeks to approximate the real situation and certainly can match the Marxist patterns, but one could, by increasing the number of variables at will, verify almost any other hypothesis. *If the Marxist conception is coherent, as we argue here, it is expected to work in a situation in which the values of the inputs and the values of the outputs remain stable even if this situation is idealized*. This is what we show by our algorithmic development based on simultaneous equations applied to a system in transient equilibrium. As I. Rubin pointed out (8):

> Economic life is a sea of fluctuating motion. It is not possible to observe the state of equilibrium in the distribution of labor among the various branches of production at any one moment. But without such a theoretically conceived state of equilibrium, the character and direction of the fluctuating movement cannot be explained.

The TRPF hypothesis related to the organic composition of capital is confirmed for the models we present. We show that the Okishio theorem applies, but contrary to the usual interpretation, not only it does not challenge the TRPF but it also explains its initiation. In this context, it is interesting to note that the organic composition of capital can change in opposite directions depending on whether it is expressed in price or in value. As our three-branch example with fixed capital illustrates (Chapter E), it is actually the law of value that governs the TRPF. These conclusions could be drawn because our approach enables us to study the economic system as a whole. This approach is consistent with Serge Latouche's obvious and long-standing proposal about the TRPF (34):

> On the other hand, if the evolution of productivity modifies the values of the elements of constant capital as well as variable capital, it is necessary to explain the whole economic system and no longer to reason about what for Marx could appear to be a "representative branch", because any modification in a branch has repercussions throughout the economic system. This new way of formulating the problem would undoubtedly clarify the somewhat mysterious "mechanism" that is the tendency of the rate of profit to fall. [23]

[22] As an analogy, the mass of an elementary particle, rather than a substance contained in it, arises from its interaction with a scalar field called the Higgs field (33).

[23] Translated from the French version (34).

Our work shows that the Marxist conception of the transformation of value into market production prices is highly coherent once the allocation of capital in each branch is considered to be part of the solution, as much as the determination of the coefficients of transformation. The idealized experiment in which fixed capital and profit are zero is useful for this demonstration: it shows that the allocation of capital cannot be an exogenous datum as it is determined by the equilibrium between the commodity production and the fulfillment of the solvent need. An important consequence is that the contradictions of capitalism cannot be reduced to the single problem of surplus sharing. As Marx pointed out, the capitalist first helps to "create what is to be subtracted" (35). The conflict between classes originates from the sphere of production and cannot be solved by solely considering the mode of distribution. A corollary is that it is not enough to change the distribution of wealth to overcome capitalism.

The Marxist conception that postulates that any value originates from a quantity of labour and that all profit comes from human surplus labour (thus from human exploitation) is therefore not only compatible with classical economic models with multiple branches, but in our opinion contains an unequalled explanatory power for the crises of capitalism. It is the organic composition of capital in value and not the composition that appears to us at first sight (in price), that governs the TRPF at a constant rate of exploitation. Samuelson (36) could have realized this if he had not erased the table of values. [24]

The algorithm we provide for solving the transformation problem (see Appendix) enables a swift determination of a set of exact (not approximate) solutions. This algorithm also works in the case of branches with zero fixed capital or in the case of different rates of profit or surplus value between branches. A runtime of this algorithm coded in LabVIEW language is provided as an additional file.

[24] In Ref (36), P. A. Samuelson writes « In summary, "transforming" from values to prices can be described logically as the following procedure: "(1) Write down the value relations; (2) take an eraser and rub them out; (3) finally write down the price relations thus completing the so-called transformation process."»

# APPENDIX

A beta version of a program which applies the algorithm described below, and enables the solving of the transformation problem, is supplied as an annex file from the following link:

https://hal.archives-ouvertes.fr/hal-03458603

## 1. Algorithms for solving the transformation

The algorithm is described for the three-branch case. It can be generalized to a higher number of branches. Figure 10 shows the steps to obtain the system of equations in x.

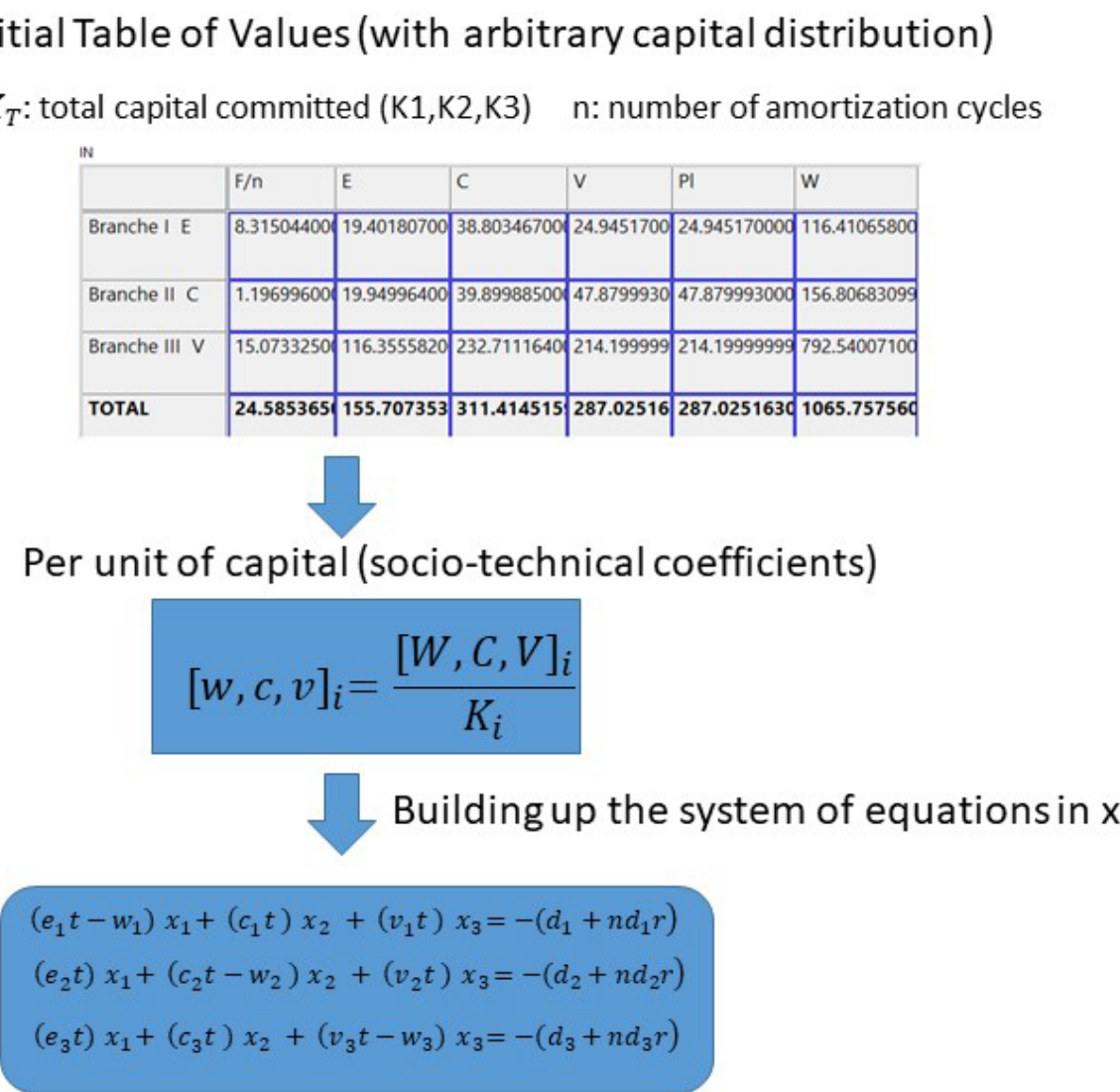

| | F/n | E | C | V | Pl | W |
|---|---|---|---|---|---|---|
| Branche I E | 8.31504400( | 19.40180700 | 38.80346700( | 24.9451700 | 24.945170000 | 116.41065800 |
| Branche II C | 1.19699600( | 19.94996400 | 39.89988500( | 47.8799930 | 47.879993000 | 156.80683099 |
| Branche III V | 15.0733250( | 116.3555820 | 232.7111640( | 214.199999 | 214.19999999 | 792.54007100 |
| **TOTAL** | **24.585365(** | **155.707353** | **311.414515** | **287.02516** | **287.0251630** | **1065.75756(** |

**Figure 10**

When initializing the value of $r$, the system of equations in x provides a solution $X^* = (x_1, x_2, x_3)_r$ on the basis of which the system in K can be solved. Since this system has an infinite number of solutions, we set the value of one of the $K_i$, for example $K_3$ ($K_3$ in the case of branches 1 and 2 having different organic compositions) to solve the following system:

$$K_1 w_1 (1 - x_1) + K_2 w_2 (1 - x_2) = -K_3 w_3 (1 - x_3)$$

$$K_1 + K_2 = K_T - K_3$$

The determinant is

$$D = \begin{vmatrix} \mathrm{w1}\ (1 - \mathrm{x1}) & \mathrm{w2}\ (1 - \mathrm{x2}) \\ 1 & 1 \end{vmatrix}$$

Case D = 0

If the organic compositions of branches 1 and 2 are identical, then we need only choose among them the branch whose capital amount is set to get back to the case D ≠ 0. If the three branches are of identical organic compositions, then the prices are identical to the values and the transformation coefficients are unity. In this particular case there is only one way to allocate profit and meet the solvent social need (demand) and the allocation $(\boldsymbol{K_1}, \boldsymbol{K_2}, \boldsymbol{K_3})$ is the unique solution of the system of the following three equations:

$$K_1[(\boldsymbol{w_1} - \boldsymbol{e_1} - \boldsymbol{d_1})] - \boldsymbol{K_2 e_2} - \boldsymbol{K_3 e_3} = \mathbf{0}$$

$$-\boldsymbol{K_1}[(\boldsymbol{c_1})] - \boldsymbol{K_2}(\boldsymbol{w_2} - \boldsymbol{c_2} - \boldsymbol{d_2}) - \boldsymbol{K_3 c_3} = \mathbf{0}$$

$$\boldsymbol{K_1} + \boldsymbol{K_2} + \boldsymbol{K_3} = \boldsymbol{K_T}$$

The first two equations indicate that the Energy productions of branch 1 and raw materials of branch 2 are sufficient, the third equation is the conservation of total capital.

Case D ≠ 0 :

The organic compositions of branch 1 and branch 2 are assumed to be different.

$$\mathbf{K1} = \frac{\begin{vmatrix} -\mathrm{K3\ w3}(1-\mathrm{x3}) & \mathrm{w2\ }(1-\mathrm{x2}) \\ KT - \mathrm{K3} & 1 \end{vmatrix}}{D}$$

$$\mathbf{K2} = \frac{\begin{vmatrix} \mathrm{w1}(1-\mathrm{x1}) & -\mathrm{K3\ w3}(1-\mathrm{x3})) \\ 1 & KT - \mathrm{K3} \end{vmatrix}}{D}$$

$$\mathrm{K3}\ fixed$$

Remark : If the organic compositions are identical for only two of the branches, then the branch whose capital is set to leave only two unknowns must be one of them to avoid the determinant being zero. For this reason, the organic compositions of branch 1 and branch 2 above are assumed to be different.

If one of the values found for $K_1$ or $K_2$ is negative and this for all possible values of $r$, this means that the value chosen for $K_3$ is impossible and that another one must be chosen.

If the values found for $K_1$ and $K_2$ are positive, it remains to determine $r^*$ which cancels the z-function which corresponds to the compliance with the fundamental equalities.

$$\boldsymbol{z} = \boldsymbol{K_1}[(w_1 - e_1)\, x_1 - c_1\, x_2 - v_1\, x_3 - (d_1 + pl_1)] + \boldsymbol{K_2}[(w_2 - c_2)\, x_2 - e_2\, x_1 - v_2\, x_3 - (d_2 + pl_2)] + \boldsymbol{K_3}[(w_3 - v_3)\, x_3 - c_3\, x_2 - e_3\, x_1 - (d_3 + pl_3)]$$

Starting with increasing values of $r$ from $r = 0$, we look for the first passage of $z$ from a positive to a negative value (see Chapter İ on z-function). This process is described is diagrammed in Figure 11. Note that the algorithm can be generalized to cases of rates of profit different between branches ($ri = ri + \Delta ri$).

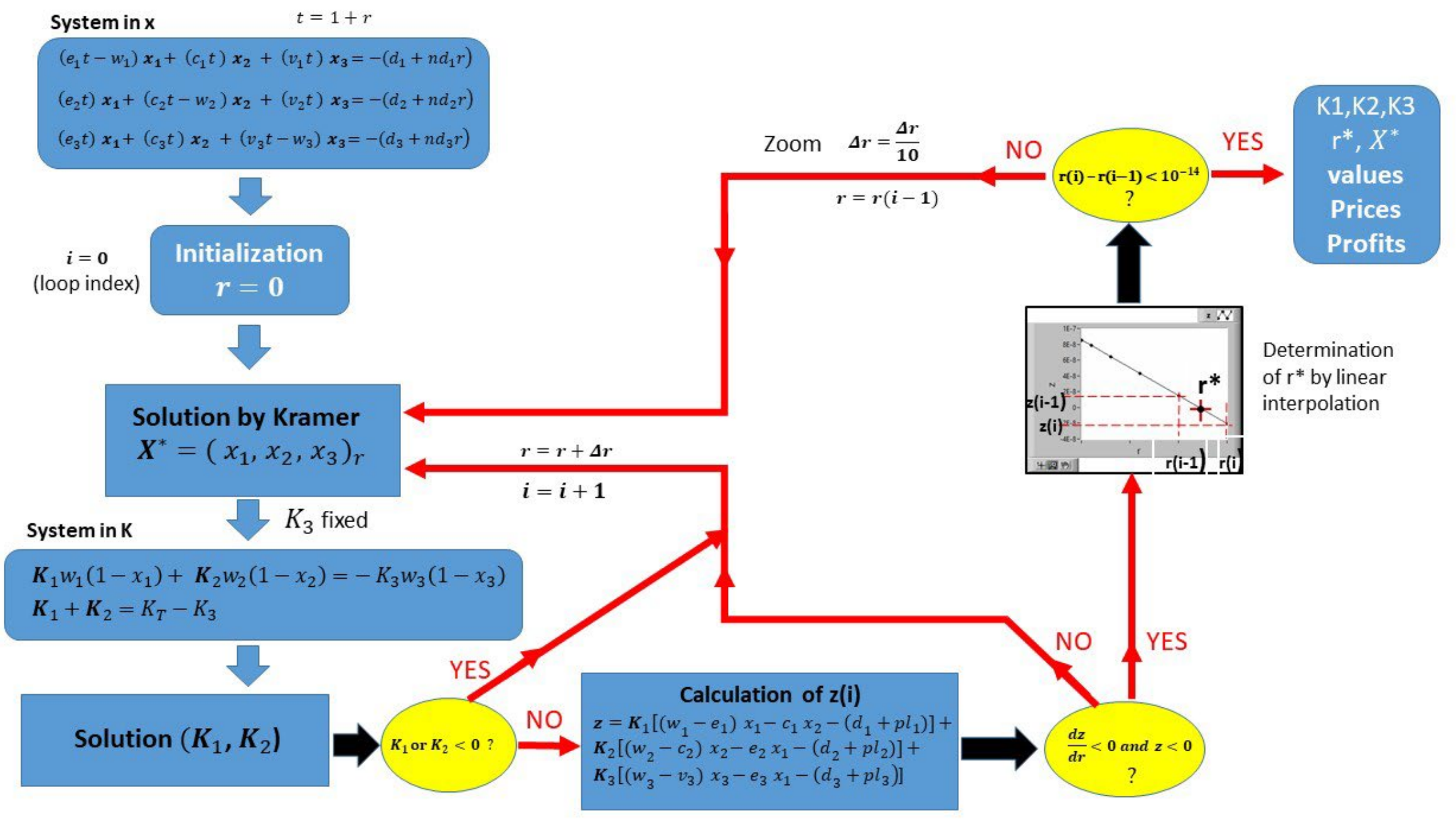

**Figure 11.** Diagram of the algorithm in the three-branch case with fixed capital different from zero. The initial data is a table of values according to (a priori) an arbitrary allocation of capital which defines the socio-technical coefficients.

## Addition of a constraint

We can add a constraint, for example that the production of branch III (V), is equal to the sum of wages plus the total surplus value.

$$K_3(w_3 - pl_3 - v_3) - K_1(pl_1 + v_1) - K_2(pl_2 + v_2) = 0$$

$$D = \begin{vmatrix} \mathrm{w1}(1-\mathrm{x1}) & \mathrm{w2}(1-\mathrm{x2}) & \mathrm{w3}(1-\mathrm{x3}) \\ 1 & 1 & 1 \\ (pl_1 + v_1) & -(pl_2 + v_2) & (w_3 - pl_3 - v_3) \end{vmatrix}$$

If D is nonzero, the solution is unique:

$$K_1 = \frac{\begin{vmatrix} 0 & \mathrm{w2}(1-\mathrm{x2}) & \mathrm{w3}(1-\mathrm{x3}) \\ KT & 1 & 1 \\ 0 & -(pl_2 + v_2) & (w_3 - pl_3 - v_3) \end{vmatrix}}{D}$$

$$K_2 = \frac{\begin{vmatrix} \mathrm{w1}(1-\mathrm{x1}) & 0 & \mathrm{w3}(1-\mathrm{x3}) \\ 1 & KT & 1 \\ (pl_1 + v_1) & 0 & (w_3 - pl_3 - v_3) \end{vmatrix}}{D}$$

$$K_3 = \frac{\begin{vmatrix} \text{w1}(1-\text{x1}) & \text{w2}(1-\text{x2}) & 0 \\ 1 & 1 & KT \\ (pl_1+v_1) & -(pl_2+v_2) & 0 \end{vmatrix}}{D}$$

For each additional branch an additional equation representing a particular constraint can be added to provide a unique solution.

## Case of zero fixed capital

This case requires a different algorithmic processing.

$$w_1 x_1 = (1+r)e_1 x_1 + \ (1+r)c_1 x_2 + (1+r)v_1 x_1$$

$$w_2 x_2 = (1+r)e_2 x_1 + \ (1+r)c_2 x_2 + (1+r)v_2 x_2$$

$$w_1 x_3 = (1+r)e_3 x_1 + \ (1+r)c_3 x_2 + (1+r)v_3 x_3$$

$$(1+r) \ \begin{bmatrix} \mathbf{e_1} & \mathbf{c_1} & \mathbf{v_1} \\ \mathbf{e_2} & \mathbf{c_2} & \mathbf{v_2} \\ \mathbf{e_3} & \mathbf{c_3} & \mathbf{v_3} \end{bmatrix} \begin{pmatrix} x_1 \\ x_2 \\ x_3 \end{pmatrix} = (w_1, w_2, w_3) \begin{pmatrix} x_1 \\ x_2 \\ x_3 \end{pmatrix}$$

$$(1+r) \ \begin{bmatrix} \mathbf{e_1}/w_1 & \mathbf{c_1}/w_1 & \mathbf{v_1}/w_1 \\ \mathbf{e_2}/w_2 & \mathbf{c_2}/w_2 & \mathbf{v_2}/w_2 \\ \mathbf{e_3}/w_3 & \mathbf{c_3}/w_3 & \mathbf{v_3}/w_3 \end{bmatrix} \begin{pmatrix} x_1 \\ x_2 \\ x_3 \end{pmatrix} = \begin{pmatrix} x_1 \\ x_2 \\ x_3 \end{pmatrix}$$

$$\begin{bmatrix} \mathbf{e_1}/w_1 & \mathbf{c_1}/w_1 & \mathbf{v_1}/w_1 \\ \mathbf{e_2}/w_2 & \mathbf{c_2}/w_2 & \mathbf{v_2}/w_2 \\ \mathbf{e_3}/w_3 & \mathbf{c_3}/w_3 & \mathbf{v_3}/w_3 \end{bmatrix} \begin{pmatrix} x_1 \\ x_2 \\ x_3 \end{pmatrix} = \frac{1}{(1+r)} \begin{pmatrix} x_1 \\ x_2 \\ x_3 \end{pmatrix}$$

$$[\boldsymbol{A}] = \begin{bmatrix} \mathbf{e_1}/w_1 & \mathbf{c_1}/w_1 & \mathbf{v_1}/w_1 \\ \mathbf{e_2}/w_2 & \mathbf{c_2}/w_2 & \mathbf{v_2}/w_2 \\ \mathbf{e_3}/w_3 & \mathbf{c_3}/w_3 & \mathbf{v_3}/w_3 \end{bmatrix}$$

The eigenvalue of the matrix is equal to the inverse of (1+r) and thus determines the rate of profit. The rate of profit depends on the values $w_i$ which are greater than 1 for the branches making a surplus value. The rate of profit thus depends on the exploitation rate.

The value of the rate of profit is also calculated as:

$$r = \frac{\text{PL}}{(\text{ E} + \text{C} \ + \ \text{V})} = \frac{\sum_i Ki \ pli}{\sum_i Ki \ (\text{ ei} + \text{ci} \ + \ \text{vi})}$$

Unlike the matrix A, the rate of profit calculated in this way depends on the values of $K_1$, $K_2$, $K_3$. But the matrix does not tell us anything about capital allocation.

It turns out that when the capital allocation meets the solvent social need (demand), the two calculations lead to the same result. The socio-technical coefficients contain the proportions of each of the commodities contained in another. Therefore, they provide information on the solvent social need.

The norm of the eigen vector must have the appropriate value for the fundamental equalities to be complied with. Thus the matrix resolution alone does not allow the transformation vector to be completely determined. There is a particular norm for which a set of values ($K_1$, $K_2$, $K_3$) is compatible with these equalities. For this set, the calculation of the rate of profit coincides with the eigenvalue of the matrix A.

$$[\boldsymbol{A}] \quad \begin{pmatrix} x_1 \\ x_2 \\ x_3 \end{pmatrix} = \frac{1}{(1+r)} \begin{pmatrix} x_1 \\ x_2 \\ x_3 \end{pmatrix}$$

We need to determine the eigen vectors of the matrix $[A]$ to find an $\boldsymbol{X}$ transformation vector of the values into prices.

$$\boldsymbol{X} = \begin{pmatrix} \boldsymbol{x_1} \\ \boldsymbol{x_2} \\ \boldsymbol{x_3} \end{pmatrix}$$

However, the result of the algorithm provides a unit eigenvector.

$$\|\boldsymbol{X}\| = \sqrt{\boldsymbol{x_1}^2 + \boldsymbol{x_2}^2 + \boldsymbol{x_3}^2} = 1$$

We therefore know the relationship between the different prices but not their absolute amounts yet.

The actual transformation vector $\boldsymbol{X}^*$ is such that:

$$\boldsymbol{X}^* = q\,\boldsymbol{X}$$

The looked-for value of $q$ is the value $q^*$ which ensures the compliance with both fundamental equalities.

From this point on, the resolution algorithm is identical to the one used with fixed capital, except that instead of varying r, it is the variable q that is incremented until we determine by linear interpolation the particular value of $q$, $q^*$ which cancels the z-function. A new vector $\boldsymbol{X}^*$ and hence a new allocation $(K_1, K_2, K_3)$ is obtained for each new value of $q$. These $K_i$ values determine at their turn a new value of the variable $r$ used in the eigenvalue equation.

$$z = W_T - \sum_i K_i q(e_i x_1 + c_i x_2 + v_i x_3)$$

with $W_T = \sum_i K_i\, w_i$

The algorithms for processes with and without fixed capital are shown in Figures 11 and 12.

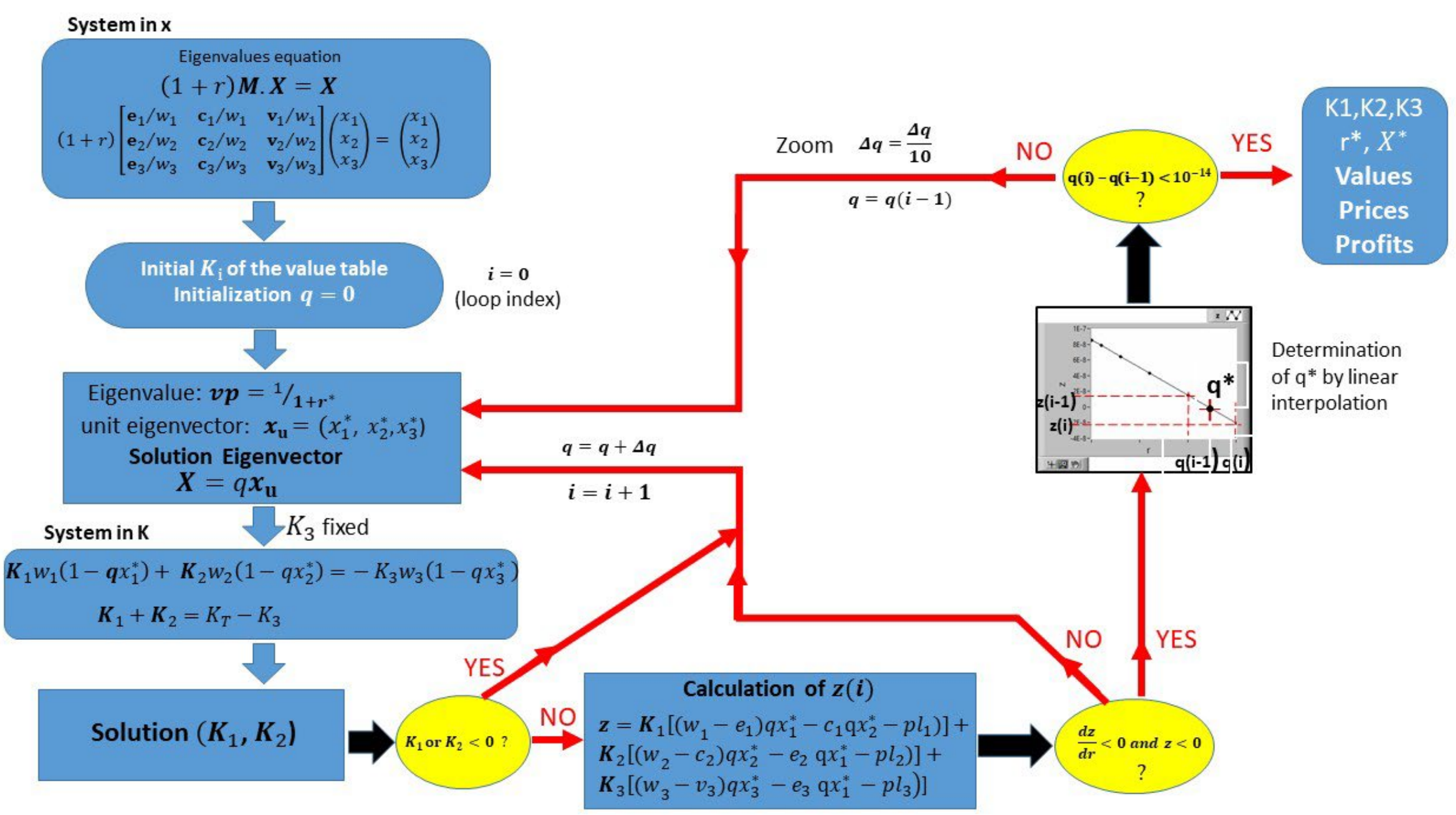

**Figure 12.** Diagram of the algorithm in the three-branch case with zero fixed capital.

## 2. Simulation parameters

### Parameters of simulation E-1-a

For this simulation we use the socio-technical coefficients from the example in Table 10 with zero fixed capital. The total committed capital, (754.147032 m.u.) is constant. The capital of branch 2 with an initial value of 385 m.u. is decremented by 0.0025 m.u. at each iteration. The rates of profit $r_1$ and $r_3$ of branches 1 and 3 are set so that: r1= r2+Δr1 et r3=r2-Δr1, with $r_2$ used as the reference. The capitals of branches 1 and 3 as well as $r_2$ are calculated by the main algorithm in such a way as to comply with the fundamental equalities and the imposed difference in rates of profit. At the first iteration the rate of profit $r_1$ is the largest with $\Delta r1$=+0.001. The factor $\Delta r1$ is then decremented at each iteration by 0.00001. The simulation reaches the point at which all three rates of profit are equal between the 99iem and 100iem iterations. The simulation ends at iteration 120 (on the graph the first iteration starts at zero).

The same simulation was repeated using a rate of profit differential this time at the start of the simulation we have: $\Delta r1 = 0.014$. The factor $\Delta r1$ is then decremented by 0.00025 and there are 72 iterations.

### Parameters of simulation E-1-b

Compared to the socio-technical coefficients of the example in Tables 10, we keep the same depreciation period of the fixed capital with n=10 cycles but we have increased the fixed capital of branch 1 by 100 monetary units (which also makes it possible to have branch 2 with an increasing rate of profit, the conclusions of the simulation not being changed otherwise). The total committed capital is thus 1100 (um), constant. The capital of branch 2 with initial value 385 (um) is decremented by 0.0025 (um) at each iteration, the rates of profit r1 and r3 of branches 1 and 3 are set so that: $r1 = r2 + \Delta r1$ et $r3 = r2 - \Delta r1$, with r2 used as the reference. The capitals of branches 1 and 3 as well as r2 are calculated by the main algorithm in such a way as to comply with the fundamental equalities and the imposed difference in rates of profit. At the first iteration the rate of profit r1 is the largest and the rate of profit r3 is the smallest. The initial value of $\Delta r1$ is 0.001. The factor $\Delta r1$ is then decremented at each iteration by 0.00001. The simulation reaches the point at which all three rates of profit are equal at the 101st iteration.

## ACKNOWLEDGEMENTS

We thank Stefan Neuwirth, Ivan Cohen and Jean-Marie Harribey for their helpful comments on earlier versions of the manuscript.